\documentstyle[12pt,epsf,aasms4]{article}
\def\be{\begin{equation}}
\def\ee{\end{equation}}

\def \D{\displaystyle}

\def\fpj{\hspace{-.7cm}}

\def\thalf{{\textstyle{\frac{1}{2}}}}

\def\ni{\noindent}

\begin{document}
\setlength\baselineskip{15pt}

\title{ EVOLUTION OF PROTONEUTRON STARS }
\author{J.A. PONS$^{1,2}$, S. REDDY$^1$, M. PRAKASH$^1$,
J.M. LATTIMER$^1$, and J.A. MIRALLES$^2$}

\affil{$^1$Department of Physics \& Astronomy\\
SUNY at Stony Brook, Stony Brook, NY 11794-3800, USA  \\
$^2$Departament  d'Astronomia i Astrof\'{\i}sica \\
Universitat de Val\`{e}ncia, 46100 Burjassot, Spain \\ }

\begin{abstract}

We study the thermal and chemical evolution during the
Kelvin-Helmholtz phase of the birth of a neutron star, employing
neutrino opacities that are consistently calculated with the
underlying equation of state (EOS).  Expressions for the diffusion
coefficients appropriate for general relativistic neutrino transport
in the equilibrium diffusion approximation are derived.  The diffusion
coefficients are evaluated using a field-theoretical finite
temperature EOS that includes the possible presence of
hyperons. The variation of the diffusion coefficients is studied as a
function of EOS and compositional parameters.  We
present results from numerical simulations of protoneutron star
cooling for internal stellar properties as well as emitted neutrino
energies and luminosities.  We discuss the influence of the initial
stellar model, the total mass, the underlying EOS, and the addition of
hyperons on the evolution of the protoneutron star and upon the
expected signal in terrestrial detectors.

We find that the differences in predicted luminosities and emitted
neutrino energies do not depend much upon the details of the initial
models or the underlying high-density EOS for early times ($t<10$ s),
provided that opacities are calculated consistently with the EOS.  The
same holds true for models which allow for the presence of hyperons,
except when the initial mass is significantly larger than the maximum
mass for cold, catalyzed matter.  For times larger than about 10
seconds, and prior to the occurrence of neutrino transparency, the
neutrino luminosities decay exponentially with a time constant that is
sensitive to the high-density properties of matter.  We also find the
average emitted neutrino energy increases during the first 5 seconds
of evolution, and then decreases nearly linearly  with time.
In general, increasing the protoneutron star mass
increases the average energy and the luminosity of neutrinos, as well
as the overall evolutionary time scale.  The influence of hyperons or
variations in the dense matter EOS is increasingly
important at later times.  {\it Metastable stars, those with hyperons
which are unstable to collapse upon deleptonization, have relatively
long evolution times, which increase the nearer the mass is to the
maximum mass supported by a cold, deleptonized star.}

\end{abstract}



\newpage

\section{INTRODUCTION}

A protoneutron (PNS) star forms in the aftermath of a successful
supernova explosion as the stellar remnant becomes gravitationally
decoupled from the expanding ejecta.  The neutrinos radiated by the
remnant are important to supernova energetics, and
possibly are essential to supernova shock revival (Wilson 1985,
Burrows, Hayes, \& Fryxell 1995) They are also crucial to supernova
nucleosynthesis, especially the r-process which may occur in the
ejecta (Woosely et al. 1994).  A careful study of the energies and
time scales of neutrino emission in all flavors is also necessary for
the interpretation of the data that will be recorded by terrestrial
neutrino detectors in the event of a supernova occurring in our Galaxy
or within several hundred kiloparsecs.  In principle, observations of
supernova neutrinos may not only allow an assessment of the remnant
mass, but may also discriminate among competing types of equations of
state (EOS) and internal neutron star compositions.

The essential microphysical ingredients that govern the macrophysical evolution
of the PNS in the so-called Kelvin-Helmholtz epoch, during which the remnant
changes from a hot and lepton-rich PNS to a cold and deleptonized neutron star,
are the (EOS) of dense matter and its associated neutrino
opacity.  Among the characteristics of matter that widely vary among EOS models
are their relative compressibilities (important in determining the theoretical
neutron star maximum mass), symmetry energies (important in determining the
typical stellar radius and in the relative $n, p, e$, and $\nu_e$ abundances)
and specific heats (important in determining the local temperatures).  These
characteristics play important roles in determining the matter's composition,
in particular the possible presence of strange components (such as hyperons, a
kaon condensate, or quark matter).  These characteristics also significantly
affect calculated neutrino opacities and diffusion time scales.

Immediately following the core bounce of a massive star, the PNS shrinks from
more than 150 km in radius to less than 20 km as neutrino emission from the
low-density outer few tenths of a solar mass of the star robs this region of
pressure support.  During this same period, from 0.1 s to 0.5 s,
substantial accretion through the supernova's shock also occurs, which adds
mass and contributes significantly to the total neutrino emission.
Nevertheless, both this stage and the much longer Kelvin-Helmholtz evolution
which follows nearly fulfill the condition of hydrostatical equilibrium
(Burrows \& Lattimer 1986; henceforth BL in the text).  The Kelvin-Helmholtz
phase, which can last for several tens of seconds, consists of two major
evolutionary stages.  First, there is a {\it deleptonization stage} in which
the excess trapped electron neutrinos diffuse from the central regions outward
through the star.  The diffusing neutrinos preferentially heat the stellar core
while decreasing the net lepton and proton fractions.  This stage is followed
by an overall {\it cooling stage} in which the entropy in the star steadily
decreases.

Since the work of BL, several authors (Burrows 1988;1990, Wilson \& Mayle 1989,
Suzuki 1989, and Keil \& Janka 1995) have performed numerical simulations of
the cooling of a PNS.  Our objective here is to improve upon these calculations
by incorporating recent developments in the EOS of dense matter and neutrino
interactions, and to perform a systematic study of the effects of varying the
neutron star mass and the characteristics of the EOS.  For this study, we will
use the diffusion approximation for neutrino transport both because it provides
a comparison to earlier works and because it allows us to easily assess how
global characteristics of the neutrino emission such as average energies,
integrated fluxes, and time scales change in response to various input
parameters.  This approach is sufficient to allow us to establish the
connections between the microphysical ingredients and the duration of the
deleptonization and cooling time scales, and to estimate the effects on neutrino
signals.  However, detailed simulations using a multigroup approach and more
exact transport near the stellar surface will be necessary to calculate the
emergent neutrino spectra precisely enough to make detailed predictions for
current and future neutrino detectors.  Such calculations are in progress.

Compared to the earlier works quoted above, several improvements are carried
out in the neutrino transport scheme. These include \\
\ni (1) the development of the
appropriate diffusion equations and a derivation of explicit expressions for
the fluxes in terms of energy, temperature-, density-, and
composition-dependent diffusion coefficients, which can be obtained from the
reaction rates of neutrinos in matter.  We present the calculations of these
coefficients using the results of Reddy, Prakash, \& Lattimer (1998)
(henceforth RPL). \\
\ni (2) the development of a transport scheme that uses the appropriate
energy-averaged diffusion coefficients. \\

The most important improvements in the EOS are that \\
\ni (1) the effects of
finite entropy are calculated without approximation, and \\
\ni (2) the
influence of the possible presence of strangeness in the form of hyperons is
explored consistently.  \\
We employ EOSs with more realistic
compressibilities and high-density symmetry energies and nucleon effective
masses as compared to BL.  Keil \& Janka (1995) (henceforth KJ) considered the
consequences of having hyperons appear in the star, but with an approximate
treatment of the effects of finite entropy and no modification to the opacities
that reflected the presence of hyperons.

Several improvements over existing works are also made both in the calculation
and in the implementation of neutrino opacities.  The opacities are governed by
charged current processes (absorption on baryons) and neutral current processes
(scattering off baryons and electrons) in dense matter.  The most significant
of these improvements are: \\ \ni (1) an exact treatment of the kinematics of
the neutrino-baryon and neutrino-electron processes including the full effects
of Pauli-blocking for arbitrary matter degeneracy, and\\ \ni (2) calculations
of the effects of strong interactions arising from mass and energy shifts in a
multicomponent system (including the presence of hyperons). \\ We have found
that these improvements each tend to increase neutrino mean free paths (RPL),
and, consequently, produce shorter emission time scales relative to the
situation in which they are ignored.

In both BL and KJ, opacities were interpolated between expressions valid for a
mixture of free Fermi gases in the degenerate and nondegenerate regimes,
respectively.  However, RPL demonstrated that interpolation
can result in significant deviations
from the exact phase-space integrals.  RPL also showed
that the exact phase-space integrals are relatively easy to compute, which
obviates the need for interpolations.

The calculations of Suzuki (1989) employed free gas opacities from Bruenn
(1985) which are valid when the matter is essentially non-degenerate and
non-interacting.  However, for the conditions of interest, neither of these
conditions are generally satisfied.  As discussed in RPL, this generally
results in an underestimate of the neutrino mean free paths.

Furthermore, to include the effects of strong interactions, previous workers
employed a constant, {\it i.e.}, density-, temperature-, and
composition-independent, reduction factor multiplying the opacity.  The
effects of degeneracy and strong interactions on the opacities were not
calculated consistently with the assumed underlying EOS.

In this paper, we focus on simulations of constant baryon mass and spherically
symmetric protoneutron stars.  We explicitly ignore effects due to
accretion, rotation, magnetic fields
and convection, which, in conjunction with multidimensional
neutrino transport, may play an important role in the early evolution of
the cores of protoneutron stars.  In many current supernova models
(Burrows 1988, Burrows \& Goshy 1993), most accretion is completed
within 0.5 s.  In the constant baryon mass
models we present here, it is assumed that the initial
protoneutron star has a high-entropy mantle, so that the outer part of the star
is initially at large radius.  We find that it takes a few tenths of a second
for the neutrinos in the high-entropy mantle to leak from the star and for the
mantle to collapse to small radii.  In effect, this largely simulates the
effects of accretion on the protoneutron star's evolution.
We find that a large fraction of the star becomes connectively unstable after a
few seconds.  The omission of convection will likely lead to an underestimate
of neutrino luminosities and average energies at these times, and should speed
up protoneutron star cooling compared to our models.  The inclusion of better
treatments of accretion and the inclusion of convection
will be studied in a later paper.

In \S 2, we develop the neutrino transport equations appropriate for
the PNS problem by using the method of moments to solve the Boltzmann
transport equations.  These are simplified by employing the diffusion
approximation, which results in expressions for the number and energy
fluxes in terms of energy, temperature-, density-, and
composition-dependent diffusion coefficients.  In \S 3, we describe
the EOS employed in this paper, which is a finite-temperature,
field-theoretical model in which the interactions among baryons are
mediated by the exchange of $\sigma$, $\omega$, and $\rho$ mesons.  It
permits us to easily vary the matters' compressibility, symmetry
energy, and baryonic specific heats, as well as allowing the
systematic inclusion of hyperons, a kaon condensate, and/or quark
matter.  In \S 4, the opacities of matter are calculated using this
EOS and the diffusion constants are determined.  Our results for PNS
evolutions are detailed in \S 5, in which we first describe a baseline
simulation of a 1.6 M$_\odot$ baryon mass, nucleonic EOS, PNS.  We
proceed to examine the consequences of varying the entropy and lepton
content of the initial model, the inclusion of hyperons, and
variations in the initial mass and the EOS.  This section also
contains a discussion of the expected luminosities and identification
of connectively unstable zones.  Our conclusions and outlook are
contained in \S 6.


\section{EVOLUTION EQUATIONS}

\subsection{ Neutrino Transport Equations}
The equations that govern the transport of energy and lepton number are
obtained from the Boltzmann equation (BE) for massless particles
given by (Lindquist 1966)
\begin{equation}
p^{\beta}{\left( {\frac {\partial f}{\partial x^{\beta}}} -
{{\Gamma}_{\beta\gamma}^{\alpha}}  p^{\gamma}
{\frac {\partial f}{\partial p^{\alpha}}} \right)} =
{\left( \frac {df}{d\tau} \right)}_{coll}\,,
\end{equation}
where $f$ is the invariant neutrino distribution function, $p^{\alpha}$ are the
components of the neutrino 4-momentum with respect to the coordinate basis,
${\Gamma}^{\alpha}_{\beta \gamma}$ are the Christoffel symbols for the metric.
(For the most part, we will use units setting $\hbar=1$, $c=1$).
The indices $\alpha, \beta,$ and $\gamma$ take
the values $t,r,\theta,$ and $\Phi$.  The collision term on the right hand side
is due to neutrino-matter interactions, and is most easily evaluated in a
frame comoving with the matter.  This entails rewriting the BE in terms of the
4--momentum components of the neutrino with respect to the matter $p^a$ instead
of the components in the coordinate basis.  The comoving basis $\{
{\bf{e}}_a \}$ satisfies ${\bf e}_a \cdot {\bf e}_b = {\eta}_{ab}$, with
${\bf e_0 = u}$ representing the 4--velocity of matter. Using the metric
\begin{equation}
\label{metric}
ds^2=-e^{2\phi}dt^2+e^{2\Lambda}dr^2+r^2d\theta^2+r^2\sin^2\theta \,d\Phi^2\,,
\end{equation}
and taking ${\bf u} = (\gamma e^{-\phi}, \gamma v e^{-\Lambda}, 0 ,0)$,
where $v$ is the fluid velocity and the Lorentz factor
$\gamma = {(1-v^2)}^{-1/2}$, the BE in the comoving basis takes the form
\begin{equation}
p^{b}{\left( {e^{\beta}_b} {\frac {\partial f}{\partial x^{\beta}}} -
{\Gamma}^a_{bc}  p^c {\frac {\partial f}{\partial p^a}} \right)} =
{\left( \frac {df}{d\tau} \right)}_{coll}  ,
\end{equation}
where ${\Gamma}^a_{bc}$
($a,b,c\equiv 0,1,2,3$) are the Ricci rotation coefficients.
The non-zero Ricci coefficients in the spherically symmetric case
are
\begin{eqnarray}
{\Gamma}^1_{00} ={\Gamma}^0_{01}= {e^{-\phi} e^{-\Lambda}}
{\left( \frac{\partial (\gamma e^{\phi})}{\partial r} +
\frac{\partial (\gamma v e^{\Lambda})}{\partial t} \right) }\,, \quad
{\Gamma}^0_{11} = {\Gamma}^1_{10} =
{e^{-\phi} e^{-\Lambda}}
{\left( \frac{\partial (\gamma e^{\Lambda})}{\partial t} +
\frac{\partial (\gamma v e^{\phi} )}{\partial r} \right)}, \nonumber \\
{\Gamma}^2_{20} ={\Gamma}^0_{22}=
{\Gamma}^3_{30} ={\Gamma}^0_{33}=
-v{\Gamma}^1_{22} = v{\Gamma}^2_{21}=
-v{\Gamma}^1_{33} =v{\Gamma}^3_{31}= {{\gamma v e^{-\Lambda} } \over{r}} \,,
{\Gamma}^2_{33} =-{\Gamma}^3_{32}= - \frac{\cot \theta}{r} \,.
\end{eqnarray}
The time $t$ refers to that measured by an observer located at infinity.
In spherical symmetry, the components of the neutrino 4-momentum are
$$ p^a = \left(\omega, \omega{\mu},
\omega{(1-{\mu}^2)}^{1/2}\cos {\Phi}, \omega{(1-{\mu}^2)}^{1/2}\sin
{\Phi}\right) \,, $$
where $\mu$ is the cosine of the angle the neutrino momentum makes with respect
to the radial direction and $\omega$ is the neutrino energy measured by
a comoving observer.
Then, the general relativistic BE in the spherically symmetric case reads
(Thorne 1981)
\begin{eqnarray}
\label{BE}
\omega({e^t_0 + \mu e^t_1}) \frac{\partial f}{\partial t} &+&
\omega({e^r_0 + \mu e^r_1}) \frac{\partial f}{\partial r}
- \omega^2 {\left( {\mu} \Gamma^1_{00} + {\mu}^2 \Gamma^1_{10} +
{(1- {\mu}^2)} \Gamma^2_{20} \right)} \frac{\partial f}{\partial \omega}
\nonumber \\
&-& \omega {(1- {\mu}^2)} {\left( \Gamma^1_{00}
+ \Gamma^1_{22} + {\mu} \Gamma^1_{10}
- {\mu} \Gamma^2_{20} \right)} \frac{\partial f}{\partial \mu}
= {\left( \frac {df}{d \tau} \right)}_{coll} \,.
\end{eqnarray}

A method often used to simplify the BE is to work with its angular
moments (Thorne 1981). This is done by applying the operator
$$\frac{1}{2} \int_{-1}^{+1} d{\mu}\, {\mu}^{i} \,, \quad i=0,1,2,\cdots $$
to  equation (\ref{BE}).
Defining the $i^{\rm th}$ moment of the distribution function by
$$M_{i}=\frac{1}{2} \int_{-1}^{+1} d{\mu}\, {\mu}^{i} f \,,$$
the equations corresponding to the first two moments ($i=0$ and 1) are:
\begin{eqnarray}
\label{mom0}
{\omega} \left({e^t_0} \frac{\partial M_0}{\partial t} \right.
&+&  \left. {e^r_0} \frac{\partial M_0}{\partial r}
+  {e^t_1} \frac{\partial M_1}{\partial t}
+  {e^r_1} \frac{\partial M_1}{\partial r} \right)
\nonumber \\
&-& {\omega}^2 {\left( \Gamma^1_{00} \frac{\partial M_1}{\partial {\omega}}
+ {( \Gamma^1_{10}-\Gamma^2_{20} )} \frac{\partial M_2}{\partial {\omega}}
+ \Gamma^2_{20} \frac{\partial M_0}{\partial {\omega}} \right)}
\nonumber \\
&+& {\omega} \left( (\Gamma^1_{10} - \Gamma^2_{20})(M_0 - 3 M_2)
-2 (\Gamma^1_{00} + \Gamma^1_{22}) M_1 \right)
= Q_0
\end{eqnarray}
\begin{eqnarray}
\label{mom1}
{\omega} \left( {e^t_0} \frac{\partial M_1}{\partial t} \right.
&+&  \left. {e^r_0} \frac{\partial M_1}{\partial r}
+  {e^t_1} \frac{\partial M_2}{\partial t}
+  {e^r_1} \frac{\partial M_2}{\partial r} \right)
\nonumber \\
&-& {\omega}^2 {\left( \Gamma^1_{00} \frac{\partial M_2}{\partial {\omega}}
+ {( \Gamma^1_{10}-\Gamma^2_{20} )} \frac{\partial M_3}{\partial {\omega}}
+ \Gamma^2_{20} \frac{\partial M_1}{\partial {\omega}} \right)}
\nonumber \\
&+& {\omega} \left( 2 (\Gamma^1_{10} - \Gamma^2_{20}) (M_1-2 M_3)
+ (\Gamma^1_{00} + \Gamma^1_{22})(M_0 - 3 M_2) \right)
= Q_1 \,,
\end{eqnarray}
where
\begin{equation}
\label{q0}
Q_0 = \frac{1}{2} \int_{-1}^{+1} d{\mu}\,
{\left(\frac {df}{d \tau}\right)}_{coll} \, \quad {\rm and} \quad
\label{q1}
Q_1 = \frac{1}{2} \int_{-1}^{+1}d{\mu}\,
\mu {\left(\frac {df}{d \tau}\right)}_{coll} \,.
\end{equation}
Multiplying the $0^{\rm th}$-moment equation (\ref{mom0})
successively by  $(\omega/2\pi^2)$
and $(\omega^2/2 \pi^2)$, and integrating over the neutrino energy, one
obtains the energy-averaged neutrino number and energy transport equations:
\begin{eqnarray}
\label{vnumber}
{e^t_0} \frac{\partial N_{\nu}}{\partial t}
&+&  {e^r_0} \frac{\partial N_{\nu}}{\partial r}
+  {e^t_1} \frac{\partial F_{\nu}}{\partial t}
+  {e^r_1} \frac{\partial F_{\nu}}{\partial r}
\nonumber \\
&+& (\Gamma^1_{10}+ 2 \Gamma^2_{20}) N_{\nu}
+ (\Gamma^1_{00}- 2 \Gamma^1_{22})  F_{\nu} = S_N\,,
\end{eqnarray}
\begin{eqnarray}
\label{enumber}
{e^t_0} \frac{\partial J_{\nu}}{\partial t}
&+&  {e^r_0} \frac{\partial J_{\nu}}{\partial r}
+  {e^t_1} \frac{\partial H_{\nu}}{\partial t}
+  {e^r_1} \frac{\partial H_{\nu}}{\partial r}
\nonumber \\
&+& (\Gamma^1_{10} + 3 \Gamma^2_{20})  J_{\nu}
+ (2\Gamma^1_{00} - 2\Gamma^1_{22})  H_{\nu}
+ (\Gamma^1_{10} - \Gamma^2_{20})  P_{\nu} = S_E \,.
\end{eqnarray}
\noindent Above,
\begin{equation}
\label{defs1}
N_{\nu}= \int_0^{\infty} \frac{d\omega}{2 \pi^2} \, M_0 \omega^2 \,, \quad
F_{\nu}= \int_0^{\infty} \frac{d\omega}{2 \pi^2} \, M_1 \omega^2 \,, \quad
S_N= \int_0^{\infty} \frac{d\omega}{2 \pi^2} \, Q_0 \omega \nonumber
\end{equation}
\begin{equation}
\label{defs2}
J_{\nu}= \int_0^{\infty} \frac{d\omega}{2 \pi^2} \, M_0 \omega^3 \,, \quad
H_{\nu}= \int_0^{\infty} \frac{d\omega}{2 \pi^2} \, M_1 \omega^3 \,, \quad
P_{\nu}= \int_0^{\infty} \frac{d\omega}{2 \pi^2} \, M_2 \omega^3 \,, \quad
S_E= \int_0^{\infty} \frac{d\omega}{2 \pi^2} \,
Q_0 \omega^2 \,. \nonumber
\end{equation}
$N_{\nu}$, $F_{\nu}$, and $S_N$ are the number density, number flux and number
source term, respectively, while $J_{\nu}$, $H_{\nu}$, $P_{\nu}$, and $S_E$ are
the neutrino energy density, energy flux, pressure, and the energy
source term, respectively.

For spherically symmetric situations,  equations (\ref{vnumber}) and
(\ref{enumber}) are the transport equations  which contain full general
relativistic effects including fluid velocity terms. These equations must be
solved together with the hydrodynamical equations to study dynamical problems
such as core collapse supernova. For the PNS problem, however, fluid velocities
are small enough so that hydrostatic equilibrium is nearly fulfilled and the
evolution is quasi-static.

Substituting $\Gamma^a_{bc}$ in equations (\ref{vnumber}) and
(\ref{enumber}), and
utilizing  the continuity equation with the assumption of a quasi-static
evolution (that is, neglecting terms of order $v$),  we arrive at the
well-known neutrino transport equations (BL) in a stationary metric
in the form used so far in protoneutron star evolution:
\be
\label{number}
\frac{\partial ({N_{\nu}/n_B})}{\partial t} +
 {\frac{\partial (e^{\phi} 4 \pi r^2 F_{\nu})}{\partial a}}
= e^\phi \frac{S_N}{n_B}
\ee
\be
\label{energy}
\frac{\partial ({J_{\nu}/n_B})}{\partial t} + P_{\nu} \frac{\partial ({1/n_B})}{\partial t} +
e^{-\phi} {\frac{\partial (e^{2 \phi} 4 \pi r^2 H_{\nu})}{\partial a}}
= e^\phi \frac{S_E}{n_B} \,,
\ee
where $n_B$ is the baryon number density and $a$ is the enclosed baryon number
inside a sphere of radius $r$.


\subsection{The Equilibrium Diffusion Approximation  }

At high density and for temperatures above several MeV, the source terms
$Q_0$ and $Q_1$ in   equations (\ref{mom0}) and (\ref{mom1})
are sufficiently strong to ensure that neutrinos are in thermal and
chemical equilibrium with the ambient matter. Thus, the neutrino
distribution function in these regions is both nearly Fermi-Dirac and
isotropic.  We can approximate the distribution function as an expansion in
terms of Legendre polynomials
to $O(\mu)$, which is known as the diffusion approximation.  Explicitly,
\begin{equation}
f(\omega,\mu)= f_0(\omega) +  \mu f_1(\omega) \,, \quad
f_0 = [1+e^{\left(\frac{\omega-\mu_\nu}{kT}\right)}]^{-1} \,,
\end{equation}
where $f_0$ is the Fermi--Dirac distribution function at equilibrium
($T=T_{mat}$, $\mu_{\nu}=\mu_{\nu}^{eq}$), with
$\omega$ and $\mu_\nu$ being the neutrino energy and chemical potential,
respectively.

The main goal is to obtain a relation for $f_1$ in terms of $f_0$.
In this approximation, the
moments $M_i$ of $f$ which appear in equations (\ref{mom0})
and (\ref{mom1}) are
\begin{equation}
M_0= f_0 \,,\,\, M_1= \frac{1}{3}f_1 \,,\,\,
M_2= \frac{1}{3}f_0 \,,\,\, \quad {\rm and} \quad
M_3= \frac{1}{5}f_1 \,.
\end{equation}
We substitute these into equation (\ref{mom1}) and again neglect all
fluid velocity
dependent terms (i.e., ${e^r_0},{e^t_1}$, and $\Gamma^2_{20})$. Furthermore,
since the transport is driven by slowly varying spatial gradients in
temperature and chemical potential, the time derivative of $f_1$
has also been neglected.  One then finds
\begin{equation}
\label{mom1s}
{e^{-\Lambda}} \left( \frac{\partial f_0}{\partial r}
- {\omega}{\frac{\partial \phi}{\partial r}}
\frac{\partial f_0}{\partial \omega} \right) = 3 \frac{Q_1}{\omega}    \,.
\end{equation}

The interaction between neutrinos and matter is dominated by scattering (from
both
baryons and leptons), absorption (on baryons) and emission (due to electron
capture) processes; thus the collision term $Q_1$ is given by
\begin{equation}
\label{coll}
{\left(\frac{df}{d\tau}\right)}_{coll} =
\omega \left( j_a(1-f) - \frac{f}{\lambda_a} +
j_s(1-f) - \frac{f}{\lambda_s} \right) \,,
\end{equation}
\noindent
where $j_a$ is the emissivity, $\lambda_a$ is the absorptivity, and $j_s$ and
$\lambda_s$ represent the scattering contributions to the source term.  These
are
\begin{equation}
\label{scat1}
j_s = \frac{1}{{(2 \pi)}^3} \int_{0}^{\infty}
d\omega'\, \omega'^{2} \int_{-1}^1 d\mu' \int_0^{2\pi} d\Phi  \,
f(\omega',\mu') R^{in}_s(\omega,\omega', \cos\theta)\,,
\end{equation}
\begin{equation}
\label{scat2}
\frac{1}{\lambda_s} = \frac{1}{{(2 \pi)}^3} \int_{0}^{\infty}
d\omega'\, \omega'^{2} \int_{-1}^1 d\mu' \int_0^{2\pi} d\Phi  \,
[1-f(\omega',\mu')] R^{out}_s(\omega,\omega', \cos\theta) \,,
\end{equation}
where $\theta$ is the angle between the ingoing and outgoing neutrino.  From
the reciprocity relations between a reaction rate and the rate of its inverse
process (the principle of detailed balance), the emissivities and
absorptivities ($R^{in}_s$ and $R^{out}_s$ in the scattering kernels) are
related according to
\begin{equation}
\frac{1}{\lambda_a(\omega)}=e^{\beta(\omega-\mu_{\nu}^{eq})} j_a(\omega)
\quad {\rm and} \quad {R^{in}_s}=e^{\beta(\omega'-\omega)} R^{out}_s \,.
\end{equation}
The different elementary processes which contribute to equation (\ref{coll}) are
described in \S4.

To arrive at the final form for the diffusion equation and the appropriate
definitions for the diffusion constants we expand the scattering kernel in
terms of its Legendre moments:
\begin{equation}
R^{out}_l = \int_{-1}^1 d\cos{\theta}\, P_l (\cos\theta)
R^{out}_s(\omega, \omega', \cos{\theta} ) \,.
\end{equation}
Then, one finds
\begin{equation}
Q_0 = \omega
\left( j_a(1-f_0(\omega)) - \frac{f_0(\omega)}{\lambda_a}
+ \kappa^s_0 \right)
\quad \,, \quad
\label{q1eq}
Q_1 = -\omega\frac{f_1(\omega)}{3}
\left( j_a + \frac{1}{\lambda_a} + \kappa^s_1 \right)\,,
\end{equation}
where the scattering contributions to $Q_0$ and $Q_1$ are given by
\begin{eqnarray}
\kappa^s_0 = \frac{1}{{(2 \pi)}^2} \int_0^{\infty} d\omega'\,{\omega'}^2
\left\{
\left[ f_0(\omega') e^{\beta(\omega'-\omega)} - f_0(\omega)
+ (1-e^{\beta(\omega'-\omega)}) f_0(\omega') f_0(\omega) \right] R^{out}_0
\right. \nonumber \\
\left.
+ (1-e^{\beta(\omega'-\omega)}) \frac{f_1(\omega') f_1(\omega)}{9} R^{out}_1
\right\}\,,
\end{eqnarray}
\begin{eqnarray}
\kappa^s_1 = \frac{1}{{(2 \pi)}^2} \int_0^{\infty} d\omega'\,{\omega'}^2
\left\{
\left[ 1 - f_0(\omega')(1-e^{\beta(\omega'-\omega)}) \right] R^{out}_0
\right. \nonumber \\
\left.
- \frac{f_1(\omega')}{f_1(\omega)} \left[ e^{\beta(\omega'-\omega)}
+f_0(\omega)(1-e^{\beta(\omega'-\omega)}) \right] R^{out}_1
\right\} \,.
\end{eqnarray}
In the case that the scattering is isoenergetic, $\kappa_1^s$ may be further
simplified:
\begin{eqnarray}
\kappa^s_1 = \frac{\omega^2}{{(2 \pi)}^2}
\int_{-1}^1 d\cos{\theta}\, (1-\cos{\theta})  \,
R^{out}(\omega, \omega, \cos{\theta}) \,,
\end{eqnarray}
which is the well-known {\it transport opacity} used in many applications.
If we substitute the above expression for $Q_1$ into
equation (\ref{mom1s}), we obtain the relation between
$f_0$ and $f_1$ that we are looking for:
\begin{equation}
\label{f1}
f_1 = - D(\omega)
\left[ {e^{-\Lambda}} \frac{\partial f_0}{\partial r}
- {\omega} {e^{-\Lambda} \frac{\partial \phi}{\partial r}}
{\frac{\partial f_0}{\partial \omega}} \right] \,.
\end{equation}
\noindent
The explicit form of the diffusion coefficient $D$ appearing above is given by
\begin{equation}
D(\omega) = {\left( j+\frac{1}{\lambda_a}+\kappa^s_1 \right)}^{-1} \,.
\end{equation}

Because in most situations of interest to us neutrino scattering is dominated
by scattering off baryons, the scattering processes are nearly elastic and the
implicit dependence on $f_1$ in the diffusion coefficient may be safely
ignored. However, should electron scattering or the inelastic
contribution to baryon scattering become important, one is forced to solve a
more complex integro-differential equation than the familiar Fick's law.

To obtain explicitly the equations that determine the lepton number and
energy fluxes we note that
\begin{equation}
\frac{\partial f_0}{\partial r} =
-\left( T \frac{\partial \eta_{\nu}}{\partial r} + \frac{\omega}{T}
\frac{\partial T}{\partial r}\right)
\frac{\partial f_0}{\partial \omega}~,
\end{equation}
where $ \eta_{\nu}=\mu_{\nu}/T $ is the neutrino degeneracy parameter.
Substituting the above relation in equation (\ref{f1}), we obtain
\begin{equation}
f_1 = - D(\omega) e^{-\Lambda} \left[
T \frac{\partial \eta}{\partial r}
+ \frac{\omega}{T e^{\phi}} \frac{\partial (T e^{\phi})}{\partial r} \right]
\left(- \, \frac{\partial f_0}{\partial \omega} \right)\,.
\end{equation}

In line with the definition of the fluxes given before in
equations (\ref{defs1}) and (\ref{defs2}), the energy-integrated
lepton and energy fluxes are
\begin{eqnarray}
F_{\nu}&=&- \, \frac{e^{-\Lambda} e^{-\phi}T^2}{6 \pi^2}
\left[ D_3 \frac{\partial (T e^{\phi})}{\partial r} +
(T e^{\phi}) D_2 \frac{\partial \eta}{\partial r}  \right] \nonumber \\
H_{\nu}&=&- \, \frac{e^{-\Lambda} e^{-\phi}T^3}{6 \pi^2}
\left[ D_4 \frac{\partial (T e^{\phi})}{\partial r} +
(T e^{\phi}) D_3 \frac{\partial \eta}{\partial r}  \right] \,.
\label{fluxes}
\end{eqnarray}
The coefficients $D_2$, $D_3$, and $D_4$ are related to the energy-dependent
diffusion coefficient $D(\omega)$ through
\be
D_n = \int_0^\infty dx~x^n D(\omega)f_0(\omega)(1-f_0(\omega))~,
\label{d2d3}
\ee where $x=\omega/T$.  Note that these diffusion coefficients arise
naturally from the transport equations. They depend only on the
microphysics of the neutrino-matter interactions (see \S4 for
details). The fluxes appearing in the above equations are for one
particle species. To include all six neutrino types, we redefine the
diffusion coefficients in equation (\ref{fluxes}):
\begin{eqnarray}
D_2=D_2^{\nu_e}+D_2^{\bar{\nu}_e}\,, \quad
D_3=D_3^{\nu_e}-D_3^{\bar{\nu}_e}\,, \quad
D_3=D_4^{\nu_e}+D_4^{\bar{\nu}_e}+4 D_4^{\nu_\mu}\,.
\end{eqnarray}
\subsection{Numerical Notes}
We generally assume in this paper that the total baryon number in each
simulation is conserved, that is, we do not include the effects of accretion.
These effects will be incorporated in future work.  The enclosed baryon number
$a$ is then a convenient Lagrangian variable.  The equations to be solved
split naturally into a transport part, which has a strong time
dependence, and a structure part, in which evolution is much slower.
We choose to solve these two parts separately in an iterative fashion which is
equivalent to solving them together.

Explicitly, the structure equations are
\begin{equation}
{{\partial r}\over{\partial a}} = \frac{1}{4 \pi r^2 n_B e^{\Lambda}}
\,, \quad
\frac{\partial m}{\partial a} = \frac{\rho}{n_B e^{\Lambda}}
\label{a-struc}
\end{equation}
\begin{eqnarray}
\frac{\partial \phi}{\partial a} = \frac{e^{\Lambda}}{4\pi r^4 n_B}
{\left( m + 4\pi r^3 P \right)} \,, \quad
\frac{\partial P}{\partial a} = - (\rho + P)
\frac{e^{\Lambda}}{4\pi r^4 n_B}
{\left( m + 4\pi r^3 P \right)} \,.
\label{b-struc}
\end{eqnarray}
The quantities $m$, $\rho$, and $P$ include
contributions from the leptons.  To
obtain the equations employed in the transport, we combine
equation (\ref{number}) and the corresponding equation for the electron
fraction
\be
\frac{\partial Y_e}{\partial t}=-e^\phi\frac{S_N}{n_B} \ \ee to obtain \be
\label{a-number}
\frac{\partial Y_L}{\partial t} +
e^{-\phi} {\frac{\partial (e^{\phi} 4 \pi r^2 F_{\nu})}{\partial a}}
= 0 \,.
\ee
Similarly, we combine equation (\ref{energy}) with the matter energy equation
\be
\frac {dU}{dt} + P \frac{d({1/n_B})}{dt} = - e^\phi \frac{S_E}{n_B} \,,
\ee
where $U$ is the specific internal energy
and use the first law of thermodynamics to obtain
\be
e^\phi T\frac{\partial s}{\partial t} + e^\phi\mu_{\nu} \frac{\partial Y_L}{\partial t}
+ e^{-\phi} {\frac{\partial e^{2 \phi} 4 \pi r^2 H_{\nu}}{\partial a}}
= 0 \,.
\label{a-energy}
\ee
In regions of small optical depth, diffusion codes can
predict fluxes that exceed the black-body limit. To prevent this, we correct
the fluxes by a flux limiter $3\Lambda (\xi)$, where $\xi_F =F_{\nu}/N_{\nu}$
and $\xi_H=H_{\nu}/J_{\nu}$ for the number and energy fluxes, respectively.
For the functional form of $\Lambda(\xi)$, we tested both the Levermore \&
Pomraning (1981) and the Bowers \& Wilson (1982) flux-limiters. Only the
outermost few mass shells are affected by this choice, and we have not found it
to make an appreciable difference in the overall evolution or in the mean
neutrino energies or luminosities. For the PNS cooling
problem, as long as most of the transport is nearly diffusive, small errors in
the transport
near the surface cannot affect the global evolution. This, however, is not the
case in supernova simulations, in which the flux-limiter affects more of the
computational grid  and an appropriate treatment becomes a crucial aspect of
transport.

The structure equations (\ref{a-struc}) and (\ref{b-struc}) satisfy the
boundary conditions
\begin{eqnarray}
r(a=0) &=& 0;~~~~m(a=0)=0,
\nonumber \\
\phi(a=a_s) &=& \frac{1}{2} \log
\left[1-\frac{2 m(a=a_s)}{r(a=a_s)}\right];~~~~P(a=a_s)=P_s  \,,
\end{eqnarray}
where $a_s$ is the total number of baryons in the star,
which remains fixed during the evolution, and $P_s$  is
the surface pressure, which is chosen to be very small.
We have checked that our results are insensitive to the
exact value of $P_s$.
The transport equations (\ref{a-number}) and (\ref{a-energy}) also require
boundary conditions. Both the energy and number fluxes must
be zero at the center (this is equivalent to imposing null
gradients of the temperature and chemical potential).  At the surface, we
may write
\begin{eqnarray}
\xi_F = \alpha_n \quad {\rm and} \quad \xi_H = \alpha_e \,,
\end{eqnarray}
where $\alpha_n$ and $\alpha_e$ are constants chosen to each be
in the range $0.4-0.6$.    The results for overall neutrino number and energy
fluxes are insensitive to the exact choice of these constants,
due to the fact that the fluxes saturate near the surface and
the values of  $Y_{\nu}$ in the surface region adjusts
itself to compensate for small changes in $\alpha_n$ and $\alpha_e$.

An initial model is specified by choosing entropy $s$ and lepton fraction $Y_L$
profiles in $a$ and relaxing the star to hydrostatic equilibrium, which
determines values for the radius $r$, gravitational mass $m$, gravitational
potential $\phi$, and pressure $P$.  In general, this initial model will not be
in radiative equilibrium; however, we found that radiative equilibrium is
established within about 5 ms, a time too short to be important for the
protoneutron star evolution.

The evolution is begun by taking an initial transport time step, determining
the quantities $Y_{L}^\prime(a)$ and $s^\prime(a)$ as first estimates for the
lepton fraction and entropy at the end of the time step.  In this step, the
radius, gravitational mass and pressure profiles are kept fixed. Now using
these values of lepton number $Y_{L}^\prime(a)$ and entropy $s^\prime(a)$, we
solve the structural equations relaxing the star into hydrostatic equilibrium,
which determines $r^\prime(a)$, $m^\prime(a)$, $\phi^\prime(a)$, and
$P^\prime(a)$.   Using these primed values for $r, m, \phi$, and $P$, but the
original values of $s(a)$ and $Y_L(a)$, we retake the transport time step,
thereby determining updated  estimates, $s^{\prime\prime}(a)$ and
$Y_L^{\prime\prime}(a)$.  This is followed by another hydrostatic equilibrium
relaxation which determines $r^{\prime\prime}(a)$, $m^{\prime\prime}(a)$,
$\phi^{\prime\prime}(a)$, and $P^{\prime\prime}(a)$. This iterative procedure
is found to converge rapidly, typically in 3-4 iterations.  This method is
equivalent to using a full implicit discretization scheme to solve the
structure and transport equations simultaneously.

\subsection{Neutrino Luminosities}
The ultimate objective of an observation of neutrinos from a Galactic
supernova will be to deduce the properties of the PNS, including its
mass, composition, and the EOS.  The prediction of neutrino signals,
and the determination of how well the mass and properties of the EOS
can be estimated from them, is an important goal of our work.  The
signal depends both upon the neutrino fluxes and energy spectra.  Its
accurate determination requires the use of better transport than the
diffusive, approach used in this paper.  Thus, we have
limited our predictions to those utilizing information only from the
total neutrino luminostiy and the average emitted energy of all
neutrinos.  However, our neutrino transport calculations tracked each
flavor of neutrino and the diffusion constants were defined
appropriately.  In fact, detailed transport calculations show that the
emitted spectra are, to a fair approximation, Fermi-Dirac with a small
effective degeneracy parameter for all flavors of neutrinos (Mayle,
Wilson, \& Schramm 1987).  Furthermore, the presence of a small
degeneracy parameter gives negligible corrections to a detector's count
rate (Lattimer \& Yahil 1989).  Thus, a fair representation of the
signal in a terrestrial detector can be found from the time dependence
of the total neutrino luminosity and average neutrino energy together
with an assumption of a Fermi-Dirac spectrum with zero chemical potential.

The total neutrino luminosity is
the time rate of change of the star's gravitational mass, and is
therefore primarily a global property of the evolution.  This
luminosity, due to energy conservation, must also equal
\begin{equation}
L_\nu=e^{2 \phi} 4 \pi r^2 H_{\nu}\,
\end{equation}
at the edge of the star.  We have checked this test of energy
conservation for our code and find that it is valid to better than 1\%
at all times greater than about 5 ms, when the star comes into
radiative equilibrium.  For times greater than about 5 ms, initial
transients become quite small and the predicted luminosities should be
relatively accurate compared to full transport simulation.
We estimate the average
energy of neutrinos from the temperature $T_\nu$ of the matter at the
neutrinosphere $R_\nu$, defined to be the location in the star where
the flux factor $\xi_H=0.25$. However, since the spectrum may not be
Fermi-Dirac at the neutrinosphere, our transport scheme cannot give a
very precise value for the average energy.  Comparing with better
transport in this region, we find that the average energy
$<E_\nu>\approx3T_\nu$, where $T_\nu$ is a mass average in the
outermost zone.  Because it is a globally determined quantity, the
luminosity $L_\nu$ is necessarily more accurately determined than
either $R_\nu$ or $T_\nu$.  In lieu of more complete transport
results, we will place more weight on $L_\nu$ in the estimate of
detector signals in \S 5.7.

\section{THE EQUATION OF STATE }
The role of the EOS in determining the
structure of neutron stars has been studied extensively (see, for example,
Prakash et al.
1997 \& references therein). The masses and radii of neutron stars
depend upon the matters' compressibility, the composition of matter at
high density, and the nuclear symmetry energy.
In the PNS problem, the finite temperature aspects of the
EOS also play an important role. During the early evolution the entropy in the
central regions is moderately high $s\sim 1-2$, which correspond to
temperatures in the range $T=20-50$ MeV. In this work we employ
a finite temperature  field-theoretical model
where the interactions between baryons are mediated by the exchange of
$\sigma,\omega$, and $\rho$ mesons.  Including the leptonic contributions,
the total Lagrangian density is
given by~(Serot \& Walecka 1992),
\begin{eqnarray}
L &=& L_\ell + L_H \nonumber \\
  &=& \sum_{l}\overline{l}(-i\gamma^{\mu}\partial_{\mu}-m_l)l
   + \sum_{i} \overline{B_i}(-i\gamma^{\mu}\partial_{\mu}-g_{\omega i}
\gamma^{\mu}\omega_\mu
-g_{\rho i}\gamma^{\mu}{\bf{b}}_{\mu}\cdot{\bf t}-M_i+g_{\sigma i}\sigma)B_i
\nonumber \\
&-& \frac{1}{4}W_{\mu\nu}W^{\mu\nu}+\frac{1}{2}m_{\omega}^2\omega_{\mu}\omega^
{\mu} - \frac{1}{4}{\bf B_{\mu\nu}}{\bf
B^{\mu\nu}}+\frac{1}{2}m_{\rho}^2 b_{\mu}b^{\mu}
+ \frac{1}{2}\partial_{\mu}\sigma\partial^{\mu}\sigma -\frac{1}{2}
m_{\sigma}^2\sigma^2-U(\sigma)
\end{eqnarray}
Here, the
leptons, $l = e^-$ and $\mu^-$, and
$B$ are the Dirac spinors for baryons and $\bf t$ is the isospin
operator. The sums include baryons $i=n,p,\Lambda,\Sigma$, and $\Xi$.
The field strength tensors for the $\omega$ and
$\rho$ mesons are $W_{\mu\nu} = \partial_\mu\omega_\nu-\partial_\nu\omega_\mu$
and ${\bf B}_{\mu\nu} =  \partial_\mu{\bf b}_\nu-\partial_\nu{\bf b}_\mu$,
respectively.  The potential $U(\sigma)$ represents the self-interactions of
the scalar field and is taken to be of the form
\begin{eqnarray}
U(\sigma) =  \frac{1}{3}bM_n(g_{\sigma N}\sigma)^3 + \frac{1}{4}c(g_{\sigma
N}\sigma)^4\,.
\end{eqnarray}
Electrons and muons are included in the model as noninteracting particles,
since their interactions give small contributions compared to those of
their free Fermi gas parts.

In the mean field approximation, the partition function (denoted by $Z_H$) for
the hadronic degrees of freedom is given by
\begin{eqnarray}
\ln Z_H =\beta V\left[\thalf m_{\omega}^2\omega_0^2+\thalf
m_{\rho}^2b_0^2 - \thalf m_{\sigma}^2\sigma^2-U(\sigma)\right]
+ 2V\sum_i \int\frac{d^3k}{(2\pi)^3} \,\ln\left(1+{\rm e}
^{-\beta(E^*_i-\nu_i)}\right)\,,
\label{hyp1}
\end{eqnarray}
where $\beta = T^{-1}$ and $V$ is the volume (we take $k_B=1$).
The contribution of antibaryons is
not significant for the thermodynamics of interest here, but is
straightforwardly
included in equation (\ref{hyp1}). Here, the effective baryon masses
$M_i^*=M_i-g_{\sigma i}\sigma$ and
$E^*_i=\sqrt{k^2+M^{*2}_i}$. The chemical potentials are given by
\begin{equation}
\mu_i = \nu_i+g_{\omega i}\omega_0+g_{\rho i}t_{3i}b_0\;,\label{hyp2}
\end{equation}
where $t_{3i}$ is the third component of isospin for the baryon. Note
that particles with $t_{3i}=0$, such as the $\Lambda$ and $\Sigma^0$
do not couple to the $\rho$.  The effective chemical potential
$\nu_i$ sets the scale of the temperature dependence of the thermodynamical
functions.

Using $Z_H$, the thermodynamic quantities can be obtained in the standard way.
The pressure $P_H=TV^{-1}\ln Z_H$, the number density for species $B$, and the
energy density $\varepsilon_H$ are given by
\begin{eqnarray}
n_i~&=&~2\int\frac{d^3k}{(2\pi)^3}
\left({\rm e}^{\beta(E^*_i-\nu_i)}+1\right)^{\!-1}\;,\nonumber\\
\varepsilon_H~&=&~\thalf m_{\sigma}^2\sigma^2+U(\sigma)+
\thalf m_{\omega}^2\omega_0^2+\thalf m_{\rho}^2 b_0^2
+2\sum_i
\int\frac{d^3k}{(2\pi)^3}
E_i^*\left({\rm e}^{\beta(E^*_i-\nu_i)}+1\right)^{\!-1}\;.\label{hyp4}
\end{eqnarray}
The entropy density is then given by
$s_H=\beta(\varepsilon_H+P_H-\sum_i\mu_i n_i)$.
The meson fields are obtained by extremization of the partition function,
which yields the equations
\begin{eqnarray}
&&\fpj m_{\omega}^2\omega_0=\sum_i g_{\omega i} n_i\quad;\quad
m_{\rho}^2b_0=\sum_i g_{\rho i}t_{3B}n_i\;,\nonumber\\
&&\fpj m_{\sigma}^2\sigma=-\frac{dU(\sigma)}{d\sigma}
+\sum_i g_{\sigma i} ~~2\hspace{-1mm}\int\!\frac{d^3k}{(2\pi)^3}
\frac{M_i^*}{E_i^*}\left({\rm e}^{\beta(E^*_i-\nu_i)}+1\right)^{\!-1} \,.
\label{hyp5}
\end{eqnarray}
The total partition function $Z_{total}=Z_HZ_L$, where $Z_L$ is the standard
noninteracting partition function of the leptons.
The additional conditions needed to obtain a solution are provided by the
charge neutrality requirement, and, when neutrinos are not
trapped, the set of equilibrium chemical
potential relations required by the general condition
\begin{eqnarray}
\mu_i = b_i\mu_n - q_i\mu_l\,,
\label{beta}
\end{eqnarray}
where $b_i$ is the baryon number of particle $i$ and $q_i$ is its
charge.
For example, when $\ell=e^-$, this implies the equalities
\begin{eqnarray}
\mu_{\Lambda} = \mu_{\Sigma^0} = \mu_{\Xi^0} = \mu_n \,,~~~
\mu_{\Sigma^-} = \mu_{\Xi^-} = \mu_n+\mu_e \,,~~~{\rm and}~~~
\mu_p = \mu_{\Sigma^+} = \mu_n - \mu_e \,.
\label{murel}
\end{eqnarray}
In the case that the neutrinos are trapped, equation
(\ref{beta}) is replaced by
\begin{eqnarray}
\mu_i = b_i\mu_n - q_i(\mu_l-\mu_{\nu_\ell})\,.
\label{tbeta}
\end{eqnarray}
The new equalities are then obtained by the replacement
$\mu_e \rightarrow \mu_e - \mu_{\nu_e}$ in equation (\ref{murel}).  The
introduction of additional variables, the neutrino  chemical potentials,
requires additional constraints, which we supply by fixing the lepton
fractions, $Y_{L\ell}$, appropriate for conditions prevailing in the
evolution of the protoneutron star.  The contribution to pressure from
neutrinos of a given species is given by their free Fermi gas expressions.

In the nucleon sector, the constants $ g_{\sigma N}, g_{\omega N}, g_{\rho N},
b$, and $c$ are determined by reproducing the nuclear matter equilibrium
density $n_0$, and the binding energy per nucleon, the symmetry energy, the
compression modulus $K_0$, and the nucleon Dirac effective mass $M^*$ at $n_0$.
The parameters of the models are listed in Table 1 and for the most part are
taken from Glendenning \& Moszkowski (1991).  Model GM1 is a stiff EOS
($K_0=300$ MeV) while the model GM3 is softer with $K_0=240$ MeV. To study the
possible role played by the nuclear symmetry energy we also investigate the
model GM4, which is similar to GM3 but with a smaller symmetry energy.

In addition to models containing only nucleonic degrees of freedom (GM1np \&
GM3np) we investigate models that allow for the presence of hyperons (GM1npH \&
GM3npH). The hyperon coupling constants may be determined by reproducing the
binding energy of the $\Lambda$ hyperon in nuclear matter ~(Glendenning \&
Moszkowski 1991). Parameterizing the hyperon-meson couplings in terms of
nucleon-meson couplings through
\begin{eqnarray}
x_{\sigma H}=g_{\sigma H}/g_{\sigma N},~~~
x_{\omega H}=g_{\omega H}/g_{\omega N}
,~~~x_{\rho H}=g_{\rho H}/g_{\rho N} \,,
\end{eqnarray}
the $\Lambda$ binding energy at nuclear density is given by
\begin{eqnarray}
(B/A)_\Lambda = -28 = x_{\omega \Lambda} g_{\omega N} \omega_0
- x_{\sigma \Lambda} g_{\sigma N} \sigma_0\,,
\end{eqnarray}
in units of MeV.  Thus, a particular choice of  $x_{\sigma
\Lambda}$ determines $x_{\omega \Lambda}$ uniquely.   To keep the number of
parameters small, the coupling constant
ratios for all the different hyperons are assumed to be the same.
In model GM1, the hyperon couplings are given by
\begin{eqnarray}
x_\sigma = x_{\sigma\Lambda} = x_{\sigma\Sigma} = x_{\sigma\Xi} = 0.6\,,~~~{\rm
and}~~~
x_\omega &=& x_{\omega\Lambda} = x_{\omega\Sigma} = x_{\omega\Xi} = 0.653 \,.
\end{eqnarray}
The $\rho$-coupling is of less consequence and is taken to be of similar order,
i.e. $x_\rho = x_\sigma$\,. In GM3, $x_{\sigma}=0.6$ , $x_{\omega}=0.659$
and $x_{\rho}=0.6$.

\begin{figure}
\begin{center}
\epsfxsize=6.in
\epsfysize=7.in
\epsffile{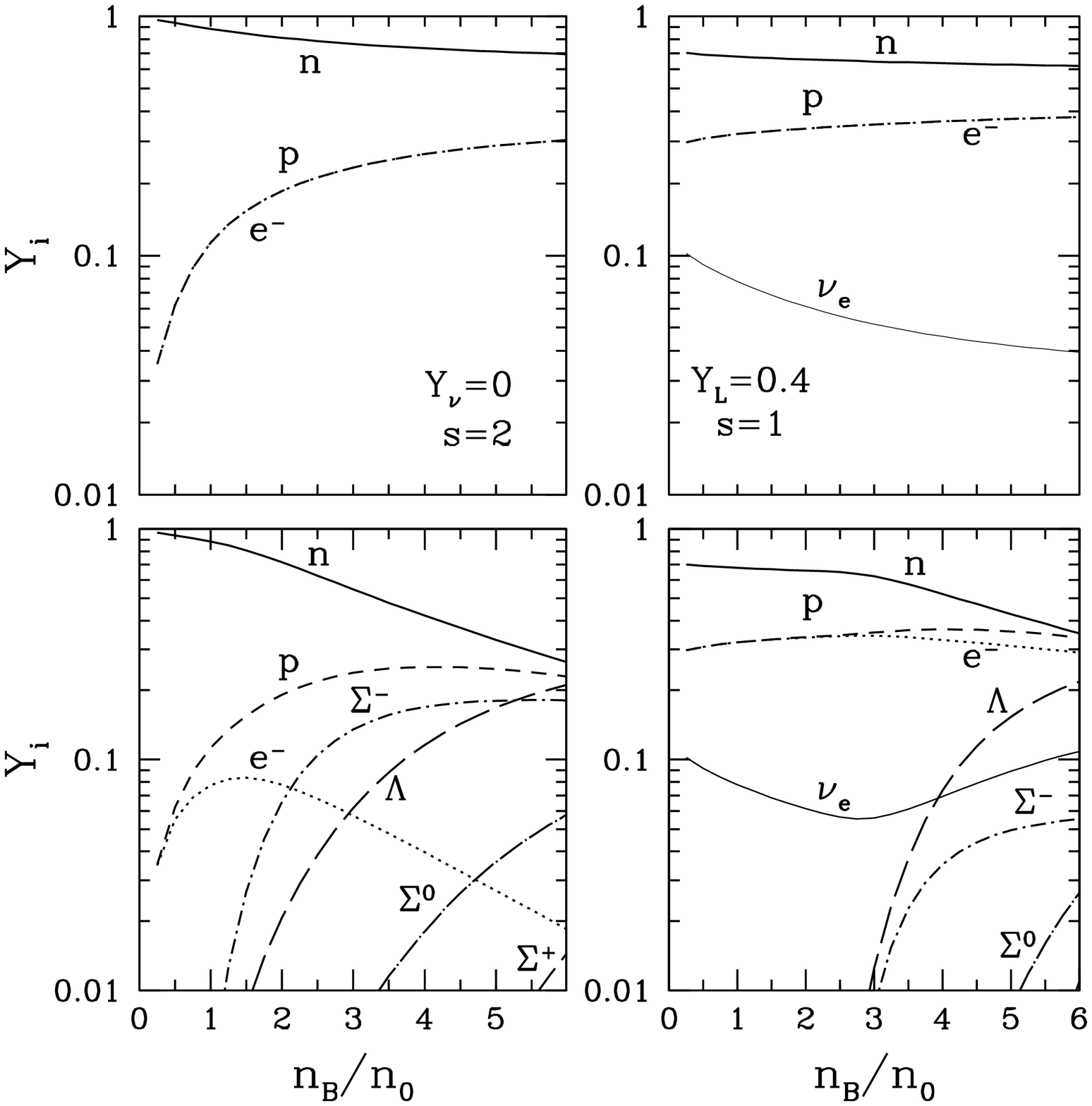}
\end{center}
\caption{}
{\label{composition}}
\end{figure}

In Figure~\ref{composition} the particle fractions for models GM3np and GM3npH
are shown with and without trapped neutrinos.  The left panels show the
particle fractions in hot and neutrino-free matter (mimics the ambient
conditions subsequent to deleptonization) and the right panels show the
particle fractions in neutrino trapped matter (ambient conditions in the core
at birth).  Considering matter with nucleons-only first, note that the major
effect of trapping is to make the electron (and hence the proton) concentration
high compared to the case in which matter is neutrino free. Compared to the
free gas case the proton fraction is appreciably larger in matter with strong
interactions due to the larger symmetry energies.  At fixed $Y_L$, this acts to
decrease the electron neutrino fraction at high density (see top right panel).
During the deleptonization phase, the EOS affects the evolution mainly through
the density dependent nuclear symmetry energy as it directly influences
$\lambda_{\nu_e}$ (see \S4) and the magnitude of the spatial gradient $\partial
\mu_\nu/ \partial r$ (see \S 2). Subsequent to deleptonization, the proton
fractions decreases, but is still high enough to enable the charged current
reactions to occur for most relevant densities and temperatures of interest
(see \S 4).

Although the threshold densities for the appearance of hyperons are sensitive
to the hyperon coupling constants (Knorren, Prakash, \& Ellis 1997),
the higher entropies encountered during the early evolution generally
results in hyperon fractions which are larger compared to the case of zero
temperature matter (bottom left panel).  In lepton-rich matter (bottom right
panel), the appearance of hyperons is delayed to higher densities due to the
relative dispositions of the various chemical potentials. With the appearance
of the hyperons, the neutrino population begins to increase with density, in
contrast to the monotonic decrease exhibited in the hyperon free case. Since
the negatively charged hyperons effectively replace the electrons at high
density, the final lepton content of stars with hyperons is smaller relative to
those in nucleons-only matter. Correspondingly, we should expect to see a
larger lepton number radiated from stars with hyperons (see \S5).

\begin{figure}
\begin{center}
\epsfxsize=6.in
\epsfysize=7.in
\epsffile{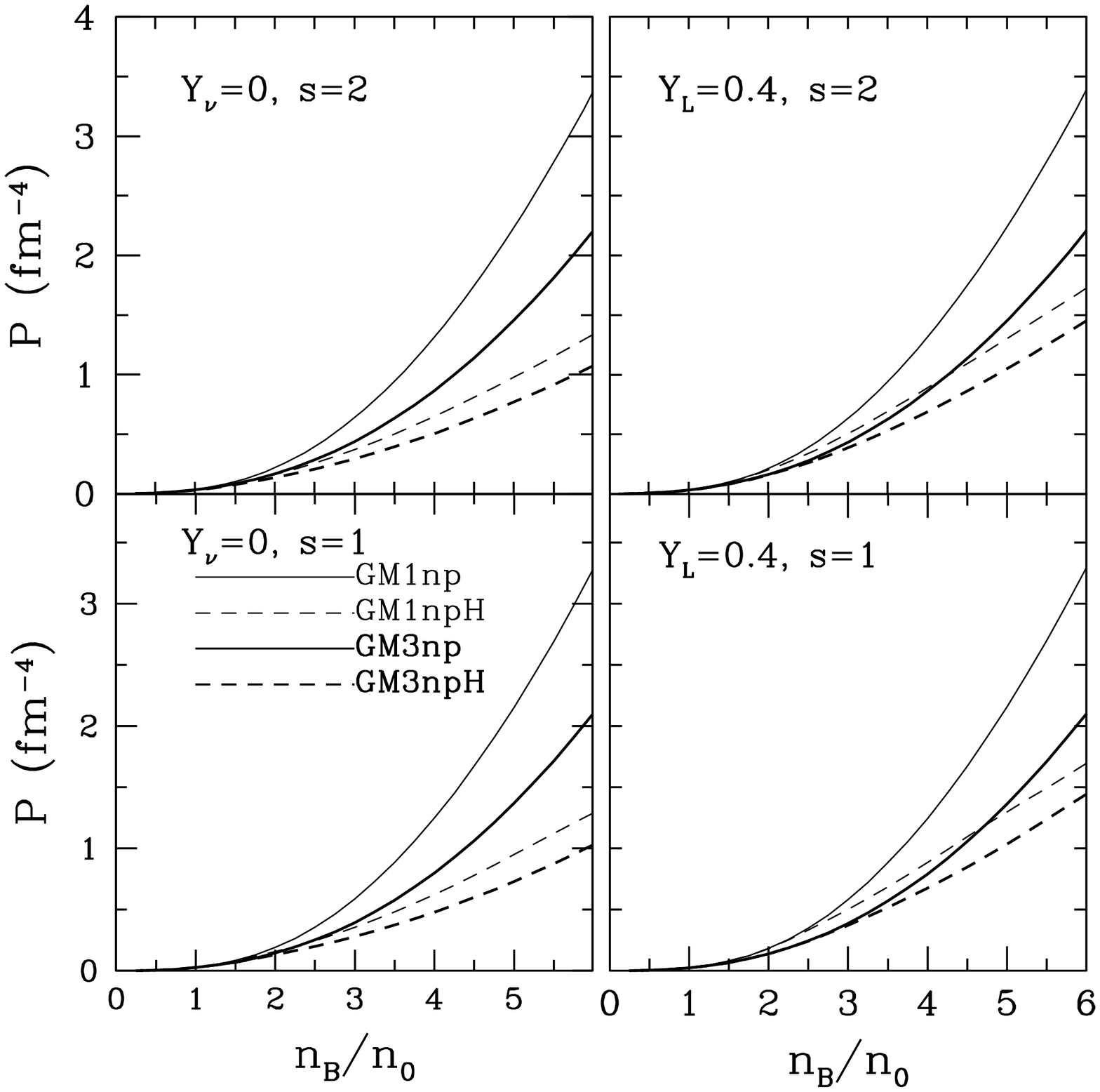}
\end{center}
\caption{}
{\label{pressure}}
\end{figure}

In Figure~\ref{pressure} the variation of pressure with baryon density is
shown.   The upper panels show the results for $s=2$ and the lower panels for
$s=1$.   Thermal effects contribute only modestly  to the total pressure.  The
pressure is dominated from  effects of interactions and degeneracy at high
density.  The left panels in Figure~\ref{pressure} show results for
$Y_{\nu_e}=0$, and the right panels for a fixed lepton fraction of $Y_L=0.4$.
While the differences between lepton rich and lepton poor matter for the np
models are negligible, significant differences arise in models with hyperons.
When the electron neutrino fraction is large, their presence inhibits the
appearance of hyperons (see also Figure~\ref{composition}; note  however, that
a large hyperonic component at high density  in the neutrino free case
significantly softens the EOS (Prakash et al. 1997)). This softening with
decreasing lepton number will lead to compression during the deleptonization
epoch.  Hence, there exists a window of initial masses for which the star
becomes unstable to gravitational collapse during deleptonization. The maximum
masses supported at low temperature and for lepton poor matter are
significantly smaller than for hot and lepton rich matter.

\begin{figure}
\begin{center}
\epsfxsize=6.in
\epsfysize=7.in
\epsffile{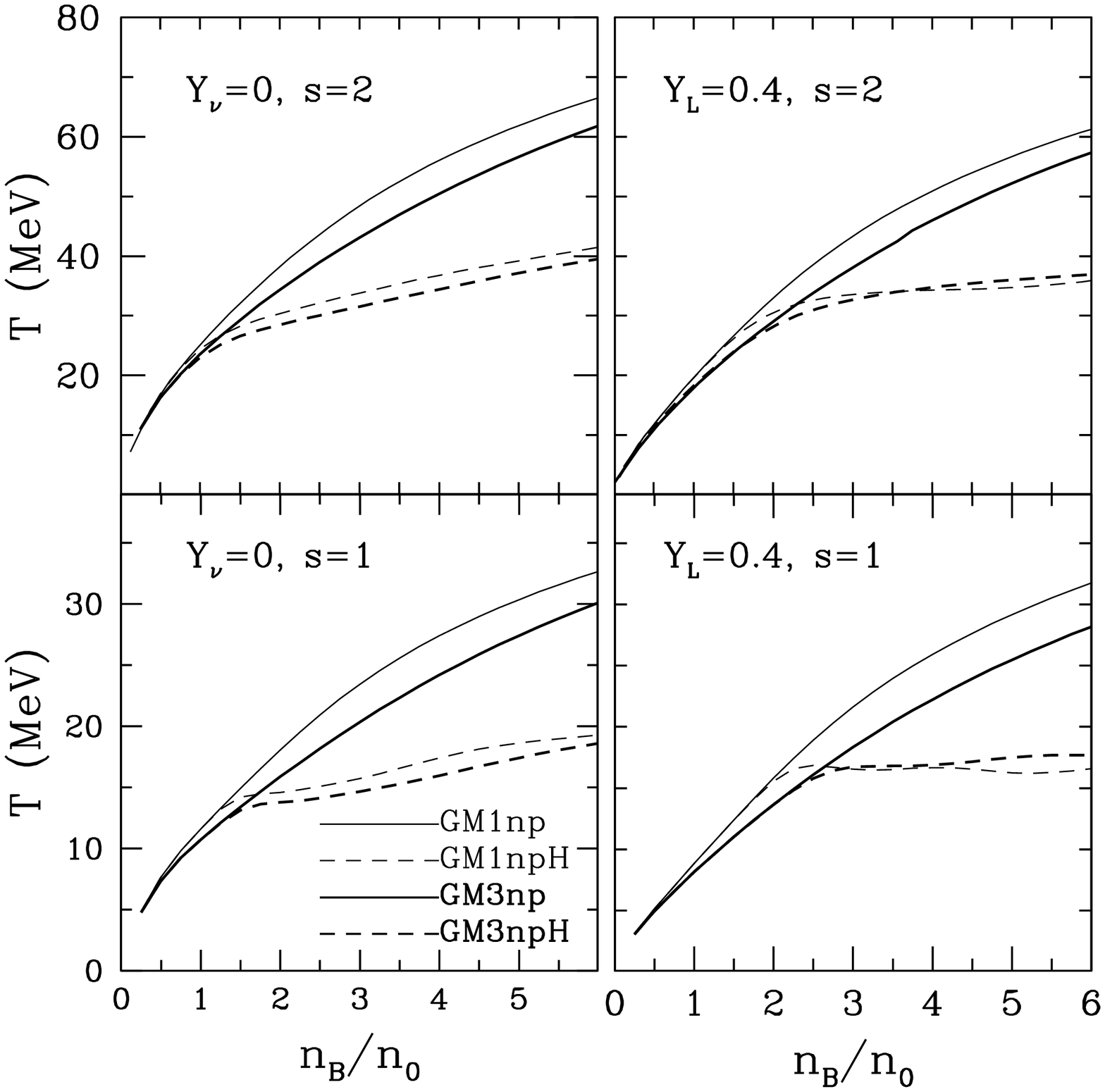}
\end{center}
\caption{}
{\label{entropy}}
\end{figure}

Finite temperature properties of matter at high density influence the diffusion
of neutrinos especially through the specific heat. Neutrino mean free paths are
strongly temperature dependent (see \S4) and the ambient matter temperature
controls the diffusion of neutrinos to a large extent. In Figure~\ref{entropy},
we show the matter temperature at fixed entropy per baryon as a function of the
baryon density to contrast the behavior observed in models with and without
hyperons. The trends observed in the np models are similar while the
differences between np and npH models  are very significant. The appearance
of additional fermionic degrees of freedom in the form of hyperons (we may
expect similar behavior in models with quarks (Prakash, Cooke, \& Lattimer
1995)) increases the low temperature
specific heat. This generally favors lower temperatures
during the evolution for the npH models, all else being equal. This is
because, in charge neutral
and beta-equilibrated matter, bulk of the entropy resides in the
baryons (Prakash et al. 1997). Under degenerate situations
($T/E_{F_i} >>1$), this result may be understood from the relation that
connects the temperature and baryonic entropy (or specific heat) through the
concentrations ($Y_i=n_i/n$):
\begin{eqnarray}
\frac {\pi^2T}{s} = \frac {(3\pi^2n_B)^{2/3}} {\sum_i Y_i^{1/3}
{\sqrt {M_i^{\star^2}+k_{F_i}^2}} }
\Rightarrow
\left\{
 \begin{array}{ll}
  \D  \frac {(3\pi^2n_B)^{2/3}} {\sum_i Y_i^{1/3}M_i^\star}
     &	{\rm for} \;\; M_i^\star >> k_{F_i} \\
  \D  \frac {(3\pi^2n_B)^{1/3}} {\sum_i Y_i^{2/3}}
     &	{\rm for} \;\; M_i^\star << k_{F_i}    \,,
 \end{array}
\right.
\end{eqnarray}
where $k_{F_i} = (3\pi^2n_BY_i)^{2/3}$ are the Fermi momenta.  For the
temperatures of interest here, and particularly with increasing density, the
above relation provides an accurate representation of the exact results for
entropies per baryon even up to $s=2$. The behavior with density of both the
concentrations and effective masses controls the temperatures for a fixed $s$.
In the absence of any variation with $M_i^\star$, a system with more components
at a given baryon density has a smaller temperature than a system with fewer
components (recall that $\sum_i Y_i =1$).  Effective masses that rapidly drop
with density oppose this behavior and lead to larger temperatures.
The fact that the temperatures are smaller in the presence of
hyperons than those in nucleons-only matter attests to the predominant effect
of concentrations. Note that the differences between the temperatures for the
cases with and without hyperons are larger than those obtained by using
different models for the individual cases.

\begin{figure}
\begin{center}
\epsfxsize=6.in
\epsfysize=7.in
\epsffile{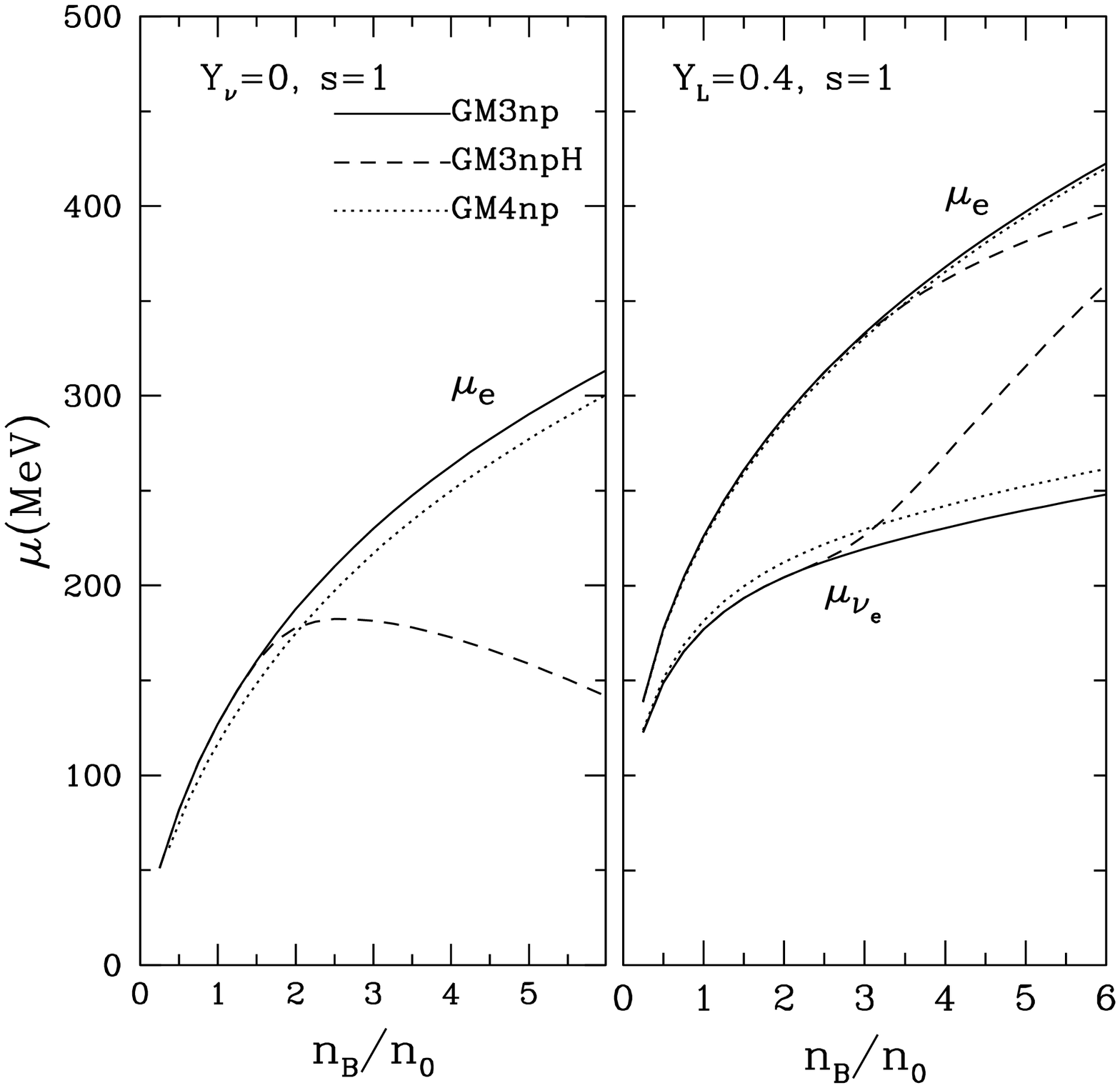}
\end{center}
\caption{}
{\label{mus}}
\end{figure}

However, as we
have seen from Figure~\ref{pressure}, the appearance of hyperons also leads to
higher central densities due to softening; this acts to compensate for the
higher specific heat in npH models. In addition, the larger neutrino chemical
potentials (see Figure~\ref{mus}) in npH models generate larger entropies
during the deleptonization epoch; this also acts to increase the central
temperature in these models. Results discussed in \S5 clearly show that despite
the significant differences seen in Figure~\ref{entropy} the maximum central
temperatures reached in the both np and npH models are nearly equal due to
the above mentioned feedbacks.

The lepton chemical potentials influence the deleptonization epoch
(see \S5) and are shown in Figure~\ref{mus}. For np models
a lower nuclear symmetry energy favors a larger $\nu_e$ fraction
and has little effect on the $e^-$ fraction at $Y_l=0.4$. Models with
hyperons lead to significantly larger $\mu_{\nu_e}$ and lower $\mu_e$, both of
which influence the diffusion of electron neutrinos. The electron
chemical potentials in neutrino free matter are reduced to a greater extent
by changes in composition and symmetry energy as there are no neutrinos to
compensate for changes in $\hat{\mu}=\mu_n-\mu_p$. The impact these
modifications will have on the evolution are discussed in \S5.

\begin{figure}
\begin{center}
\epsfxsize=6.in
\epsfysize=7.in
\epsffile{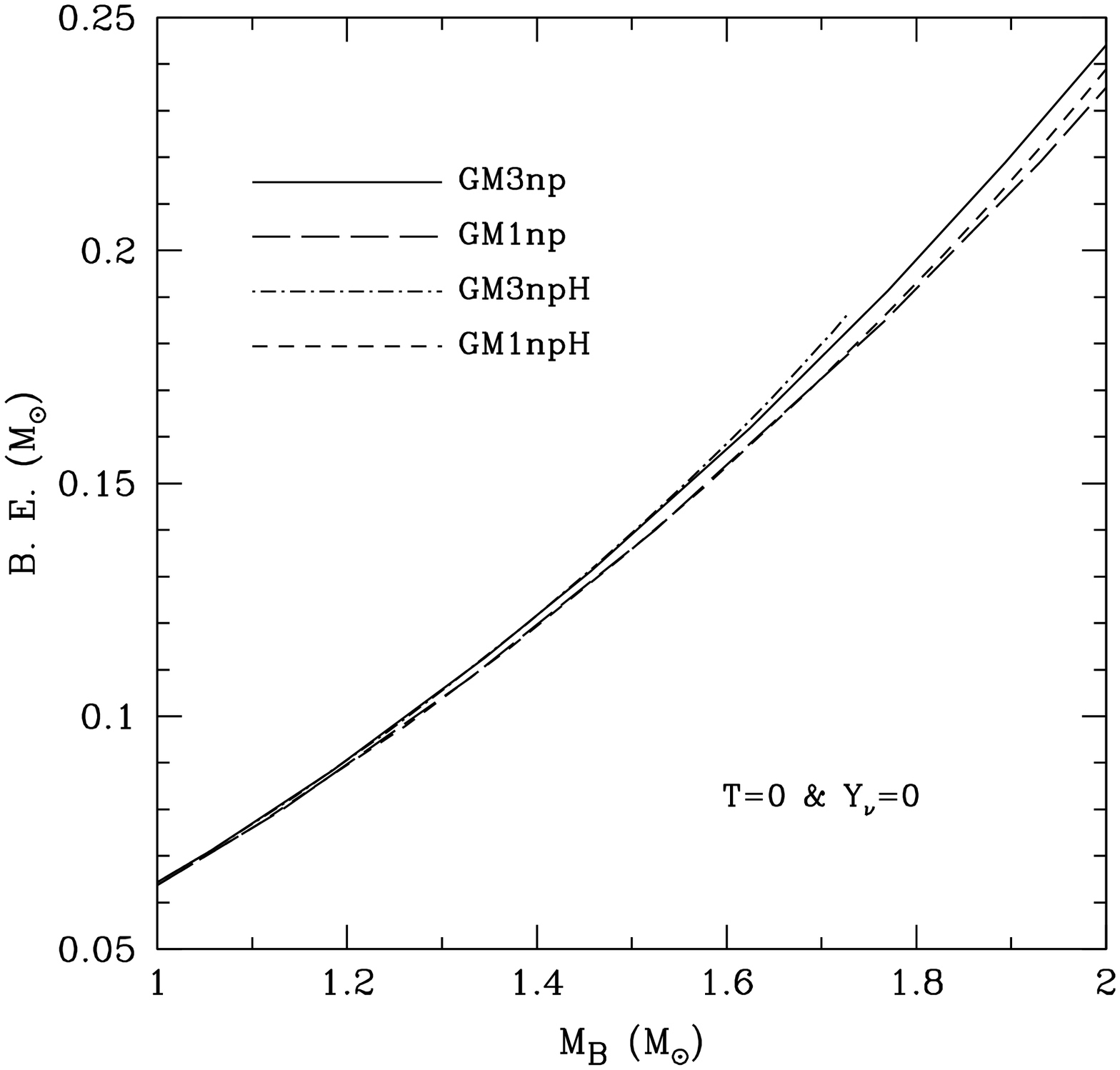}
\end{center}
\caption{}
{\label{binding}}
\end{figure}

We turn now to some aspects of the global energetics.  The binding energy,
which is the difference between baryonic and gravitational masses,
is an important observational parameter because
nearly $99\%$ of it appears as radiated neutrino energy.  The rate of change of
binding is essentially the total neutrino luminosity.  In
Figure~\ref{binding}, we display the binding energy as a function the
baryonic mass for the
models considered here. As found by Lattimer \& Yahil (1989) and Prakash
et al. (1997), there exists a
universal empirical relationship between the binding energy and the mass:
\be
B.E. = (M_B-M_G)c^2 \simeq (0.065\pm 0.01)
\left(\frac{M_B}{{\rm M}_\odot}\right)^2 {\rm M}_\odot \,.
\label{universal}
\ee
This universality is not altered by the presence of significant softening due
to the presence of hyperons, kaons or quarks. Since nearly all of the
binding energy is released in the form of neutrinos, {\em an accurate
measurement of the total neutrino energy will lead to a good estimate of the
remnant mass}.  However, as the results of Figure~\ref{binding} show, it will
not be possible to distinguish the various EOSs from the total binding energy
alone.

Table 2 shows the maximum gravitational masses for the models studied. The
three columns roughly depict the three important stages of the evolution of a
protoneutron star. At birth the PNS has $s=1$ and $Y_L=0.4$, after
deleptonization the entropy has reached its maximum ($s\sim 2$), and finally
cools down to its final cold catalyzed state.

\section{NEUTRINO-MATTER INTERACTION RATES }

One of the important microphysical inputs in PNS simulations is the
neutrino opacity at supra-nuclear density~
(Bruenn 1985; Mezacappa \& Bruenn 1993; Burrows \& Lattimer 1986;
Wilson \& Mayle 1989; Suzuki \& Sato 1992; Keil 1994; Keil \& Janka 1995).
However, calculations of neutrino
opacities in dense matter have received relatively little attention
compared  to other physical inputs such as the EOS.  Although it was
realized over a decade ago that the effects due to
degeneracy and strong interactions significantly alter the neutrino
mean free paths, it is only recently that detailed calculations have
become available (Reddy \& Prakash 1997; RPL 1998; Reddy et al. 1997, Prakash
\& Reddy 1997,  Burrows \& Sawyer, 1998a,b).  The weak interaction rates in hot
and dense matter are modified due to many in-medium effects.  The most
important of these are: \\

\ni (1) {\it Composition}:   The neutrino mean free paths depend
sensitively on the composition which is sensitive to the nature of strong
interactions.  First, the different
degeneracies of the different Fermions determines the single-pair
response due to Pauli blocking. Second, neutrinos couple differently
to different baryonic species; consequently, the net rates will depend
on the individual concentrations. \\
\ni (2) {\it In-medium dispersion relations}:
At high density, the single-particle spectra are
significantly modified from their noninteracting forms due to effects
of strong interactions.  Interacting matter features
smaller effective baryon masses and energy shifts relative to non-interacting
matter. \\
\ni (3) {\it Correlations}: Repulsive particle-hole interactions
and Coulomb interactions
generally result in a screened dielectric response and also lead to
collective excitations in matter.  These effects may be
calculated using the Random Phase Approximation (RPA),  in which ring
diagrams are summed to all orders. Model calculations
(Sawyer 1975,1989; Iwamoto \& Pethick 1982; Horowitz \& Wehrberger 1991a,b;
1992,
Raffelt \& Seckel 1995, Sigl 1995, Reddy et al. 1997, Prakash \& Reddy 1997,
Burrows \& Sawyer
1998a,b) indicate that at high density the neutrino
cross sections are suppressed relative to  the case in which these
effects are ignored.   In addition, these correlations enhance the
average energy transfer in neutrino-nucleon collisions.  Improvements in
determining the many-body dynamic form factor and assessing the role
of particle-particle interactions in dense matter at finite
temperature are  necessary before the full effects of many-body
correlations may be ascertained. \\
\ni (4) {\it Axial charge renormalization}: In  dense matter,
the axial charge of the baryons is renormalized (Wilkinson 1973, Rho 1974,
Brown \& Rho 1991), which alters the
neutrino-baryon couplings from their vacuum values. Since the
axial contribution to the scattering and absorption reactions is typically
three times larger than the vector contributions, small changes in the
axial vector coupling constants significantly affect the cross sections. \\
\ni (5) {\it  Multi-Pair excitations}:
Neutrinos can also excite many-particle states in an interacting
system, inverse bremsstrahlung being an example of a two-particle
excitation (Hannestad \& Raffelt 1998).  These excitations provide an efficient
means of transferring energy between the neutrinos and baryons which are
potentially significant in low-density matter.  However, multigroup neutrino
transport will be needed to fully include this affect.  In addition, such
calculations require source terms for neutrino processes such as
bremsstrahlung and neutrino pair production.
The latter process has been accurately treated by Pons, Miralles, \&
Iba\~nez (1998). \\

The relative importance of the various effects described above on neutrino
transport is only beginning to be studied systematically. As a first step, we
will  focus on effects due to modifications (1) and (2) above, since RPL have
shown how to calculate them consistently with the underlying EOS.  Work on the
other modifications listed above is currently in progress and will be
incorporated in a future publication. In what follows, we recapitulate some of
the
essential results for the evaluation of the neutrino mean free paths and
diffusion constants.

The scattering and absorption reactions that contribute to the neutrino
opacity are
\begin{eqnarray}
\nu_e+B&\rightarrow& e^-+B' \,,\quad \quad
\bar{\nu}_e+B\rightarrow e^++B' \,,\\
\nu_X+B&\rightarrow& \nu_X+B' \,,\quad \quad
\nu_X+e^-\rightarrow \nu_X +e^-  \,,
\end{eqnarray}
where the scattering reactions are common to all neutrino species and the
dominant source of opacity for the electron neutrinos is due to the charged
reaction.  The total cross  section per unit volume for the scattering reaction
$\nu+ B \rightarrow \nu' + B'$ and  absorption reaction  $\nu+ B
\rightarrow e^-+ B'$ are given by
\begin{eqnarray}
\frac{\sigma(E)}{V}=
\frac{G_F^2}{2\pi^2E^2} \int_{-\infty}^{E}
dq_0~ \frac{(1-f(E'))}{1-\exp(-z)}
\int_{|q_0|}^{2E-q_0} dq~q~q_{\mu}^2~ [AR_1+R_2+BR_3]\,.
\label{cross}
\end{eqnarray}
where $E$ and $E'$ are incoming and outgoing lepton
energies, and $q_\mu=(q_0,\vec{q})$ is the four momentum transfer to the
baryons so that $q_{\mu}^2=2 E E' (\cos\theta-1)$, where $\theta$ is the
angle between the incoming and outgoing lepton. The factor
$(1-\exp (-z))$ arises due to the principle of detailed balance, and
$z=(q_0+\mu_B-\mu_{B'})/T$. The response functions $R_1$, $R_2$, and $R_3$
are functions of the energy and momentum  transfers $q_0=E-E'$
and $\vec{q}=\vec{k}-\vec{k'}$,
and strongly depend
on the properties of matter, such as the individual particle degeneracies and
in-medium modifications to the baryon propagators. The density and temperature
dependent baryon effective masses and chemical potentials required for their
evaluation are obtained directly from the EOS. Explicitly, the response
functions are written in terms of the neutrino baryon coupling constants
${\cal V}$ and ${\cal A}$, and the polarization functions
\begin{eqnarray}
R_1&=&({\cal V}^2+{\cal A}^2) ~[{\rm Im}~ \Pi^R_L(q_0,q)+{\rm Im}~
\Pi^R_T(q_0,q)]\\ R_2&=&({\cal V}^2 + {\cal A}^2)~{\rm Im}~
\Pi^R_T(q_0,q) - {\cal A}^2~{\rm Im}~ \Pi^R_A(q_0,q)\\ R_3&=&2{\cal
V}{\cal A} ~{\rm Im}~ \Pi^R_{VA}(q_0,q)  \,.
\end{eqnarray}
In the case of absorption ${\cal V}=Cg_V$ and ${\cal A}=Cg_A$,
where $C$ is the Cabibbo factor,
$g_V$ and $g_A$ are the vector and axial-vector couplings.
For scattering reactions   ${\cal V}=g_V/2$ and ${\cal A}=g_A/2$,
respectively.  Numerical values are taken from Tables I and II of RPL.
The imaginary part of the various polarization functions
$\Pi$ may be found in (Horowitz \& Wehrberger 1991, 1992;
RPL 1998).  Comparing with equation (\ref{scat2}) we identify
\begin{equation}
R^{out}_s = {4 G_F^2} \frac {(\cos\theta-1)}{1-e^{-z}} \left[ A R_1 +
R_2 + B R_3 \right]
\end{equation}
\begin{eqnarray}
\frac{1}{\lambda_a} =  \frac{G_F^2}{\pi^2} \int_{0}^{\infty} dE_e \,
E_e^2 \, \left[1-f_{eq}(E_e)\right] \int_{-1}^{+1} d\cos{\theta} \,
\frac{(\cos\theta-1)}{1-e^{-z}} \left[ A R_1 + R_2 + B R_3 \right]
\end{eqnarray}
Note that the above formalism accounts for the effects arising due
to  matter degeneracy and in-medium modifications in the baryon dispersion
relations exactly. As stated earlier, ignoring the effects due to correlations,
the total scattering kernel $R^{out}_s$ is simply the incoherent sum over
all possible scattering process. Similarly, the  absorption mean free path
$\lambda_a$ is the sum over all possible absorption reactions.

\begin{figure}
\begin{center}
\epsfxsize=6.in
\epsfysize=7.in
\epsffile{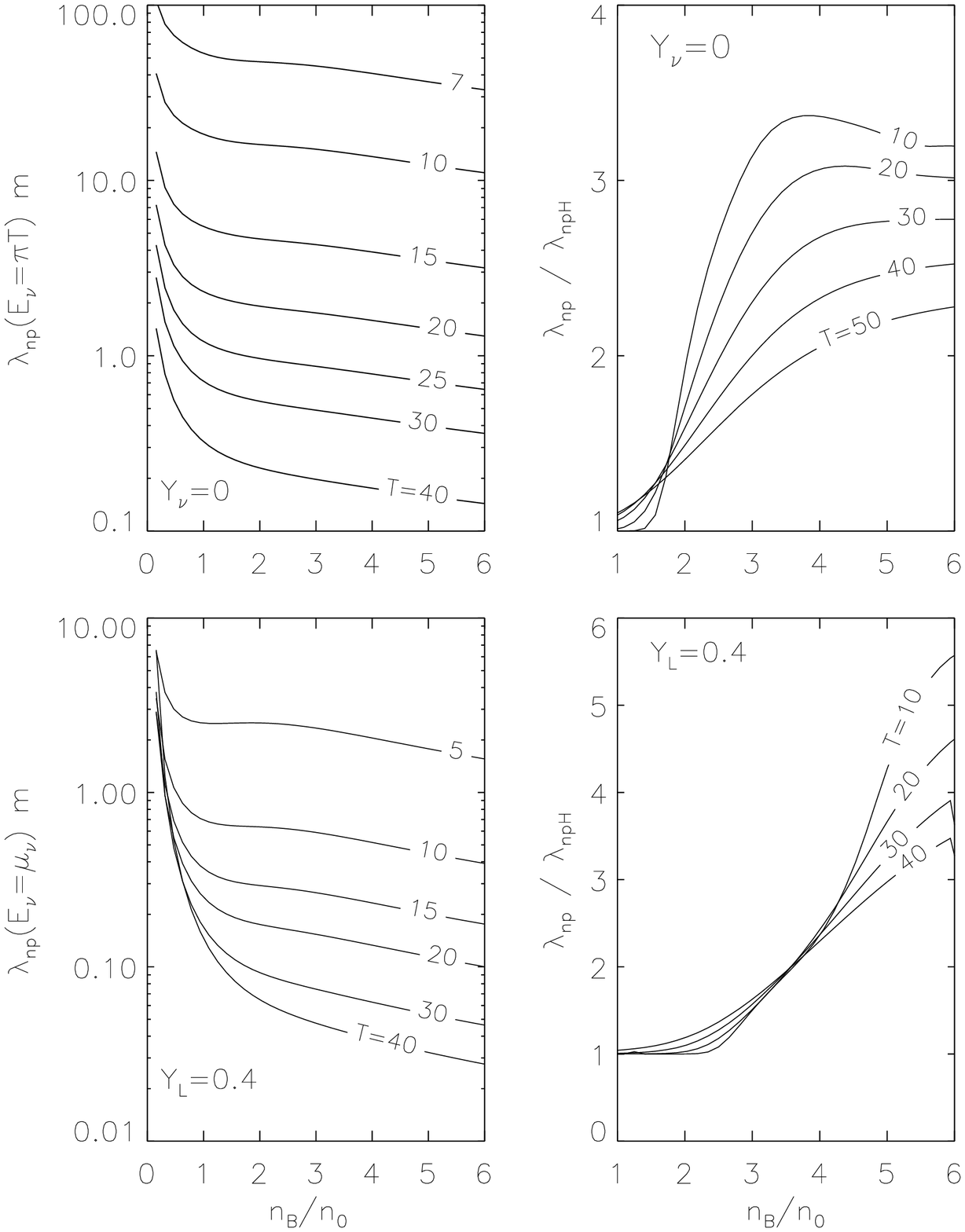}
\end{center}
\caption{}
{\label{sig}}
\end{figure}

Under degenerate conditions even modest changes to the composition
significantly alter the neutrino scattering and absorption mean free paths.
In Figure~\ref{sig}, the neutrino scattering and absorption cross sections are
shown for models GM3np and GM3npH relevant to the deleptonization and
cooling epochs. The top panels show the scattering cross sections common to
all neutrino species in neutrino free matter. The scattering cross section per
unit volume for thermal neutrinos ($E_{\nu}=\pi T$) is shown in the left
panel for various temperatures. To study the influence of hyperons,
the ratio of the $\sigma_{npH}/ \sigma_{np}$ is shown in the right panels.
The presence of hyperons significantly increase the scattering cross sections,
by a factor $\sim (2- 3)$. Similar results for the absorption cross sections
are shown in the lower panels for $Y_L=0.4$.  Again we notice a significant
enhancement (right panel) when hyperons appear, the factor here could be as
large as 5.

\begin{figure}
\begin{center}
\epsfxsize=6.in
\epsfysize=7.in
\epsffile{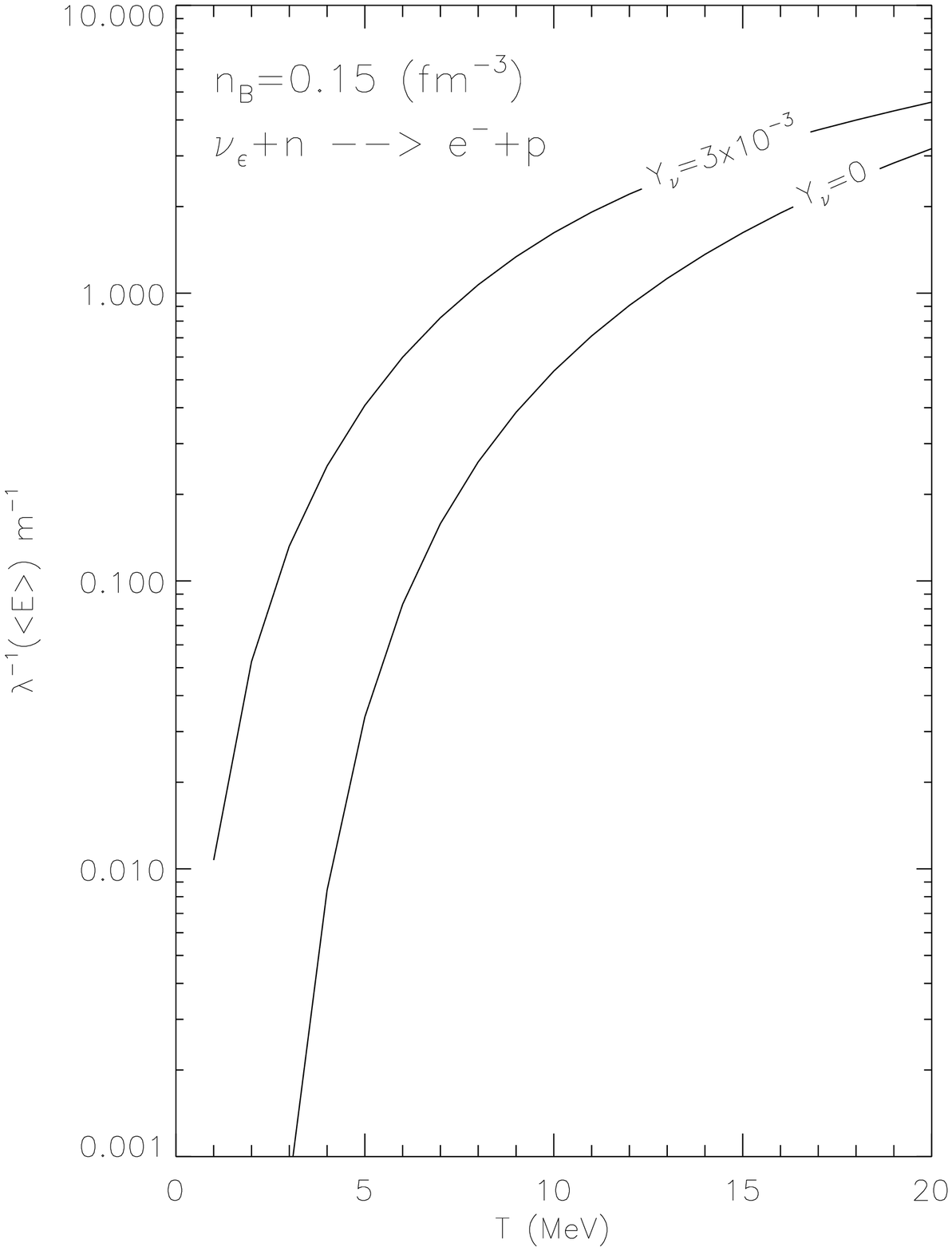}
\end{center}
\caption{}
{\label{csig}}
\end{figure}

During the deleptonization stage, lepton number transport is sensitive
to charged current reactions which dominate scattering reactions. At
zero temperature, charged current reactions $\nu + n \leftrightarrow e
+ p$ depend sensitively on the proton fraction $Y_p$~(Lattimer et al.
1991).  Kinematic restrictions require $Y_p$ to be larger than
$11-14\%$ (direct Urca threshold).  At early times, a finite neutrino
chemical potential favors a large $Y_p$ throughout the star, which
enables these reactions to proceed without any hindrance. Toward the
late stages, however, $Y_p$ decreases with decreasing $\mu_{\nu}$ and
charged current reactions may be naively expected to become
inoperative. The threshold density for the charged current reaction
when $\mu_{\nu}=0$ and $T=0$ depends sensitively on the density
dependence of the nuclear symmetry energy.  In field-theoretical
models, in which the symmetry energy is largely given by contributions
due to $\rho$-meson exchange, the critical density is typically
$n_B=2\sim3 n_0$. However, finite temperatures favor larger $Y_p$'s
and increase the average neutrino energy enabling the charged current
reactions to proceed even below these densities.  Figure~ \ref{csig}
shows that this is the case even at relatively low temperatures
($T\sim 3-5$) MeV for a baryon density $n_B=0.15~{\rm fm}^{-3}$. The
sharp rise with temperature, which occurs even for $Y_{\nu}=0$,
clearly indicates that this reaction dominates the $\nu_e$ opacity
even during the late deleptonization era.  Thus, charged current
reactions cannot be simply turned off when the neutrino chemical
potential becomes small enough as was done in prior PNS simulations
(BL).

\begin{figure}
\begin{center}
\epsfxsize=6.in
\epsfysize=7.in
\epsffile{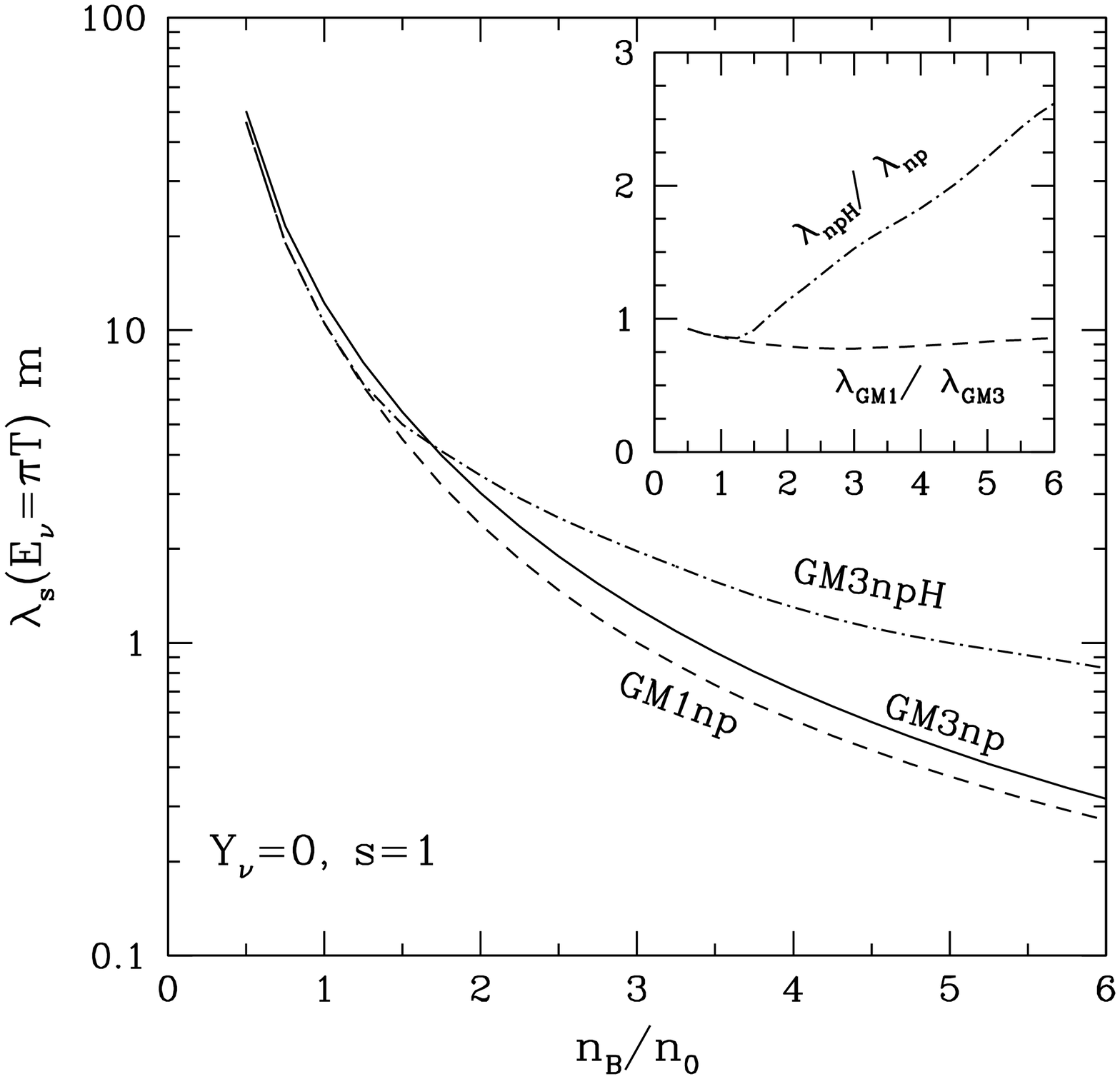}
\end{center}
\caption{}
{\label{ssig}}
\end{figure}

The EOS and neutrino mean free paths are intimately related, which is best
illustrated by  comparing the results shown in Figure~\ref{sig} with those
shown in Figure~\ref{ssig}.  Composition and the baryon effective masses
influence both the neutrino mean free paths and the matter's specific heat.
Hyperons decrease the neutrino mean free paths at constant temperature
(Figure~\ref{sig}). This trend is reversed at constant entropy due to the
significantly lower temperatures favored in npH matter. Similar effects are
apparent  when we compare np models with different baryon effective  masses. At
a constant temperature, the larger effective mass in model GM3np favors
larger cross sections, while at constant entropy this trend is again
reversed due to the lower temperatures favored by the larger specific heat.

The diffusion coefficients are calculated using equation (\ref{d2d3}) with the
cross sections discussed above.  Earlier works (BL \& KJ)
made  simplifying assumptions to define the diffusion
coefficients including the use of interpolation to bridge the degenerate and
nondegenerate limits.  But in partially degenerate matter, effects due to Fermi
motion and Pauli blocking are substantial. The effects of density-dependent
baryon effective masses, which are significant, were also ignored.  In this
work, we numerically integrate equation (\ref{d2d3}) over the neutrino energies
by employing the exact cross sections.  The diffusion coefficients $D_2,D_3$,
and $D_4$ are functions of $n_B$, $T$, and $Y_{\nu_e}$.


\section{RESULTS AND DISCUSSION}

\subsection{Baseline Model}
We have performed simulations for several dense matter models, initial
conditions and baryonic masses. To begin, we describe in detail the general
features of a reference model, corresponding to the dense matter EOS containing
nucleons and leptons (model GM3np), and for a baryon mass $M_B=1.6$ M$_\odot$. We
shall initially concentrate on the temporal evolution of the various
thermodynamic quantities, such as $s$, $T$, $n_B$, $Y_e$, $Y_{\nu}$ and
$\mu_{\nu}$ inside the PNS (see Figure~\ref{baseline}). Our overall results are
qualitatively similar to those obtained in earlier works (BL; KJ), but some
interesting differences exist nevertheless.

The initial entropy and $Y_L$ profiles were taken from the profiles
obtained in detailed core collapse calculations of an $M_B=1.08$ M$_\odot$ star
(Wilson \& Mayle 1989) within a hundredth of a second of core bounce.  For
larger mass PNSs, the $s$ and $Y_L$ profiles were scaled in terms of the
relative enclosed baryon mass. At early times  $Y_L\sim 0.36$ and $s\sim 1$ in
the central regions, while the shocked outer mantle which begins at about 0.5
M$_\odot$ (for the 1.08 M$_\odot$ star) is at high entropy ($s\ge8$)  and
contains smaller lepton fractions.  The outermost regions, which contains
subnuclear density matter, and tends to nearly symmetric matter at low density,
hence, it contains a large positive gradient in electron concentration.  By
scaling the results for 1.08 M$_\odot$ for larger masses, we attempted to
simulate the profiles expected for larger collapsing stars.  The masses of both
the inner core with roughly constant $s$ and $Y_L$ and the shocked, relatively
lepton-poor outer mantle should increase with total mass.

The evolution of the 1.6 M$_\odot$ star proceeds as follows. The steep neutrino
chemical potential gradients drive a radially outward flow of electron
neutrinos, causing deleptonization.   The most rapid
deleptonization occurs in the low-density outer mantle, which collapses during
the first few tenths of a second due to substantial pressure loss from neutrino
emission.  Note from Figure~\ref{baseline} that compression of the outer half
of the star leads to a significant temperature increase.  The deleptonization
of the central regions takes much longer, tens of seconds in this example.  The
central regions, as they deleptonize, are also heated due to neutrino transport
(Joule heating).  Eventually, after about 15 seconds,  the central heating is
overwhelmed by neutrino cooling and the central temperature begins a steady
decrease.  This star  took about 50 seconds to cool to below 5 MeV temperatures
in the central regions. At these low temperatures, the neutrino mean free path
has increased to values approaching that of the stellar radius and the star
became transparent to neutrinos.  Beyond this time, diffusion is no longer a
valid approximation to the transport equations; however, the neutrino
luminosities are very low (see below) and perhaps even undetectable.

Qualitatively, the simulations of BL are similar: the central entropy rises
from $s=1$ to the peak value of about $s=2$ in about 15 seconds, and neutrino
transparency occurs after about 50 s.  However, one  significant difference is
that at the beginning of the central cooling epoch, when the central
temperature reaches a maximum, there are still substantial numbers of trapped
neutrinos present ($\mu_\nu>>T$).   Consequently, in our simulations the
deleptonization continues through most of the cooling epoch as well.  We
attribute this difference to differences in the treatment of the nuclear
absorption processes.  In the calculations of BL, the absorption cross section
does not contain any explicit temperature dependence and, further, is assumed
to vanish when $Y_e$ fell below a critical value of about 1/9.  The value 1/9
corresponds to the critical proton fraction below which the direct Urca process
is prohibited at zero temperature (Lattimer et al. 1991), in the absence of
muons (both BL and this paper neglect muons).   However, as shown in
Figure~\ref{csig}, the absorption cross section does not behave as a step
function as long as the temperature is above a few MeV. In addition, BL used a
symmetry energy with almost no density dependence, with the result that
$\mu_n-\mu_p$, and consequently the equilibrium value of the proton fraction,
decreased to the critical value relatively early in the deleptonization.
Hence, in our calculation, absorption contributes to the opacities for a much
longer time than in BL, so that the decrease of $\mu_\nu$ in the central
regions is suppressed.

Despite the differences in the evolution of $\mu_\nu$, the long-timescale
cooling is dominated by mu- and tau-neutrinos. The upper panels of
Figure~\ref{lambda} shows neutrino mean free paths for an average energy,
$<E_\nu>=\mu_\nu+\pi T$, for four times during the evolution.  The mu- and
tau-neutrino mean free paths are significantly larger than those for electron
neutrinos.  Note that the mean free paths in the central regions initially
shrink as the star becomes denser and hotter, but by 15 s they begin to
increase due to the decreasing temperature.  It is also clear that the mu- and
tau-neutrino mean free paths are more sensitive to the temperature than they
are to the density, and their temperature dependence is greater than for
electron neutrinos.  The profiles for $\lambda_{\nu_\mu}$ are nearly mirror
images of the temperature profiles for each time.  Despite the fact that
$\lambda_{\nu_\mu}/\lambda_{\nu_e} > 1000$ at early times in the core, there
are so few mu- and tau-neutrinos relative to electron-neutrinos that the
transport is dominated by $\nu_e$s.  However, by 15 s,
$\lambda_{\nu_\mu}/\lambda_{\nu_e} < 100$ and the thermal evolution becomes
dominated by mu- and tau-neutrinos.

\subsection{Dependence Upon Initial Conditions}

A major uncertainty exists in the specifications of the initial entropy and
$Y_L$ profiles.  These are not accurately known because they depend on the
details of the EOS and weak interaction physics employed both during the late
evolution of the precollapse stars and the collapse of the core (Wilson \&
Mayle 1989, Swesty, Lattimer, \& Myra 1994, Mezzacappa \& Bruenn 1993). Ever
since the work of Bethe et al. (1979), it has been realized that the inner core
of the PNS should have canonical values of $s\approx1$ and $Y_L\approx0.38$.
Depending upon electron capture and beta decay rates and upon the symmetry
energy, the entropy could be 50\% larger or smaller and $Y_L$ could vary by
$\pm0.03$.  In addition, the $s$ and $Y_L$ profiles in the shocked matter at
enclosed baryon masses greater than 0.6 M$_\odot$ (in the $M_B=1.6$ M$_\odot$
star)  depend upon the strength of the outgoing shock and also, perhaps, upon
convective instabilities.

We performed  several simulations of the $M_B=1.6$ M$_\odot$ GM3np star  in
order to study the sensitivity of the early evolution to the $s$ and $Y_L$
profiles at $t=0$.  The spatial variation of various physical quantities for
$t=0, 1, 5$, and 15 seconds for models with lower (higher)  initial central
entropies are shown as dashed curves in Figure~\ref{lows} (Figure~\ref{highs}),
compared to the baseline model, which are displayed as solid lines.   These
variations do not produce dramatic effects, but the differences in
thermodynamic quantities, although damped, persist to late times.  Lowering the
initial central entropy results in a higher value for $\mu_\nu$, since the
central density is increased because of the lessening of thermal pressure.
However, by 15 s, the values of $\mu_\nu$ actually fall below that of the
baseline simulation, because the evolution is more rapid due to the lower
temperatures and relatively longer mean free paths.  Exactly the opposite
behavior occurs   when the initial central entropy is increased.

In Figure~\ref{highyl}, we show the effects of increasing the lepton
concentrations throughout the star; the central value of $Y_L$ is increased
from 0.36 to 0.40.  Since there are now more leptons to lose, the simulation
results in more heating, which contributes toward longer evolution time scales.
However, the effects are far smaller than in the case in which the core
entropies were changed.

We also considered variations of the entropy in the shocked mantle
material; this case
is shown as dotted lines in Figure~\ref{highs}. Note that although the entropy
is larger at all masses in the initial model, the temperatures are uniformly
lower.  This counterintuitive situation results from the additional expansion
of the outer stellar layers: the density at all masses are also smaller in this
case.  Nevertheless, due to the rapid
and extreme compression of the outer layers, these variations are
completely swamped and there is virtually no difference between this and the
baseline cases visible after a second.  The same is true of
simulations with  variations in the lepton concentration in the shocked mantle
material, which we do not display.
Variations of entropy and lepton fraction in these outer regions
manifest themselves as short-term (a few tenths of a second) transients in
the neutrino luminosities, which are discussed in \S 5.5.   We can conclude
that only variations of the initial conditions  in the central regions of the
PNS are relatively important in the long-term evolution.

A useful summary of the effects of variations in initial conditions is
contained in Figure~\ref{lumeini}, which displays the total neutrino luminosity
as a function of time.  Varying the central values of the entropy produce
persistent differences in the luminosity.  This is due to two factors: first,
the initial binding energies of lower entropy models is larger, so the
integrated luminosities must be smaller, and second, the evolution time scales
of the lower entropy models must be more rapid, due to the lower temperatures
and longer neutrino mean free paths.  Similar behavior is produced by varying
the central values of $Y_L$: higher values of $Y_L$ reduce the initial binding
energy and also lead to more entropy generation, higher temperatures, and
longer diffusion times.  Altering entropy and lepton profiles in the shocked
mantle has little effect except in the first few tenths of a second.

The observed variations in luminosity and average energy due to uncertainties
in initial conditions are relatively small, but nevertheless will contribute to
uncertainties in distinguishing EOS and compositional variations from an
observed neutrino signal.

\subsection{EOS Dependence}

Equations of state differ in many respects at high densities.  Most
significant for the PNS are variations in compressibilities, symmetry
energies, specific heats, and compositions.  To study these effects, we
consider three relativistic field-theoretical models with different high
density properties. For two of these models, we also consider the
possible presence of hyperons.  These models do not necessarily encompass the
entire range of possible behaviors.  For example, we could have considered
models with more extreme density dependences of the baryon effective masses,
which are important in the determination of the specific heats, or
the symmetry energy, which plays a role in determining the composition and
the neutron star radius.  However, the results of such
variations may be gauged from the results we present here.

The left panels of Figure~\ref{therm_eos} and Figure~\ref{lept_eos} compare the
evolutions of a PNS of an $M_B=1.6$ M$_\odot$ star for the
parameter sets GM3 and GM1 for nucleons-only matter for the 1.6 M$_\odot$
baryon mass star.  GM1 is a stiffer EOS, and it has a lower
value of the baryon effective mass $M^*$.  Therefore, the densities in the
GM3np models are uniformly greater than in the GM1np models.  The GM3np
temperatures are also greater; the higher densities more than compensate for
the greater specific heat compared to GM1np.  As a result of the higher
temperatures and densities, the neutrinos are more tightly trapped,
there is more entropy generation, and the evolution times are longer in the
GM3np models.  When hyperons are included, as shown in the right panels of
Figure~\ref{therm_eos} and Figure~\ref{lept_eos}, the GM1 and GM3 models
display the same relative trends.

Comparing models with and without hyperons, {\it i.e.,} the right and left
panels of Figure~\ref{therm_eos} and Figure~\ref{lept_eos}, the initial models
($t=0$ s) with hyperons show a small amount of softening.  The central density
of the hyperon stars becomes progressively larger than that of the purely
nucleon stars, largely because of the appearance of thermal hyperons.  The
central densities of the hyperon stars remains larger even at late times when the
remnant is cooling because hyperons are present.  Little difference in the
temperature evolution of these cases is apparent even after 20 s.
Nevertheless, the evolution time scale of hyperon stars is slightly larger than
for purely nucleon stars, due to the smaller mean free paths of hyperonic
matter (see Figure~\ref{lambda}).  An interesting feature is the appearance
of a trough in $Y_e$ and $Y_\nu$ near 0.75 M$_\odot$ at $t=0.5$ s.  This
feature is a short lived transient, because it does not appear in
Figure~\ref{highyl} which shows $Y_L$ for t=1 s.

The major effects on the temporal evolution are more easily seen by examining
Figure~\ref{mod_cquant}, which shows the time development of important physical
variables at the star's center.  It is clear that deleptonization continues
during the subsequent  cooling epoch. One should also note the plateau in the
decay of $Y_\nu$ and $\mu_\nu$ in the case of GM3np and GM1np.  This is due to
the density sensitivity of the neutrino mean free paths.  With hyperons, larger
central densities and higher electron neutrino energies are reached.  This
increases the neutrino opacity, temporarily reducing the loss of
neutrinos from the stellar core.

Figure~\ref{models} shows the luminosities and the average neutrino
energies as a function of time for the different EOS models, and for
the baryon masses $M_B=1.6 M_\odot$ and $M_B=1.8 M_\odot$. The
luminosities are virtually identical until about 10 s, at which point
the luminosity for models with a stiffer high density EOS decrease
more rapidly. Contrasting the models with and without hyperons, we
find a similar trend, with the hyperon models being able to sustain a
higher luminosity at late times. The average energy of the emitted
neutrinos are more sensitive to the EOS; once again a clear trend
emerges in which models with a stiffer high-density EOS favor lower
average energies. The average energy for models with hyperons is about
10-15\% larger than those observed in matter containing only nucleons.
These contrasting features are accentuated at late times and with
increasing mass.

We emphasize that for the mass (1.6 M$_\odot$) and coupling constants
chosen for this comparison relatively few hyperons are present at
early times (see left panel of Figure~\ref{hyperons}).  In fact, the
maximum concentration of hyperons at $t=0$ is about 0.01\%. However,
as the density and temperature in the star is raised during
deleptonization, the number and effect of hyperons becomes
greater. Hyperons do not appear only at the star's center; the large
temperatures in the mantle around 1 M$_\odot$ result in a higher
concentration of hyperons there than at the center from 0.25 s to 2 s.
At late times, the central density of the npH star is nearly 20\%
larger than the np star and the concentration of hyperons at the
star's center is about 13\%. For the mass (1.8 M$_\odot$), a
significantly larger number of hyperons are present as can be seen
from the results shown in the right panel of
Figure~\ref{hyperons}. The interesting feature to note is that at
until about 15 s, the hyperons fractions are quite small, indicating
that deleptonization in the inner core occurs very slowly. This is one
of the reasons why we see very small changes in the observed
luminosities between np and npH models at early times. In addition,
feedbacks between the EOS and the opacities tend to reduce the differences.

\subsection{Mass Dependence}

We considered the effect of varying the initial mass of the PNS.  For
this study we employed the GM3 parameterization, and considered both
np and npH matter. The actual baryon masses used in the simulations
are listed in Table 3. The results displayed in Figures~\ref{models},
\ref{mass_cquant}, and \ref{mass} indicate that the most significant
effect of increasing the initial mass is to increase the overall
time scales of evolution, and to increase both the total and average
energies released.  The central heating also becomes more significant as one
increases the remnant mass.  Although the central entropy shows very
little increase with increasing mass, the higher density of larger mass
models translates into significant increases in central temperature.
The combination of higher densities and temperatures lengthens the
time constants associated with heating, cooling, and deleptonization.
This naturally results in a longer duration of the plateau stage in
the evolutions of $Y_\nu$ and $\mu_\nu$ for np models.  Note that
these plateaus do not appear for npH matter for any mass.  In spite of
the larger time scales, the larger energy release, and increased
temperatures  larger mass models lead to larger luminosities at all
times.

\subsection {Metastability}

The $M_B=1.8$M$_\odot$ baryon mass star containing hyperons is metastable.
After more than 100 s, this star will collapse to a black hole since the
maximum baryon mass of the cold, catalyzed GM3npH EOS is 1.75 M$_\odot$.  The
possibility of metastability has been suggested in the context of
SN 1987A, for which no neutron star remnant has been observed (Bethe \&
Brown 1995, Prakash, Cooke, \& Lattimer 1995,
Ellis, Lattimer, \& Prakash 1996, Prakash et al. 1997).  The lack of a neutron
star in this object seems peculiar, because
a neutron star rotating with the frequency of the Crab pulsar and with a
magnetic field only a fraction of the Crab's field would be contributing
noticeably to the light curve of the expanding gaseous remnant of SN 1987A.
This is not observed.  Although one cannot yet rule out the existence of a
neutron star on the basis of Eddington-limit accretion, because the effective
opacity may be due to iron instead of hydrogen and therefore be much greater, a
black hole remnant seems a possibility.  This hypothesis can be reconciled with
the neutrino emission observed from SN 1987A if the PNS was metastable
because of the existence of hyperons or a kaon condensate
in cold, catalyzed dense matter.

For this reason, we further explored the case of a metastable PNS for the GM3
parameterization.  Figure~\ref{bhole_ct} shows the time development of the
central baryon density (top panel) and also the time to the collapse
instability as a function of baryon mass (bottom panel).  The larger the mass,
the shorter the time to instability, since the PNS doesn't have to evolve in
lepton number as much.  Above 2.005 M$_\odot$, the metatstability disappears
because the GM3npH initial model with the lepton and entropy profiles we chose
is already unstable.  Below about 1.73 M$_\odot$, there is no metastability,
since this is the maximum mass of the cold, catalyzed npH star for GM3.  These
values are listed in Table 2, which contains corresponding, but larger, values
for the case of GM1.  The signature of neutrino emission from a metastable PNS
should be identifiable and it is discussed in \S 5.7.

Figure~\ref{bhole_mr} shows the behavior of the quantity $GM_G/Rc^2$ as
functions of both the radius and the enclosed baryon mass for a GM3npH model
with $M_B=1.88$ M$_\odot$.  The time to instability for this model is slightly
more than 45 s.  Although the actual collapse to form a black hole is dynamic,
and in any event could never be seen from the outside, one could estimate that
the event horizon is likely to form where the redshift is largest.
With time, it appears that this point moves outwards in the star, so that the
event horizon can be expected to form relatively close to the star's physical
surface and, therefore, outside the star's neutrinosphere.

\subsection{Convective Instabilities }

The possibility of convection in PNS has been discussed by a number of authors
(Burrows \& Fryxell 1993, Herant et al. 1994, Keil, Janka, \& M\"uller 1996,
Mezzacappa et al. 1998).
The decreasing entropy gradient in the star's exterior is
obviously convectively unstable, and this gradient works its way into the
interior on the deleptonization time scale.  However, the complete determination
of convective instability also requires knowledge of the lepton gradients.
Explicitly, we consider the Ledoux criterion for convectively unstable regions:
the condition is $C_L(r) > 0$,  where
\begin{eqnarray}
C_L(r) = - \frac {1}{ \left( \frac{\partial P}{\partial n_B} \right)_{s,Y_L} }
 \left[ \left( \frac{\partial P}{\partial s} \right)_{n_B,Y_L} \frac {ds}{dr}
+  \left( \frac{\partial P}{\partial Y_L} \right)_{s,n_B} \frac {dY_L}{dr}
\right] \,.
\end{eqnarray}
We show in Figure~\ref{convec} the unstable regions of a PNS for the
GM3np and GM3npH models for $M_B=1.6$ M$_\odot$ as shaded regions.
Generally, the outermost few tenths of a solar mass is initially
unstable.  After a few seconds, the instability moves inwards,
reaching the center after 10--15 seconds.  We do not allow for
convection in our simulations, so the result that after a few seconds
the exterior of the star becomes stable against convection is
unreliable.  In fact, the calculations of Keil, Janka, \& M\"uller
(1996) show that the convection proceeds similarly to what our models
show, but that the exterior remains convectively unstable throughout
the evolution. Their time scales are, however, much shorter in
comparison to our findings. The influence of convection upon the
luminosities and average neutrino energies has been shown by prior
authors (Burrows \& Lattimer 1988) to be relatively significant, and
we will consider this effect in the next generation of our
calculations.
\subsection{Signals in Terrestrial Detectors}

Knowledge of the neutrino luminosities and energies permit the signal
from a supernova to be calculated.  To be specific, we will present
results in a way that makes comparison to the signal observed from SN
1987A (Bionta et al. 1987, Hirata et al. 1987) transparent.  We assume two
fiducial detectors, KII and IMB, whose masses, thresholds and efficiencies are
taken to be the same as in Lattimer \& Yahil (1989).  We focus on the
antineutrino signal, assuming that the antineutrino luminosity $L_{\bar\nu}$ is
1/6 of the total.

We assume a pure water detector for which the cross section of
antineutrinos absorbed onto protons is $\sigma(E_\nu)=\sigma_0E_\nu^2$
where $E_\nu$ is the antielectron neutrino energy in MeV and
$\sigma_0=9.3\cdot10^{-44}$ cm$^2$.  There are $n_p=6.7\cdot10^{28}$
free protons per kiloton of water.  We take the distance to SN
1987A to be $D=50$ kpc.  Assuming that the neutrinos leaving the
protoneutron star have a Fermi-Dirac spectrum with zero chemical
potential, the count rate is
\begin{equation}
{dN\over dt} = \left({R_\nu^\infty\over D}\right)^2
{c\sigma_0 n_p\over8\pi(\hbar c)^3}
{\cal M}\int_{E_{th}}^\infty E_\nu^4 f(E_\nu,T_\nu^\infty)W(E_\nu) dE_\nu \,,
\end{equation}
where
$E_{th}$ is the detector threshold, $W$ is the detector efficiency and
${\cal M}$ is
the detector mass in kilotons.  Also, $R_\nu^\infty=e^{-\phi_s}R_\nu$ and
$T_\nu^\infty=e^{\phi_s}T_\nu$, where $e^{\phi_s}=\sqrt{1-2GM_G/R_\nu c^2}$
and $M_G$  and $R_\nu$ are the gravitational mass of the neutron star and the
radius of the neutrinosphere, respectively.  Since the luminosity is calculated
more precisely than the neutrinospheric radius and temperature, we
rewrite this, substituting the antineutrino luminosity, as
\begin{equation}
{dN\over dt}={\sigma_0n_p\over 4\pi D^2}{\cal M} {G_4(E_{th},
e^{\phi_s}T_\nu)\over F_3(0)} e^{\phi_s} T_\nu L_{\bar\nu}  \,,
\end{equation}
where $F_3(0)=7\pi^4/120$ is an ordinary Fermi integral and
$G_i(E_{th},T)$ denotes a modified, truncated Fermi integral:
\begin{equation}
G_i(E_{th},T)=\int_{E_{th}}^\infty dz\, z^i W(zT)(1+e^z)^{-1} \,.
\end{equation}

In Figure~\ref{count} we display the count rate and integrated
counts for KII and IMB for the $M_B=1.6$ and $1.8$ M$_\odot$ models
for the various EOS cases studied.  The results can be easily
interpreted in terms of the average energies illustrated in
Figure~\ref{models}.  While the total luminosities remain essentially
identical for all cases of a given mass up to about 20 seconds, the
average energies show a systematic variation.  The softer the
underlying EOS, the larger the average energy, and this directly
translates into higher count rates.  The general trends, which are
explainable by the more extreme temperature sensitivity of the IMB
detector, are that more counts accumulate in KII than in IMB and the
time over which a signal is observed is also larger for KII than for
IMB.

The total variation among the EOS models in the accumulated
counts after 40 seconds is about 10\% for a given mass.  The mean
ratio of total counts between the 1.8 M$_\odot$ and 1.6 M$_\odot$
cases is about 1.5.  Therefore, it seems that for a Galactic supernova
observed in the near future, in which the expected number of counts
will be substantially larger because of its proximity and larger
detectors, that the protoneutron star mass could be measured to better than
0.05 M$_\odot$ from the total number of counts observed.  It also seems that
the more subtle differences between EOS models will be observable only if the
accumulations after about 20 seconds are large enough.

\subsection{Comparison with Earlier Work}

We turn now to a comparison of our results with the earlier works of BL and
KJ on the early evolution of protoneutron stars containing only nucleonic
matter.  There are important differences arising due to the
different input physics.  In these earlier works the
neutrino opacities were not
calculated consistently with the dense matter EOS.
Thus, it is possible to compare only some aspects of previous
calculations with our own.  These aspects include the luminosity
evolution and the average neutrino energies, as well as the evolution
of temperature and neutrino number in the interior.

BL and KJ both found that, for 1.4--1.6 M$_\odot$ star, the {\it
deleptonization} time, the time needed for the electron neutrino
number $Y_{\nu_e}$ to vanish in the star's center, was about 10 s (see
Figure 8 in BL \& Figure 5 in KJ).  In addition, they both found that the
central temperature reached a maximum at about the time at which the
neutrino number vanished.  In contrast, we generally find that these
two time scales are not similar: while the time required for the
central temperature to reach a maximum is of order 10 seconds, the
deleptonization time is much longer, at least 30 seconds.  We
attribute the more rapid deleptonization in earlier works to
inadequate treatments of the temperature dependence of the direct Urca
process, the nuclear symmetry energy and the matter's specific heat.
In our models, the larger electron neutrino cross sections at high
temperature and low neutrino numbers stem the flow of electron
neutrino number.

Comparing the calculations of BL in more detail, we find that the
total luminosity is larger during the deleptonization era.  After 20
seconds in BL, nearly 1/3 of the binding energy has yet to be radiated
(see Figure 19 in BL), while in our calculations at the same time less
than 10\% remains.  Burrows, Klein, \& Gandhi (1992) have argued that
the neutrino luminosity evolves as a power law at late times ($t$
larger than a few seconds), while it decays exponentially in ours.  As
Prakash et al. (1995) demonstrated, an effective $E^2$ behavior of the
neutrino cross sections should result in an exponential decrease of
temperature and luminosity for the star.

Another apparent difference between the calculations of BL and ours
concerns the average emitted neutrino energy: in our calculation this
peaks at about 5 s, while in BL it steadily decreases with time (see
Figure 20 in BL).  This difference could be important for the
questions of neutrino-assisted shocks and for nucleosynthesis.
However, it should be borne in mind that we estimated the average
energy by the relation $<E_\nu>=3T_\nu$, but this neglects
possible effects of neutrino degeneracy and the fact that the emergent
spectrum may not be Fermi-Dirac.  It will be necessary to employ more
accurate transport in the vicinity of the neutrinosphere before firm
conclusions can be drawn.

We see more substantial differences from the calculations of KJ.
During the first few seconds of the evolution, KJ find that the
central density decreases (see Figure 10 in KJ).  KJ attribute this
decrease to the heating which occurs during deleptonization.  Indeed,
KJ find a large increase in central temperature during the first five
seconds (see Figure 11 in KJ) which is much more rapid than in our
calculations.  The effects of rapid deleptonization and rapid
temperature increase could be explained by a charged-current mean free
path that is several times larger than in our calculation.  On the
other hand, the cooling time scales, which are controlled by neutral
current processes, are rather similar to what we have found.  This
probably also accounts for the remarkable similarities between our and
KJ's total neutrino luminosities, despite very different
delptonization times.

A decrease in central density during the deleptonization epoch could
happen if the thermal contributions to the pressure at 3--5 times the
nuclear density were a significant part of the total.  However, this
would imply that the specific heat of the matter is radically
different than what we find.

We can also compare our protoneutron star evolutions to KJ for cases
in which hyperons appear.  In contrast to their calculations, we find
only minor differences in the evolution of nucleons-only and
hyperon-bearing models at early times.  In fact, for most masses
studied, the evolutions of luminosity in our calculations for the two
cases are nearly identical for about 20 seconds.  KJ find that large
differences in the luminosity occur between np and npH cases as early
as 2--4 seconds (see Figure 9 in KJ) in the evolution.  They found a
time to collapse in the metastable case of less than 20 seconds when
the stellar mass was 0.02 M$_\odot$ larger than the critical mass (see
Table 2 in KJ).  For the same mass difference, we find a time to
collapse of more than 100 seconds.  These differences no doubt stem
from the same effects that led to the large differences apparent
between KJ's and our nucleon models.  Our relatively large
deleptonization times prevent hyperons from flooding the interior
until rather late times. This significantly increases the time spent
in the metastable phase when the mass is only marginally above the
maximium mass for cold, deleptonized matter (see Figure
\ref{bhole_ct}).

\section{Summary and Outlook}

In this paper, we presented calculations of protoneutron star
evolutions which used neutrino opacities that were consistently
determined from the underlying dense matter EOS. We
explored the sensitivity of the results to the initial stellar model,
the total mass, the underlying EOS, and the addition of hyperons.
We find that the differences in predicted luminosities and emitted
neutrino energies do not depend much upon the details of the initial
models or the underlying high-density EOS for early times ($t<20$ s),
provided that opacities are calculated consistently with the EOS.  The
same holds true for models which allow for the presence of hyperons,
except when the initial mass is significantly larger than the maximum
mass for cold, catalyzed matter.  Increasing the total mass tends to
increase luminosities and average energies at all times.  For times
larger than about 10 seconds, and prior to the occurrence of neutrino
transparency, the neutrino luminosities decay exponentially with a
time constant that is sensitive to the high-density properties of
matter.  We also find the average emitted neutrino energy increases
during the first 5 seconds of evolution, and then decreases
nearly linearly with time.
{\it Metastable stars, those with hyperons
which are unstable to collapse upon deleptonization, have relatively
long evolution times, which increase the nearer the mass is to the
maximum mass supported by a cold, deleptonized star.}

There are several interesting and possibly important physical effects
that remain to be investigated.\\ (1) In terms of input neutrino
opacities, this work provides a baseline calculation in terms of which
further improvements (Reddy et al. 1997, Prakash \& Reddy 1997,
Burrows \& Sawyer 1998a;b, Hannestad \& Raffelt 1998) such as the
effects of many-body correlations, axial charge renormalization and
multi-pair excitations may be assesed.  \\ (2) While the diffusion
approximation is valid for a significant portion of the star at early
times, neutrino transport in the semi-transparent regions is better
handled in a multigroup transport scheme. Better transport in the
semi-transparent regions will play an important role in determining
the average neutrino energies for the different neutrino species and
their energy spectra.  \\ (3) An adequate treatment of convection
coupled with neutrino transport appears to be necessary based on the
large regions we identified as being convectively unstable. \\ (4)
Other softening components in dense matter might produce effects
dissimilar to those we find when considering hyperons.  For example,
the appearance of a Bose condensate differs from the appearance of
hyperons because the condensate does not contribute to the matter's
specific heat. In addition, the presence of a condensate could produce
distinctive modifications to the neutrino opacities. 

Results of calculations incorporating these improvements will be
reported subsequently.\\

We gratefully acknowledge encouragement and suggestions from Adam Burrows. We
thank Thomas Janka and Wolfgang Keil for sharing information concerning some of
the initial models used in this work.  This work was supported in part by the
U.S. Department of Energy under grant numbers DOE/DE-FG02-88ER-40388 and
DOE/DE-FG02-87ER-40317, and by the NASA grant NAG52863.  J. Pons gratefully
acknowledges research support from the Spanish DGCYT grant PB94-0973 and
thanks J.M. Iban\~{n}ez for useful discussions.


\newpage
\vspace*{0.2in}
\centerline {REFERENCES}
\begin{tabbing}
xxxxx\=xxxxxxxxxxxxxxxxxxxxxxxxxxxxxxxxxxxxxxxxxxxxxxxxxxxxxxxxxxxxxxxxxxx\kill

\ni Bethe, G.E., \& Brown, G.E. 1995,
ApJ, 445, L129 \\

\ni Bethe, G.E., Brown, G.E., Applegate., J., \& Lattimer, J.M. 1979,
Nucl. Phys., B348, 345 \\

\ni Bionta, R. M., et al., 1987, Phys. Rev. Lett., 58, 1494 \\

\ni Bowers, R. L., \& Wilson, J. R. 1982, ApJS, 224 50, 115 \\

\ni Brown, G.E., \& Rho, M. 1991, Phys. Rev. Lett., 66, 2720 \\

\ni Bruenn, S., 1985, ApJS, 58, 771 \\

\ni Burrows, A., 1988, ApJ, 334, 891 \\

\ni -------, 1990, Ann. Rev. Nucl. Sci., 40, 181 \\

\ni  Burrows, A., \& Fryxell, B. 1993, ApJ, 418, L33 \\

\ni Burrows, A., Hayes, J., \& Fryxell, B. A. 1995, ApJ, 450, 830 \\

\ni  Burrows, A., \& Goshy, J. 1993, ApJ, 416, L75 \\

\ni Burrows, A., Klein, D., \& Gandhi, R. 1992,
Phys. Rev. D, 45, 3361 \\

\ni Burrows, A., \& Lattimer, J. M. 1986, ApJ, 307, 178 \\

\ni Burrows, A., \& Lattimer, J. M. 1988, Phys. Rep., 163, 51 \\

\ni Burrows, A., \& Sawyer, R. F. 1998a, Phys. Rev. C, In press \\

\ni Burrows, A., \& Sawyer, R. F. 1998b, Phys. Rev. Lett., (submitted) \\

\ni Ellis, P.J., Lattimer, J.M., \& Prakash, M. 1996, Comments on Nucl. \&
Part. Phys., 22, 63 \\

\ni Glendenning, N.K., \& Moszkowski, S. 1991, Phys. Rev. Lett., 67, 2414 \\

\ni Hannestad, S., \& Raffelt, G. 1998, ApJ (in press) \\

\ni Herant, M., Benz, W., Hicks, J., Fryer, C. \& Colgate, S.A. 1994,
ApJ, 435, 339 \\

\ni Hirata, K., et al., 1987, Phys. Rev. Lett., 58, 1490 \\

\ni Horowitz, C.J., \& Wehrberger, K. 1991a, Nucl. Phys., A531, 665 \\
\ni ---------, 1991b, Phys. Rev. Lett., 66, 272  \\
\ni ---------, 1992, Phys. Lett. B, 226, 236  \\

\ni Iwamoto, N., \&  Pethick, C. J. 1982 Phys. Rev. D, 25, 313 \\

\ni Janka, H. T., Keil, W., Raffelt, G., \& Seckl, D.
 1996, Phys. Rev. Lett., 76, 2621 \\

\ni Keil, W. 1994, Prog. Part. Nucl. Phys., 32, 105 \\

\ni Keil, W., \&  Janka, H.T. 1995, A\&A, 296, 145\\

\ni Keil, W., Janka, H.T., \& M\"uller, E. 1995, ApJ,
473, L111 \\

\ni Knorren, R., Prakash, M., \& Ellis, P. J. 1995, Phys. Rev. C, 52, 3470 \\

\ni Lattimer, J.M., Pethick, C.J., Prakash, M., \& Haensel, P. 1991,
Phys. Rev. Lett., 66, 2701 \\

\ni Lattimer, J.M., \& Yahill, A. 1989,
ApJ, 340, 426 \\

\ni Levermore, C. D., \& Pomraning, G. C. 1981, ApJ, 248, 321 \\

\ni Lindquist, R.W. 1966, Ann. Phys., 37, 478\\

\ni Mayle, R., Wilson, J.R., \& Schramm, D.N. 1987, ApJ, 318,
 288 \\

\ni Mezzacappa, A., \& Bruenn, S.W. 1993, ApJ, 405, 637 \\

\ni Mezzacappa, A., Calder, A. C., Bruenn, S. W., Blondin,  \\
J. M., Guidry, M. W., Strayer, M. R., \& Umar, A. S. 1998, ApJ, 495,
911 \\

\ni Pons, J.A., Miralles, J.A. \& Iba\~nez, J.-M. 1998, A\&AS,
 129, 343 \\

\ni  Prakash, M., Bombaci, I., Manju Prakash, Ellis, P.J.,
Lattimer, J.M., \& Knorren, R. 1997, \\~~Phys. Rep., 280, 1 \\

\ni Prakash, M., Cooke, J., \& Lattimer, J.M. 1995,
Phys. Rev. D, 52, 661  \\

\ni Prakash, M., \& Reddy, S. 1997, in Nuclear Astrophysics, \\~~
in Proceedings of the
International  Workshop XXVI \\~~
on Gross Properties of Nuclei and Nuclear Excitations, \\~~
Hirschegg,  Austria, Jan 11-17, \\~~
ed M. Buballa, W. N\"orenberg, J. Wambach, \& A. Wirzba,
(GSI: Darmstadt), pp. 187-200 \\

\ni Raffelt, G. \& Seckel, D. 1995, Phys. Rev. D, 52, 1780 \\

\ni Reddy, S., Pons, J., Prakash, M., \& Lattimer, J.M. 1997, in
~~Second Oak Ridge Symposium \\~~
on Atomic and Nuclear Astrophysics, Oak Ridge, Tennessee, Dec 2-6, 1997 \\

\ni Reddy, S., \&  Prakash, M. 1997, ApJ, 423, 689 \\

\ni Reddy, S., Prakash, M., \& Lattimer, J. M. 1998, Phys. Rev. D, 58, 
013009 \\

\ni Rho, M. 1974, Nucl. Phys. A, 231, 493 \\

\ni Sawyer, R. F. 1975, Phys. Rev. D, 11, 2740 \\
\ni ------------, 1989, Phys. Rev. C, 40, 865 \\
\ni ------------, 1995, Phys. Rev. Lett., 75, 2260 \\

\ni Serot, B.D., \& Walecka, J.D. 1986, in  Advances in Nuclear Physics,
Vol 16,\\
~~ed. J.W. Negele \& E. Vogt, (New York: Plenum), 1 \\

\ni Sigl, G. 1996, Phys. Rev. Lett., 76, 2625. \\

\ni Suzuki, H. 1989, Ph.D Thesis, University of Tokyo \\

\ni Suzuki, H., \& Sato, K. 1992, in The Structure and Evolution of
Neutron Stars,\\
~~ed. D. Pines, R. Tamagaki,  \& S. Tsuruta,
(New York: Addison-Wesley), 276 \\

\ni Swesty, F.D., Lattimer, J.M., \& Myra, E. 1994, ApJ, 425,
195 \\

\ni Thorne, K. S. 1981, MNRAS 194, 439 \\

\ni Wilkinson, D.H. 1973, Phys. Rev. C, 7, 930 \\

\ni Wilson, J.R., 1985, in Numerical Astrophysics, \\~~
eds. J. Centrella, J. LeBlanc, and R. Bowers (Jones and Bartlett, Boston)
p. 422 \\

\ni Wilson, J.R., \& Mayle, R.W., 1989, in The Nuclear Equation of State,\\
~~Part A,  ed. W. Greiner and H. St\"ocker (New York: Plenum), 731 \\

\ni Woosely, S.E., Wilson, J.R., Mathews, G.J., Hoffman, R.D., \& Meyer, B.S.
1994, \\~~ApJ, 433, 229 \\

\end{tabbing}

\newpage

\begin{center}
\centerline{TABLE 1}
\vspace*{0.15in}
\centerline{NUCLEON-MESON COUPLING CONSTANTS }
\begin{tabular}{|c|cccccccc|}
\hline
{Model } & $K_0$ & $a_{sym}$ & $M^*/M$ & $g_{\sigma}/m_{\sigma}$
& $g_{\omega}/m_{\omega}$ & $g_{\rho}/m_{\rho}$ & $b$ & $c$
\\ \hline
GM1 & 300 & 32.5 & 0.70 & 3.434 & 2.674 &  2.100 & 0.002950 & $-0.00107$ \\
GM3 & 240 & 32.5 & 0.78 & 3.151 & 2.195 &  2.189 & 0.008659 & $-0.002421$ \\
GM4 & 240 & 27.5 & 0.78 & 3.151 & 2.195 &  1.862 & 0.008659 & $-0.002421$ \\
\hline
\end{tabular}
\end{center}
\vspace*{0.15in}
NOTE.-- Coupling constants for the GM (Glendenning \& Moszkowski 1991) models
which are calibrated at the nuclear equilibrium density $n_0=0.153~{\rm
fm}^{-3}$ to give an energy per particle of symmetric nuclear matter of
$-16.3$ MeV.  The compression modulus $K_0$, symmetry energy $a_{sym}$ and
Dirac effective mass $M^*$ for the different models are also listed.
\begin{center}
\centerline{TABLE 2}
\vspace*{0.15in}
\centerline{MAXIMUM GRAVITATIONAL(BARYONIC) MASSES}
\begin{tabular}{|c|c|c|c|}
\hline
{Model} &  $Y_{L}=0.4~(s=1)$ & $Y_{\nu}=0~(s=2)$
&$Y_{\nu}=0~(s=0)$    \\
\hline
GM1np  & $ 2.278(2.609) $ & $ 2.367(2.748) $ & $ 2.346(2.805) $  \\
GM3np  & $ 1.946(2.181) $ & $ 2.044(2.328) $ & $ 2.005(2.345) $  \\
GM1npH & $ 2.024(2.259) $ & $ 1.793(1.991) $ & $ 1.776(2.020) $  \\
GM3npH & $ 1.768(1.944) $ & $ 1.573(1.726) $ & $ 1.544(1.732) $  \\
\hline
\end{tabular}
\end{center}
\vspace*{0.15in}
NOTE.-- Symbols are $Y_L$: the total lepton fraction, $Y_{\nu_e}$: electron
neutrino fraction, and $s$: the total entropy per baryon (in units of
Boltzmann's constant).  Model designations are as in Table 1 with np referring
to nucleons-only matter and npH referring to matter including hyperons.

\begin{center}
\centerline{TABLE 3}
\vspace*{0.15in}
\centerline{BARYONIC MASSES OF PNS EVOLUTION MODELS}
\begin{tabular}{|c|c||c|c|}
\hline
Mass Label &  Actual Mass (M$_\odot$) & Mass Label &
Actual Mass (M$_\odot$) \\
\hline
1.08 & 1.0852 & 1.84 & 1.8434\\
1.4  & 1.4038 & 1.88 & 1.8843\\
1.6  & 1.6125 & 1.90 & 1.9038\\
1.8  & 1.7981 & 1.94 & 1.9413\\
1.81 & 1.8162 & 1.98 & 1.9863\\
\hline
\end{tabular}
\end{center}

\newpage

\section*{FIGURE CAPTIONS}

\vspace*{1cm}
\noindent FIG. 1.-- Relative concentrations in lepton-poor (left panels)
and lepton-rich (right panels)
matter at finite entropy as a function of baryon density $n_B$ ($n_0$ is the
nuclear equilibrium density). Top panels show results in nucleons-only matter.
The bottom panels are for matter including hyperons.

\vspace*{1cm}
\noindent FIG. 2.-- Total pressure including contributions from leptons in
lepton-poor (left panels) and lepton-rich (right panels)
matter at finite entropy as a function of baryon density $n_B$ ($n_0$ is the
nuclear equilibrium density). The symbol np refers to matter
with nucleons-only and npH to matter including hyperons.
\vspace*{1cm}

\noindent FIG. 3.-- Temperatures attained in lepton-poor (left panels)  and
lepton-rich (right panels)  matter at finite entropy as a function of baryon
density $n_B$ ($n_0$ is the nuclear equilibrium density).  The symbol np
refers to matter  with nucleons-only and npH to matter including
hyperons.
\vspace*{1cm}

\noindent FIG. 4.-- Lepton chemical potentials  in lepton-poor (left panel)
and lepton-rich (right panel)  matter at finite entropy as a function of baryon
density $n_B$ ($n_0$ is the nuclear equilibrium density).  The symbol np
refers to matter  with nucleons-only and npH to matter including
hyperons.
\vspace*{1cm}

\noindent FIG. 5.-- Binding energy as a function of baryon mass. The symbol
np refers to matter with nucleons-only and npH to matter including
hyperons at T=0 and Y$_\nu$=0.
\vspace*{1cm}

\noindent FIG. 6.-- Neutrino mean free paths in matter with nucleons only (left
panels).   Right panels show ratios of mean free paths  in matter without and
with hyperons. Abscissa is  baryon density $n_B$ ($n_0$ is the nuclear
equilibrium density). Top panels show scattering mean free paths  common to all
neutrino species.  The bottom panels show results for electron neutrino mean
free paths where absorption reactions are included. The neutrino  content is
labelled in the different panels.
\vspace*{1cm}

\noindent FIG. 7.-- Inverse neutrino mean free paths due to charged current
reactions versus temperature.
\vspace*{1cm}

\noindent Fig. 8.-- Comparison of scattering mean free paths in neutrino poor
matter at fixed entropy  for different EOSs in matter containing nucleons and
also hyperons.
\vspace*{1cm}

\noindent FIG. 9.-- Temporal variation of thermodynamic quantites
($s$: the entropy per baryon, $T$: the temperature, $n_B$: the baryon density,
$Y_\nu$: the net electron neutrino fraction, $\mu_\nu$: electron neutrino
chemical potential, and $Y_e$: the net electron concentration) inside a star
for the baseline GM3 model for nucleons-only matter.  Labels indicate time in
seconds. The abscissa shows the enclosed baryonic mass.
\vspace*{1cm}

\noindent FIG. 10.-- Typical electron-neutrino and muon-neutrino mean free
paths.
For the electron neutrino the average energy $<E_\nu>=\mu_{\nu_e}+\pi T$, while
for the muon and tau neutrinos $<E_\nu>=\pi T$. Top panels show temporal
variations for the baseline GM3 model in nucleons-only matter and the bottom
panels show results in matter including hyperons.
\vspace*{1cm}

\begin{figure}
\begin{center}
\epsfxsize=6.in
\epsfysize=7.in
\epsffile{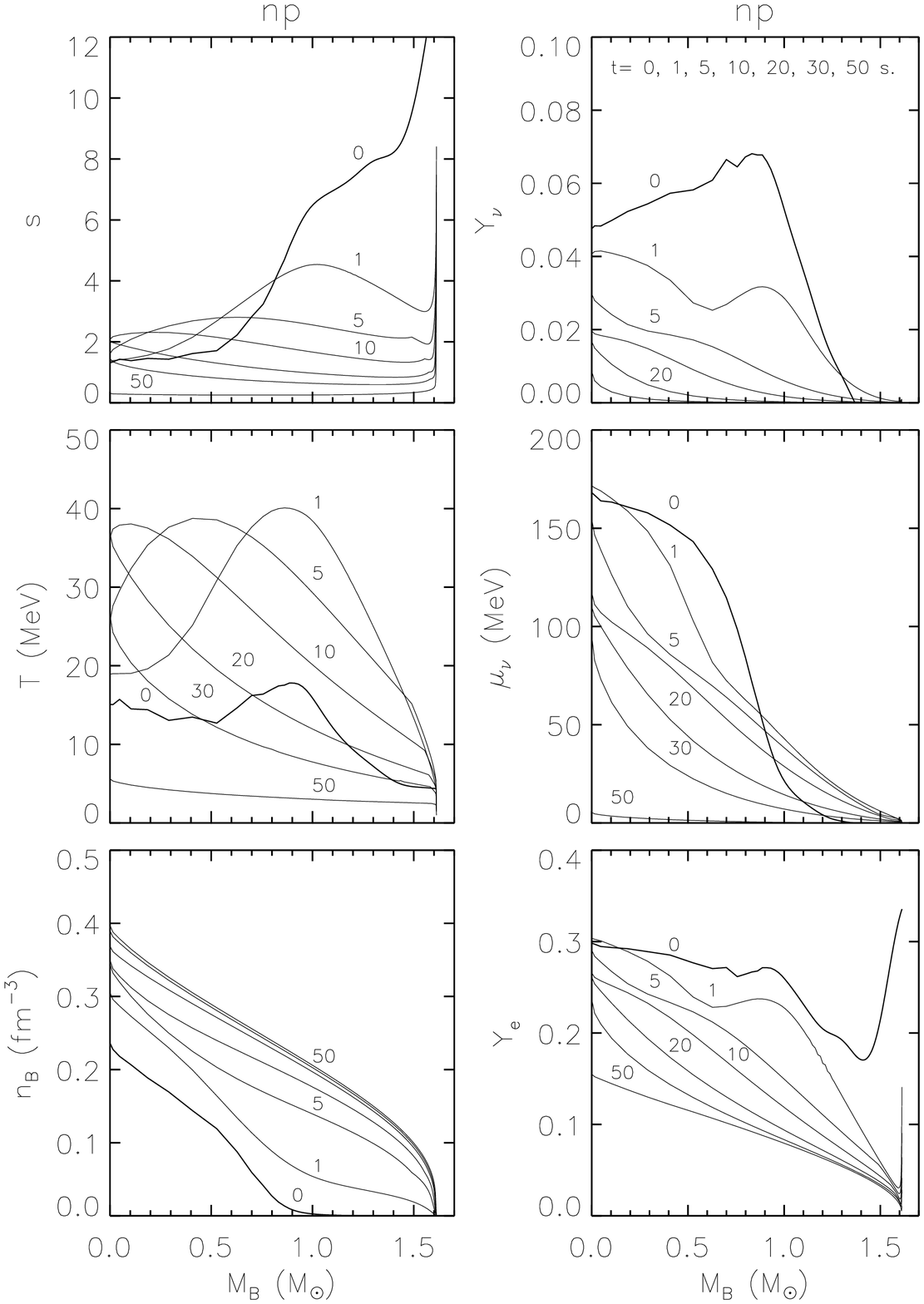}
\end{center}
\caption{}
{\label{baseline}}
\end{figure}

\begin{figure}
\begin{center}
\epsfxsize=6.in
\epsfysize=7.in
\epsffile{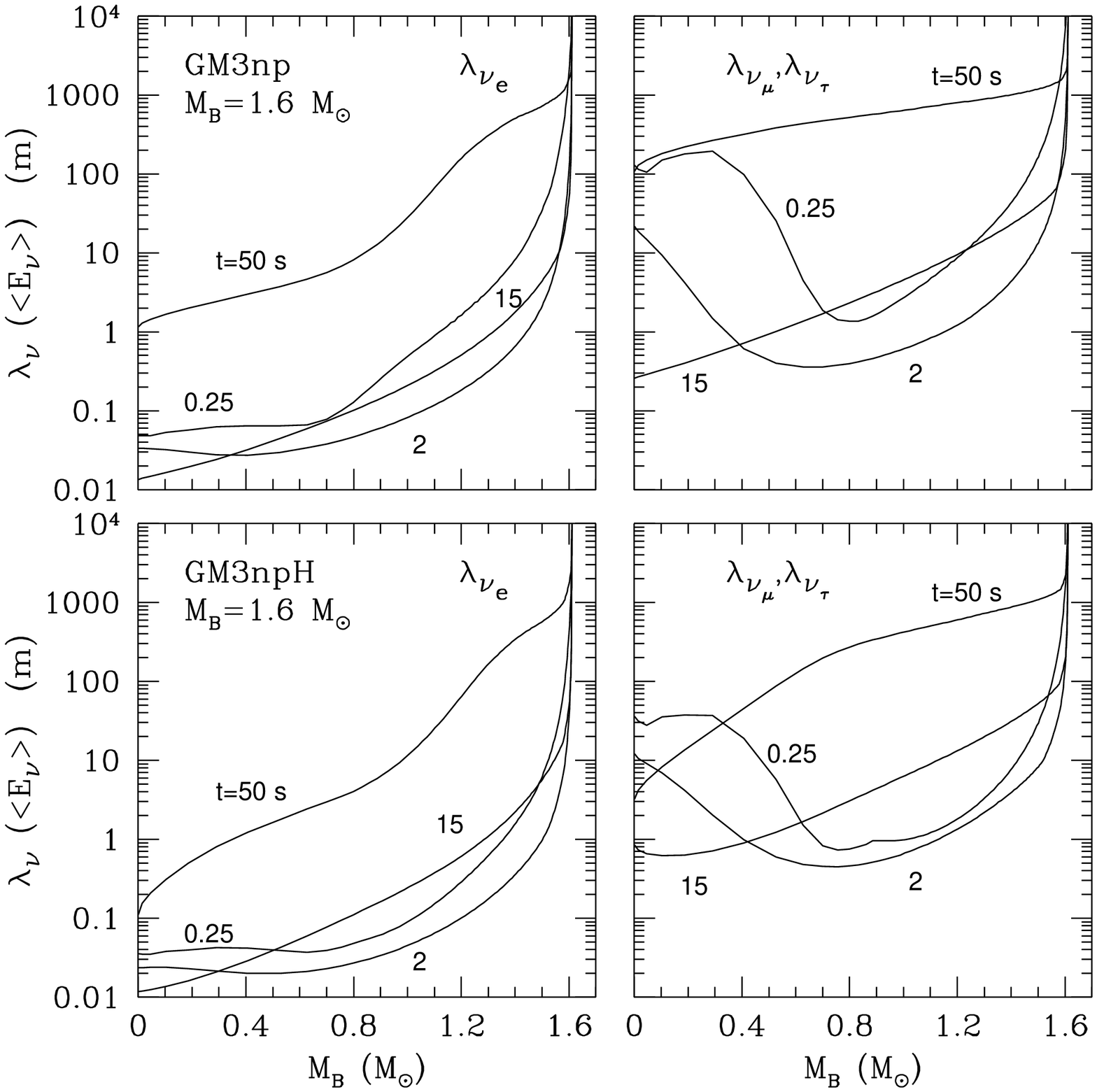}
\end{center}
\caption{}
{\label{lambda}}
\end{figure}

\begin{figure}
\begin{center}
\epsfxsize=6.in
\epsfysize=7.in
\epsffile{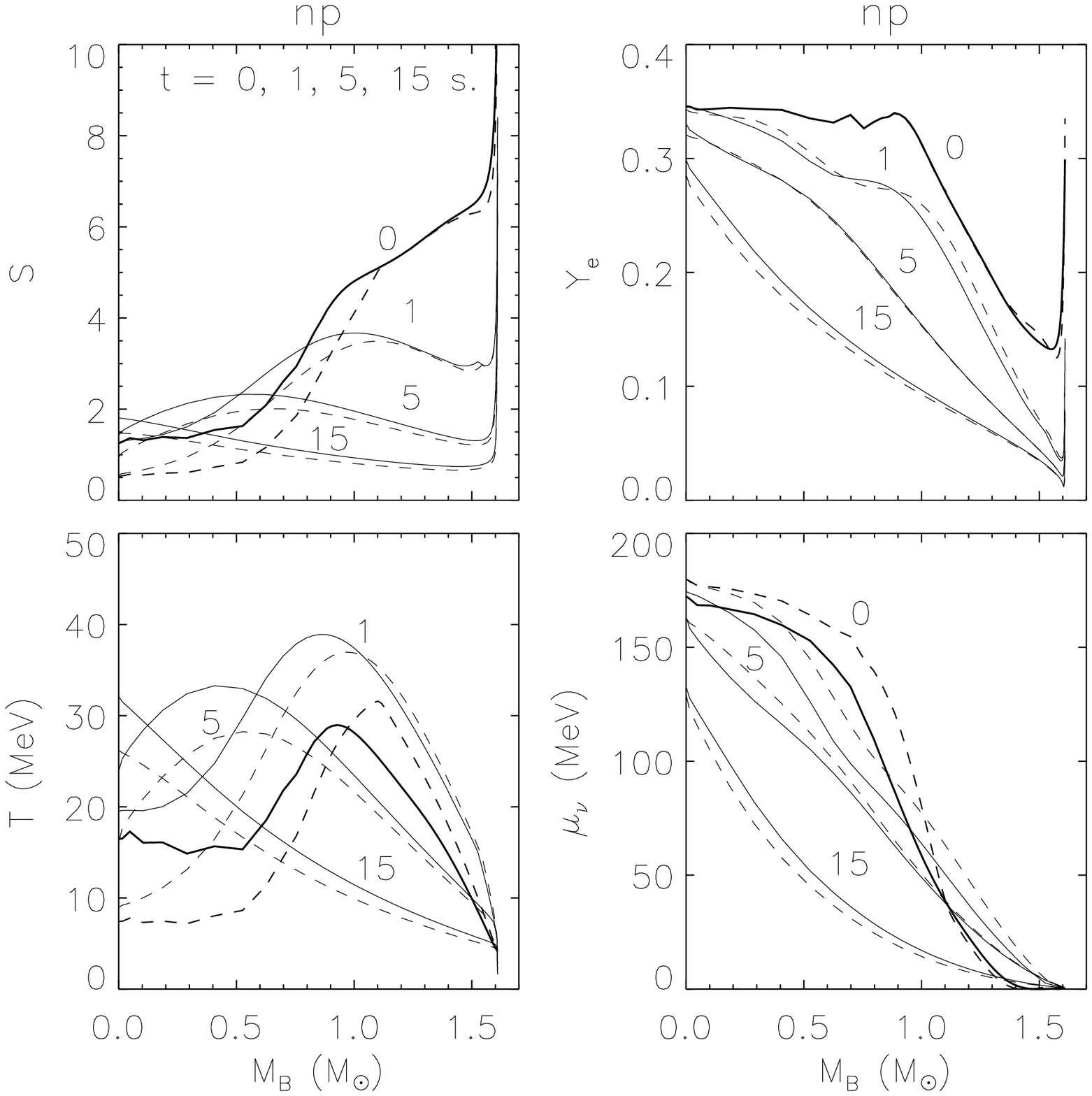}
\end{center}
\caption{}
{\label{lows}}
\end{figure}

\begin{figure}
\begin{center}
\epsfxsize=6.in
\epsfysize=7.in
\epsffile{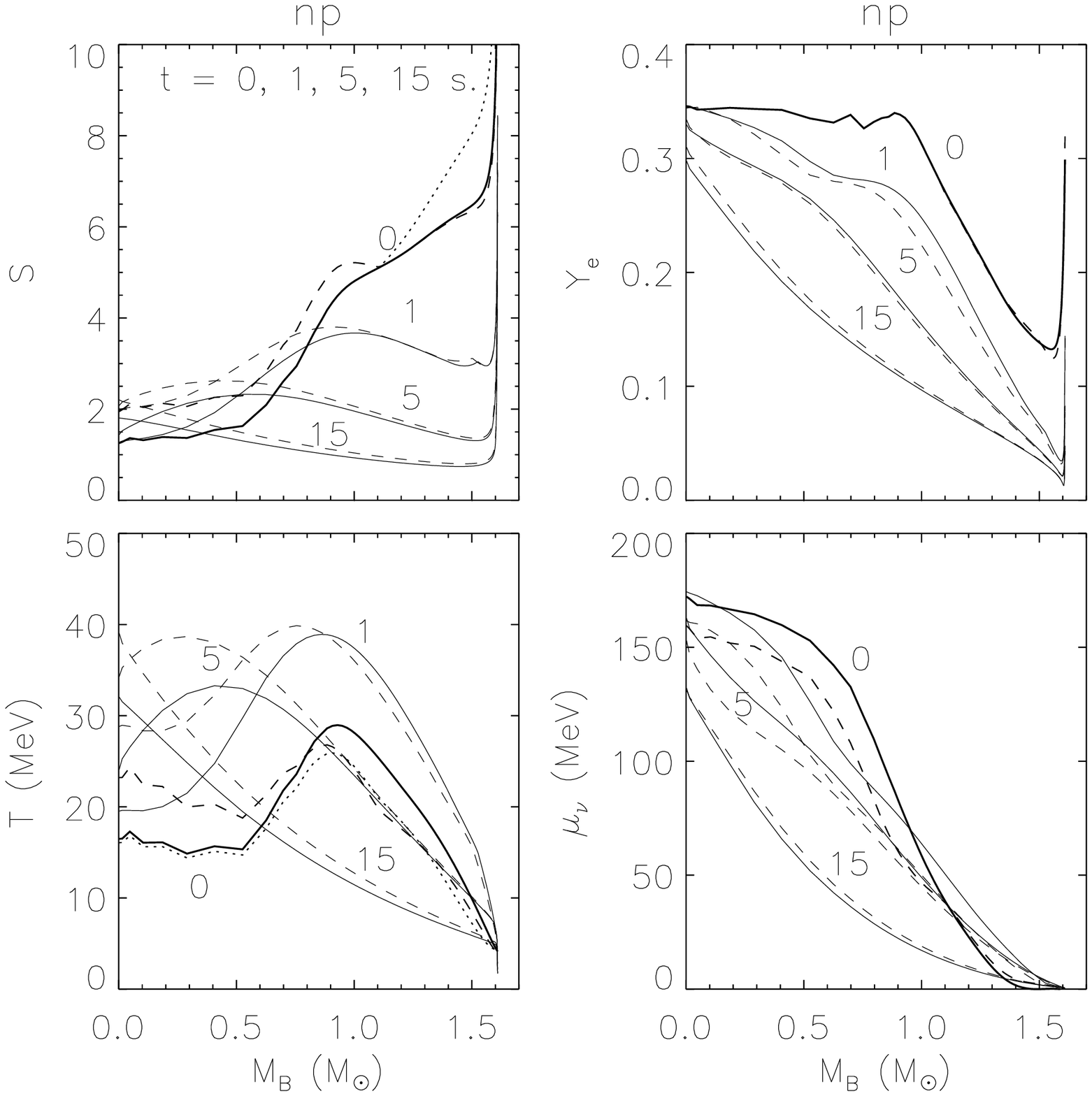}
\end{center}
\caption{}
{\label{highs}}
\end{figure}

\begin{figure}
\begin{center}
\epsfxsize=6.in
\epsfysize=7.in
\epsffile{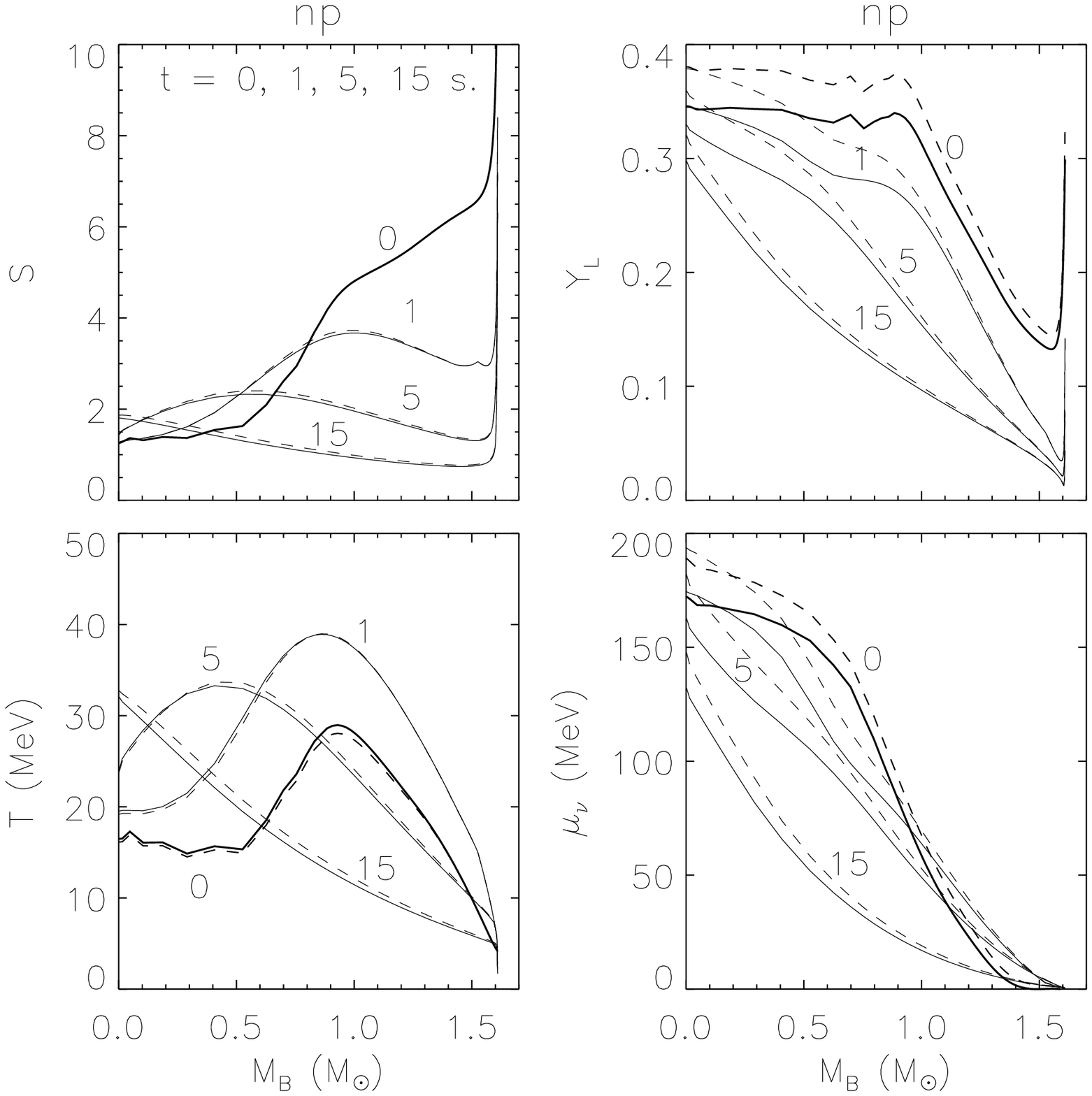}
\end{center}
\caption{}
{\label{highyl}}
\end{figure}

\begin{figure}
\begin{center}
\epsfxsize=6.in
\epsfysize=7.in
\epsffile{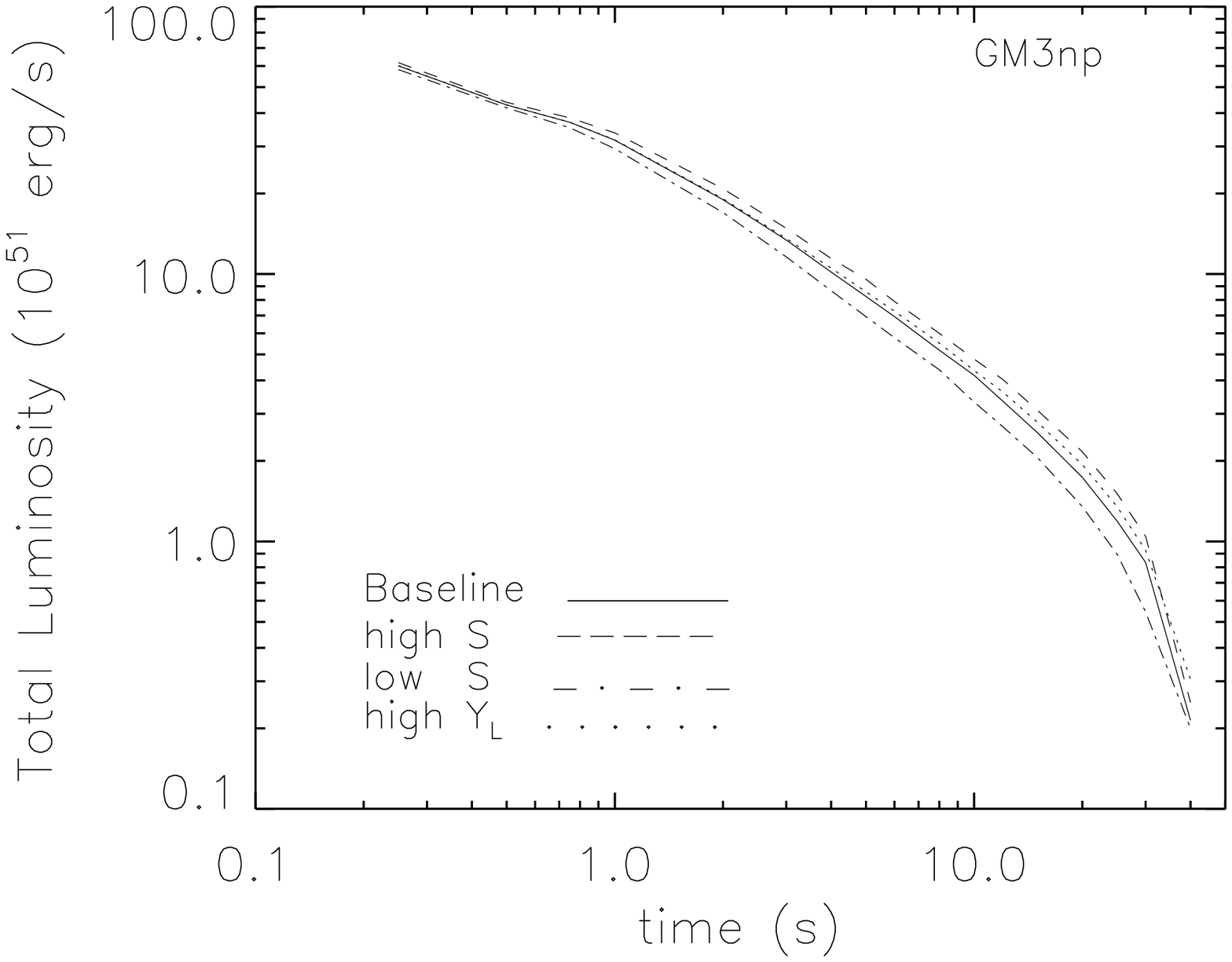}
\end{center}
\caption{}
{\label{lumeini}}
\end{figure}

\begin{figure}
\begin{center}
\epsfxsize=6.in
\epsfysize=7.in
\epsffile{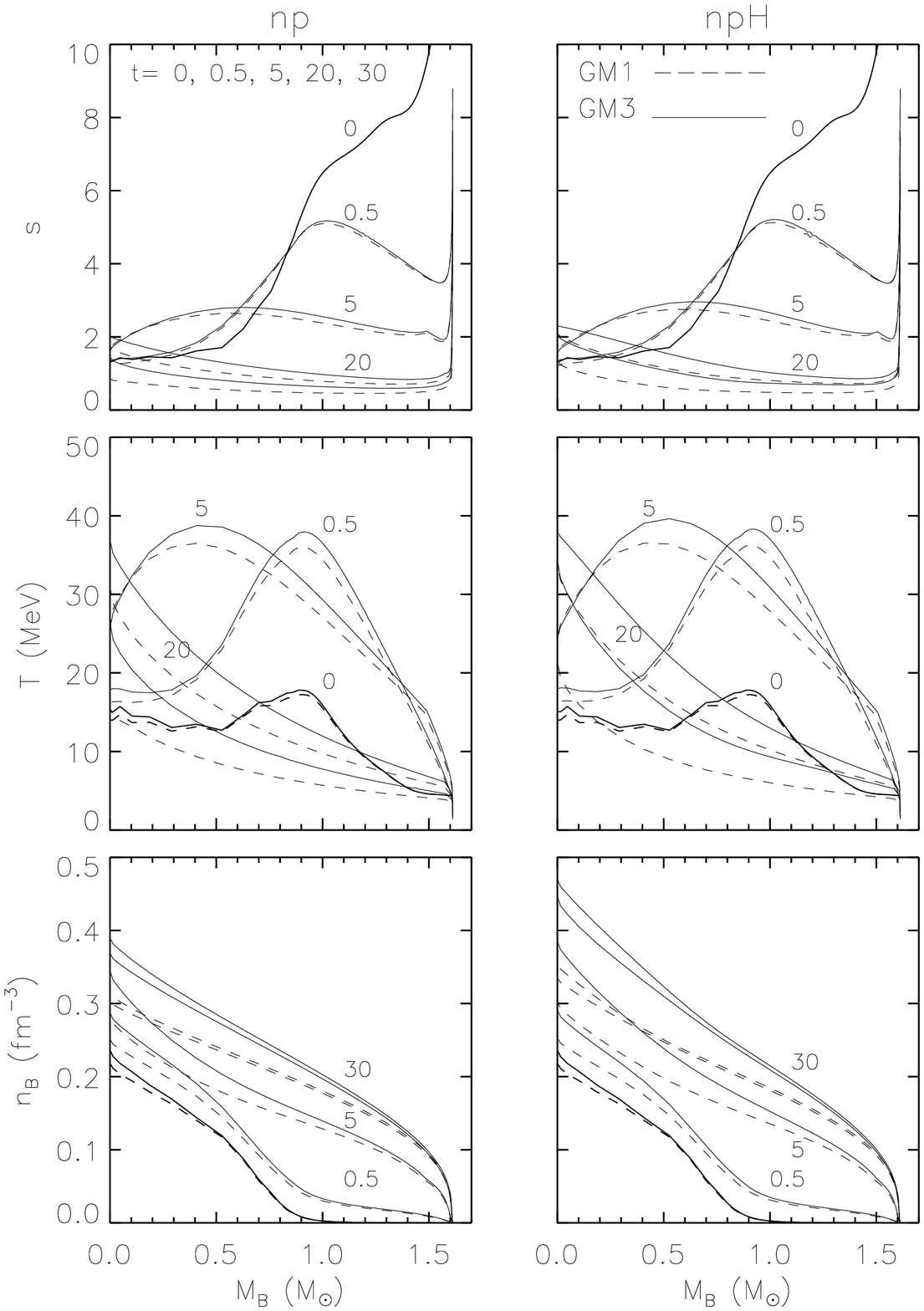}
\end{center}
\caption{}
{\label{therm_eos}}
\end{figure}

\begin{figure}
\begin{center}
\epsfxsize=6.in
\epsfysize=7.in
\epsffile{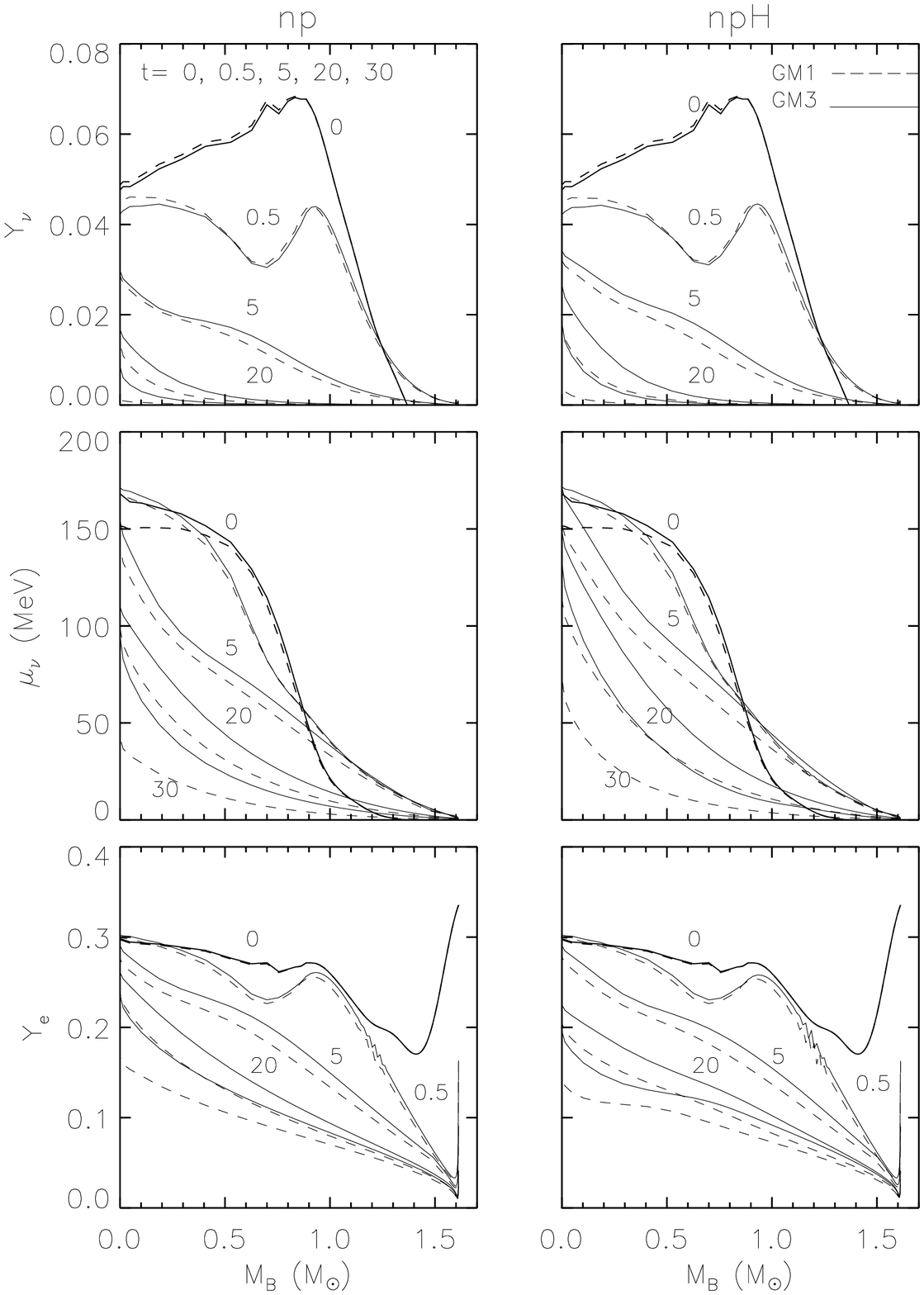}
\end{center}
\caption{}
{\label{lept_eos}}
\end{figure}

\begin{figure}
\begin{center}
\epsfxsize=6.in
\epsfysize=7.in
\epsffile{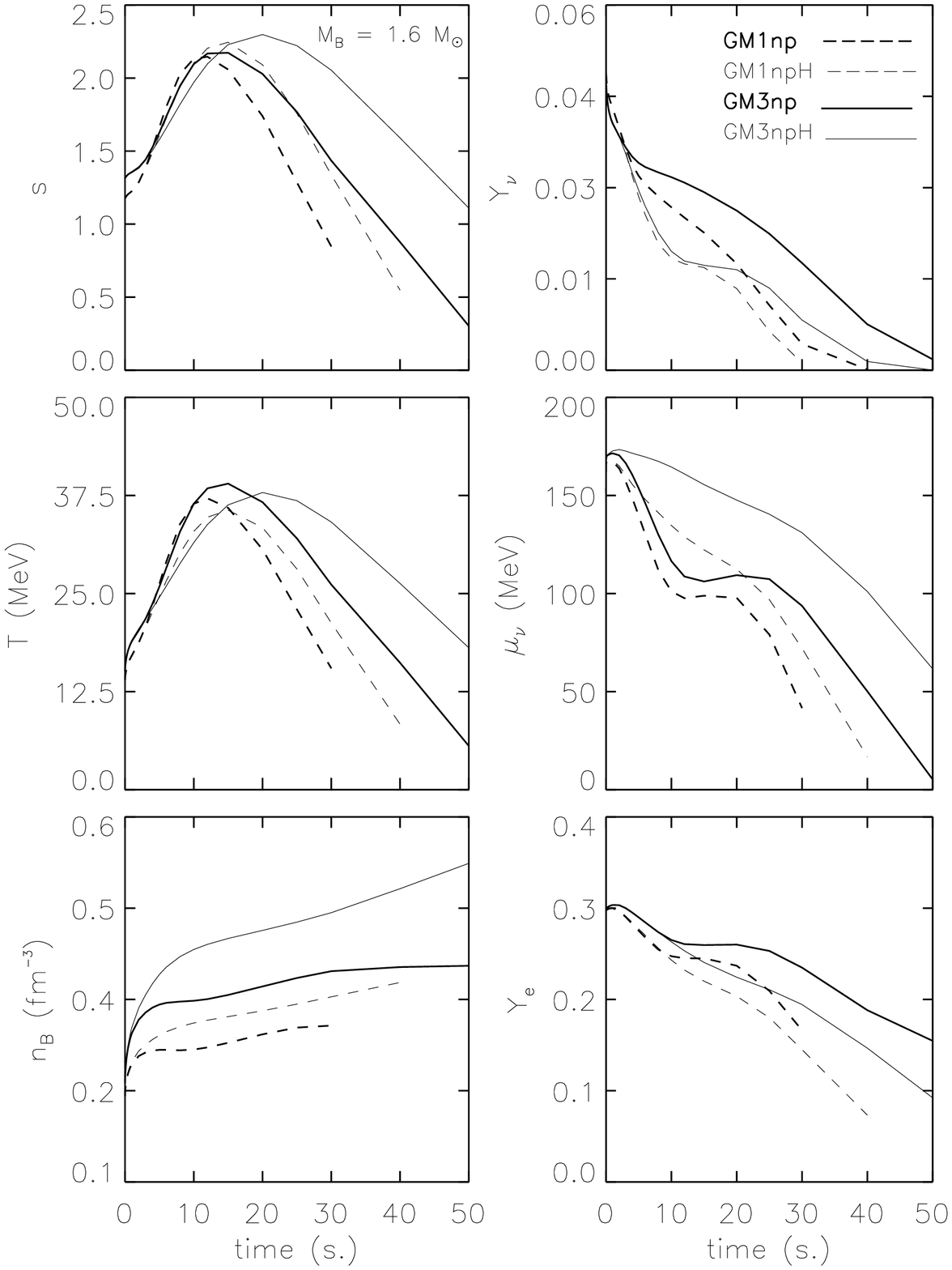}
\end{center}
\caption{}
{\label{mod_cquant}}
\end{figure}

\begin{figure}
\begin{center}
\epsfxsize=6.in
\epsfysize=7.in
\epsffile{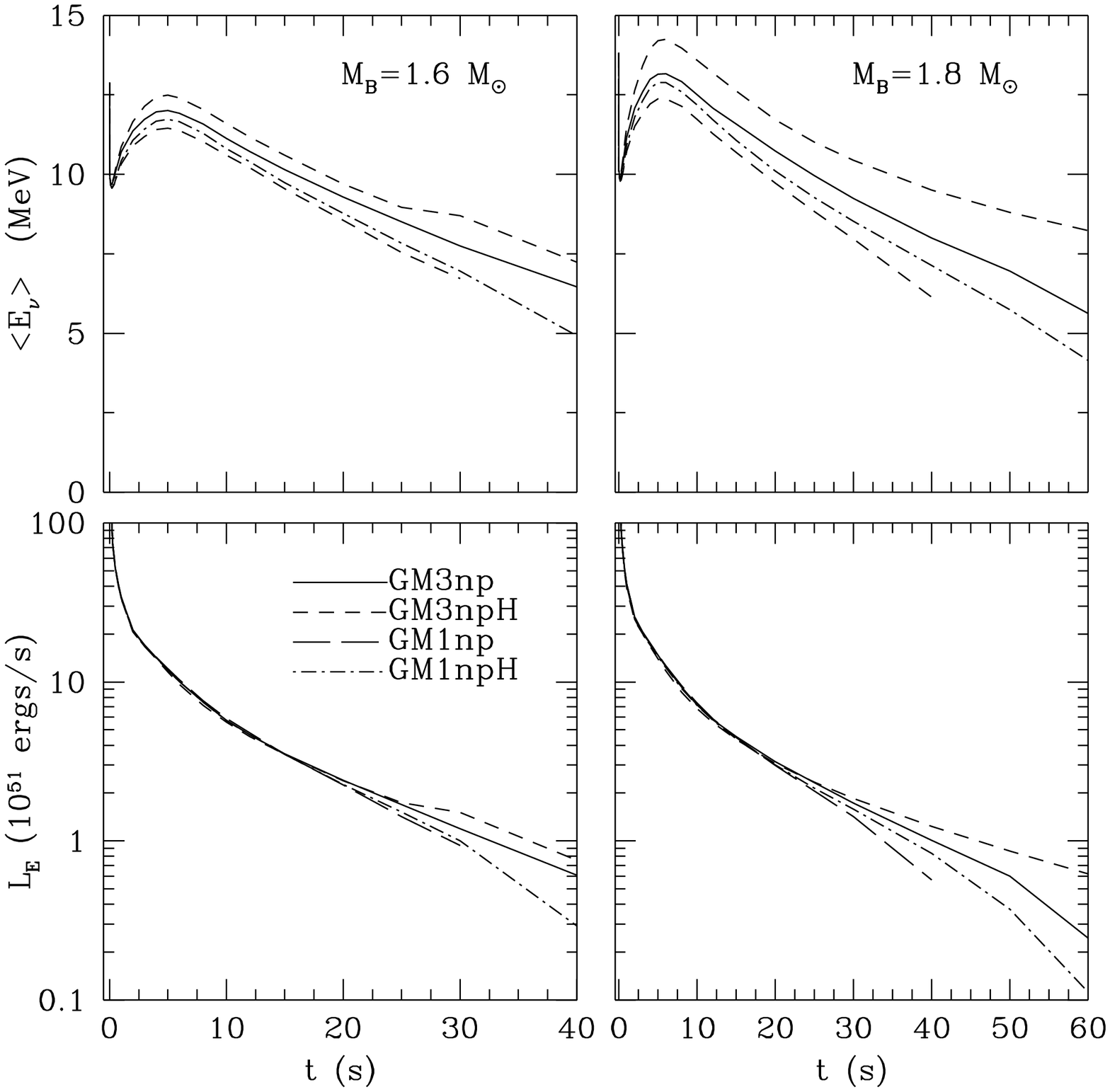}
\end{center}
\caption{}
{\label{models}}
\end{figure}

\begin{figure}
\begin{center}
\epsfxsize=6.in
\epsfysize=7.in
\epsffile{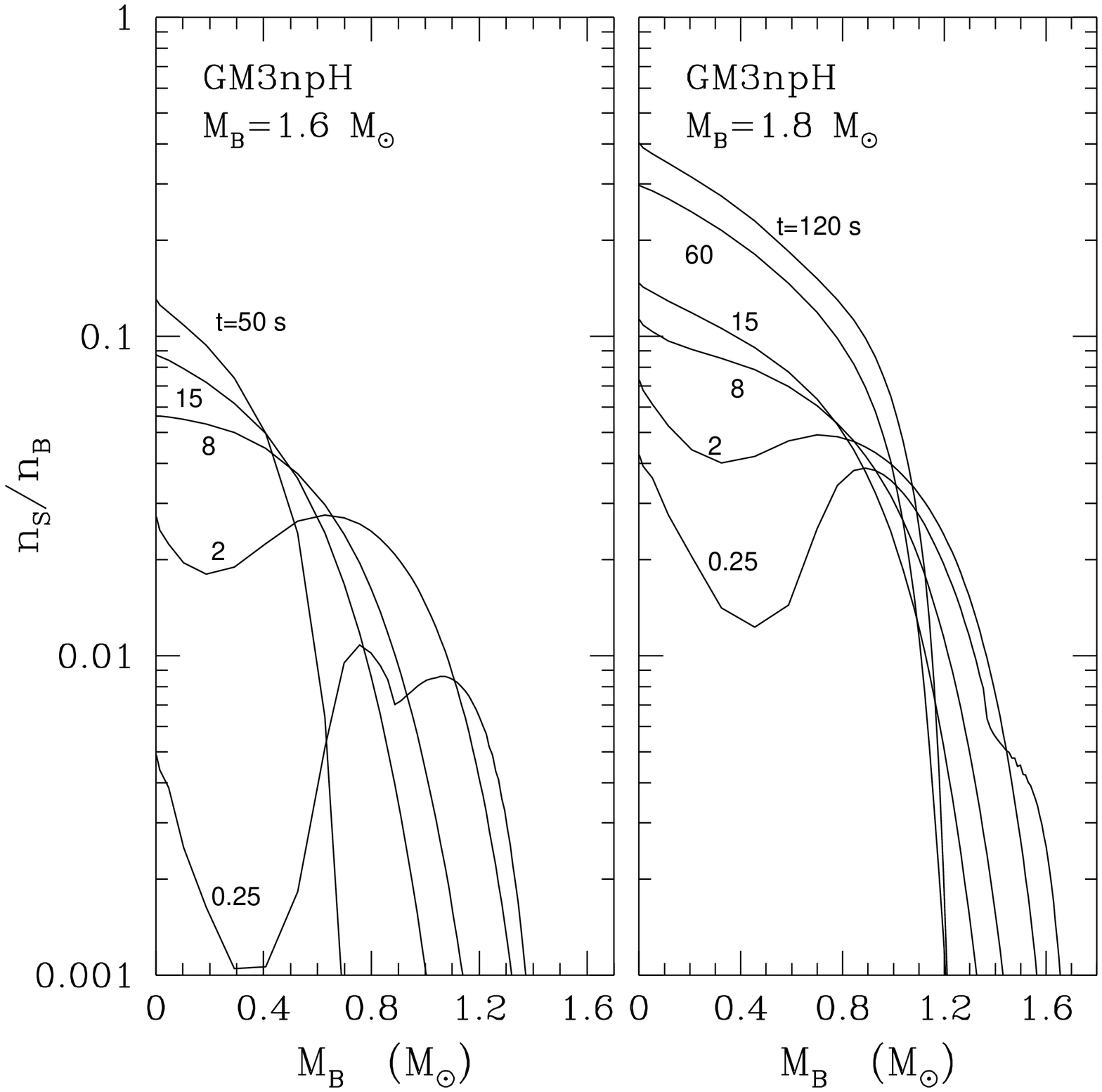}
\end{center}
\caption{}
{\label{hyperons}}
\end{figure}

\begin{figure}
\begin{center}
\epsfxsize=6.in
\epsfysize=7.in
\epsffile{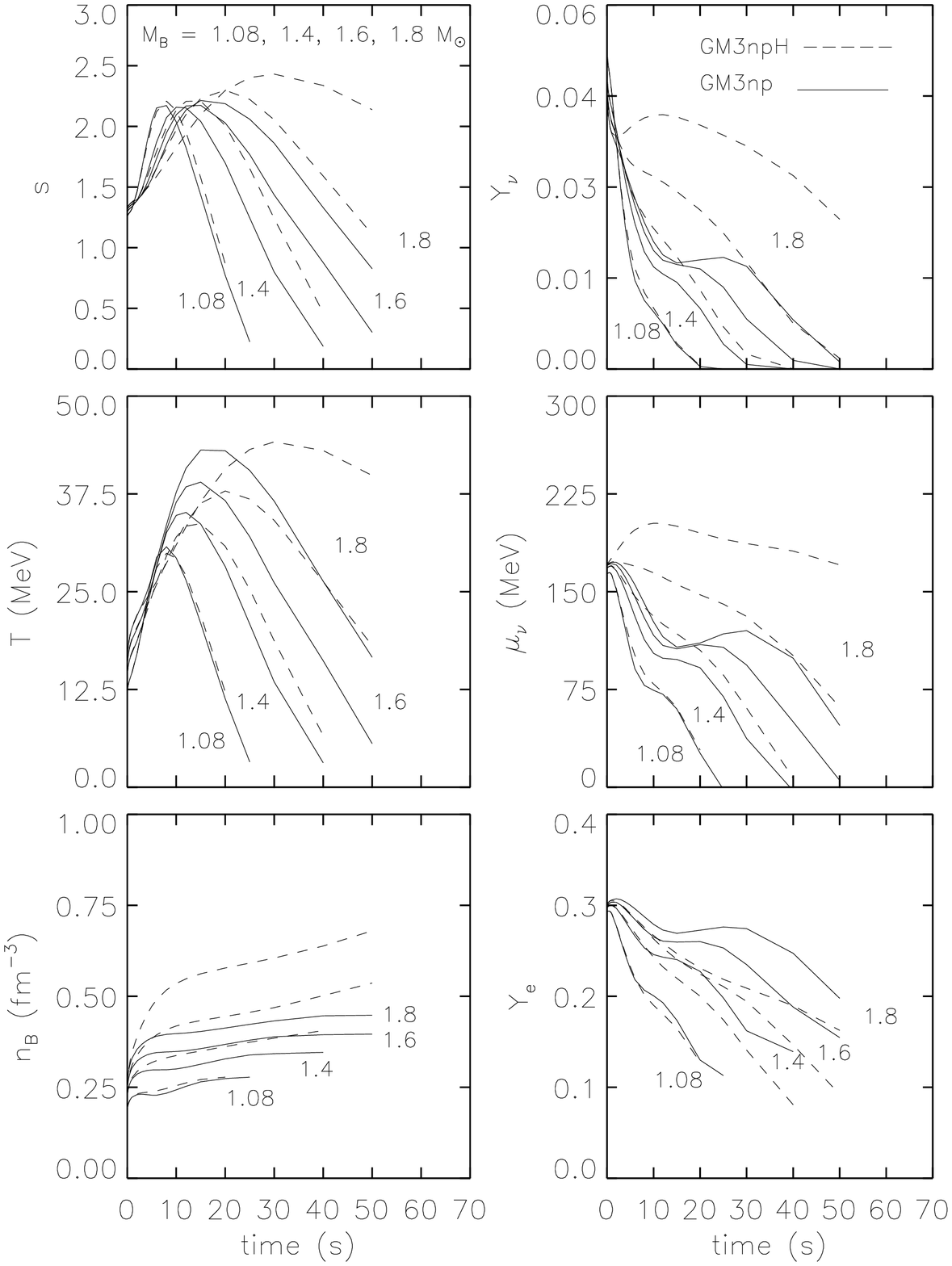}
\end{center}
\caption{}
{\label{mass_cquant}}
\end{figure}

\begin{figure}
\begin{center}
\epsfxsize=6.in
\epsfysize=7.in
\epsffile{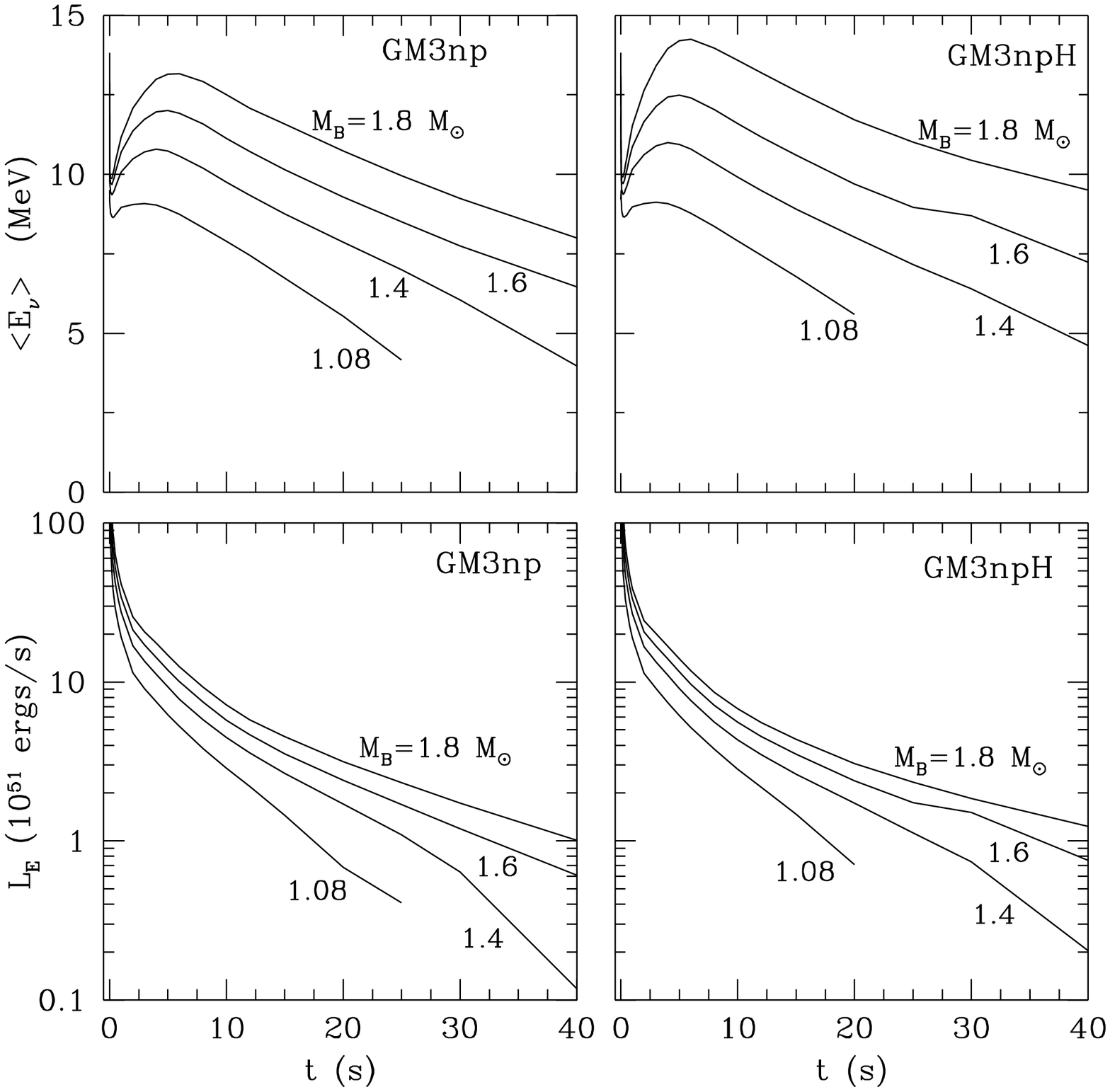}
\end{center}
\caption{}
{\label{mass}}
\end{figure}

\begin{figure}
\begin{center}
\epsfxsize=6.in
\epsfysize=7.in
\epsffile{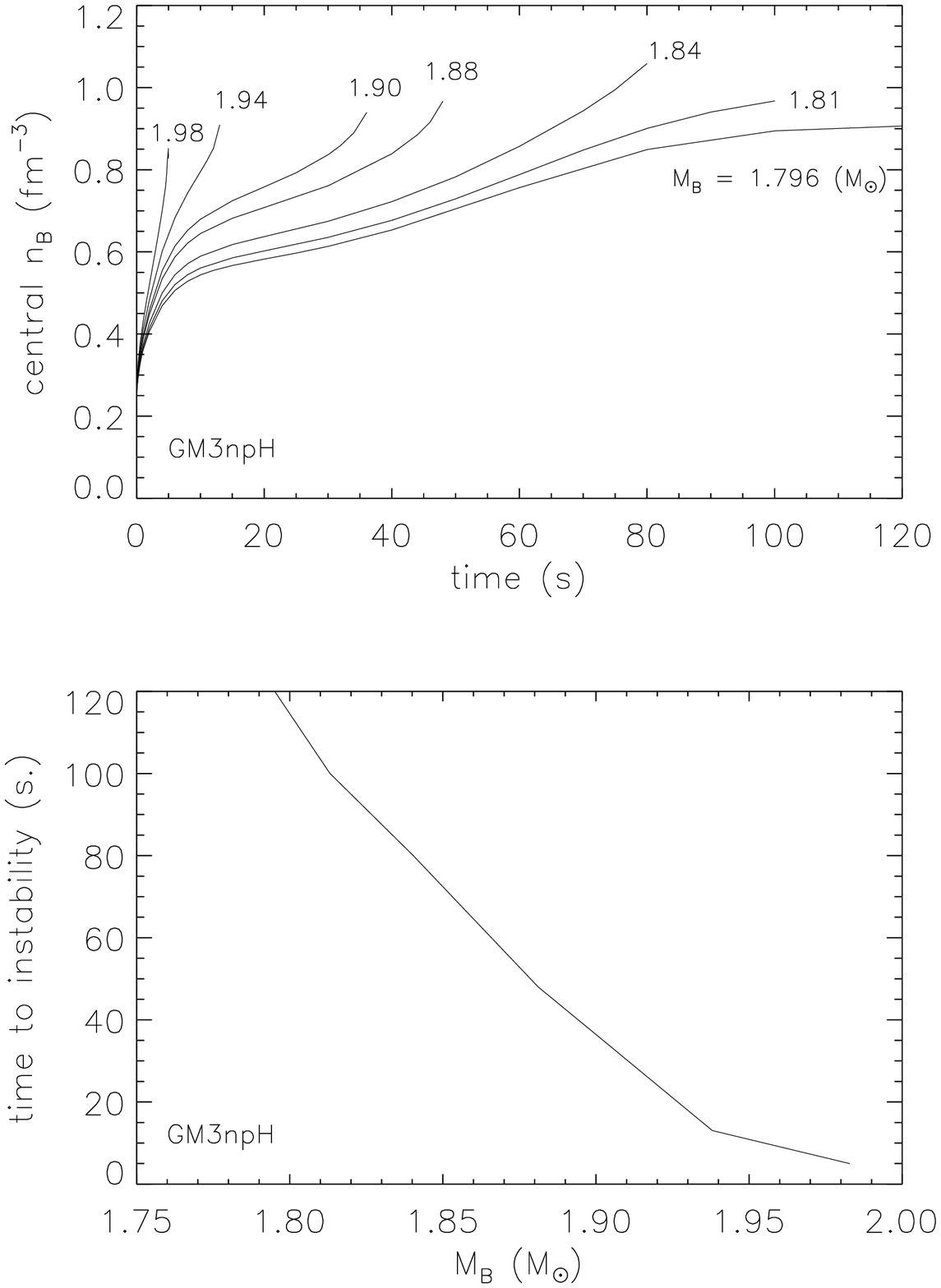}
\end{center}
\caption{}
{\label{bhole_ct}}
\end{figure}

\begin{figure}
\begin{center}
\epsfxsize=6.in
\epsfysize=7.in
\epsffile{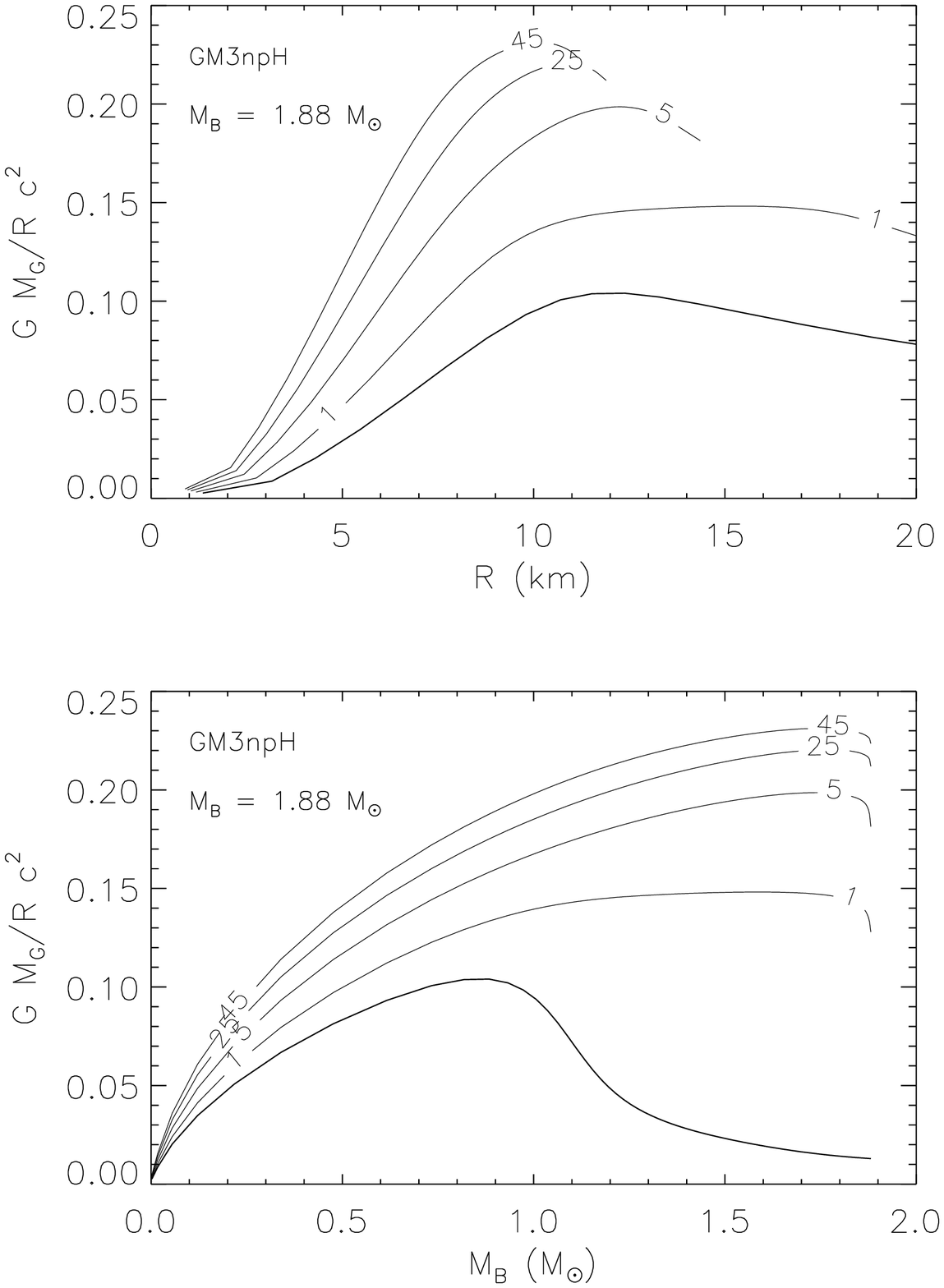}
\end{center}
\caption{}
{\label{bhole_mr}}
\end{figure}

\begin{figure}
\begin{center}
\epsfxsize=6.in
\epsfysize=7.in
\epsffile{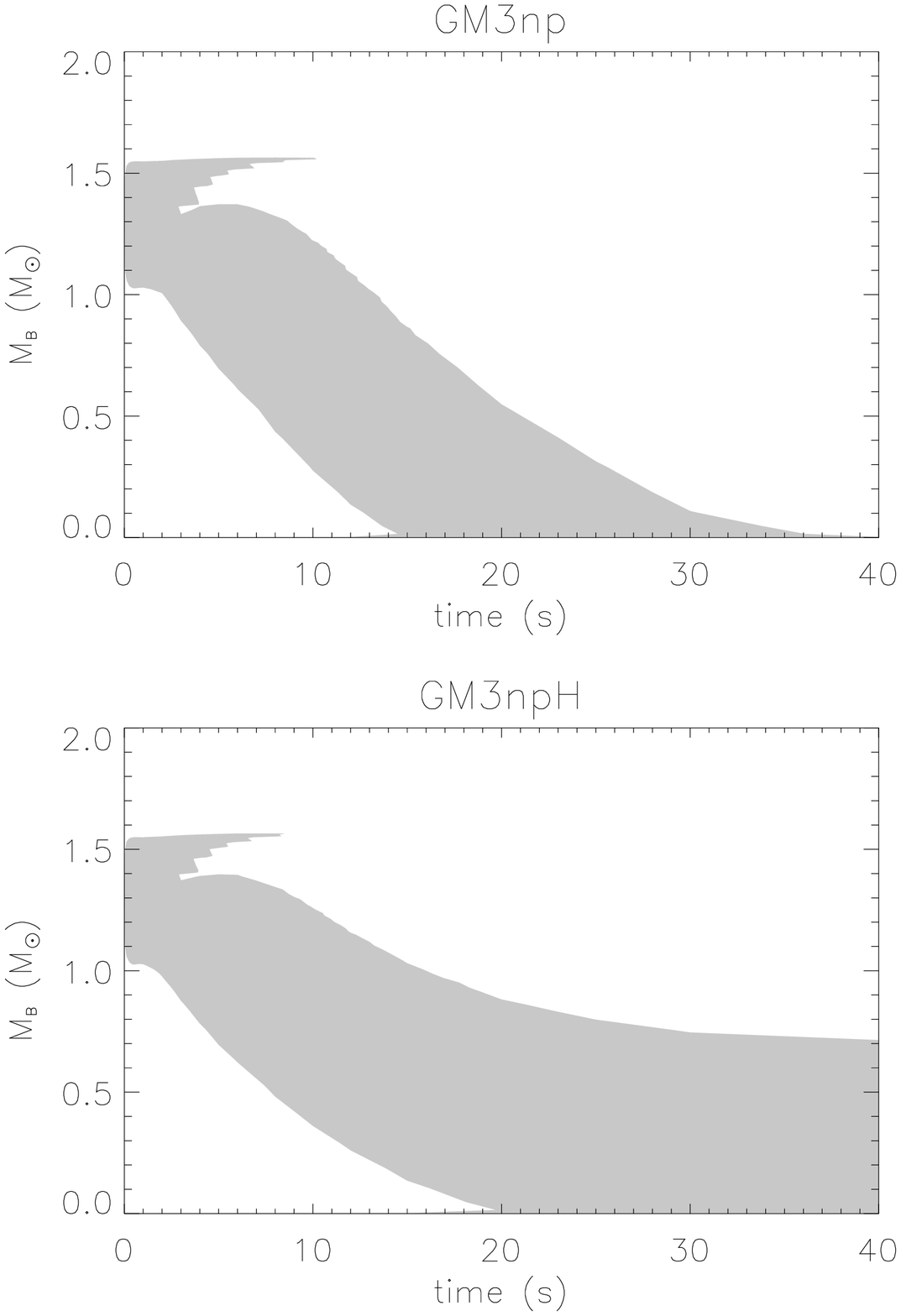}
\end{center}
\caption{}
{\label{convec}}
\end{figure}

\begin{figure}
\begin{center}
\epsfxsize=6.in
\epsfysize=7.in
\epsffile{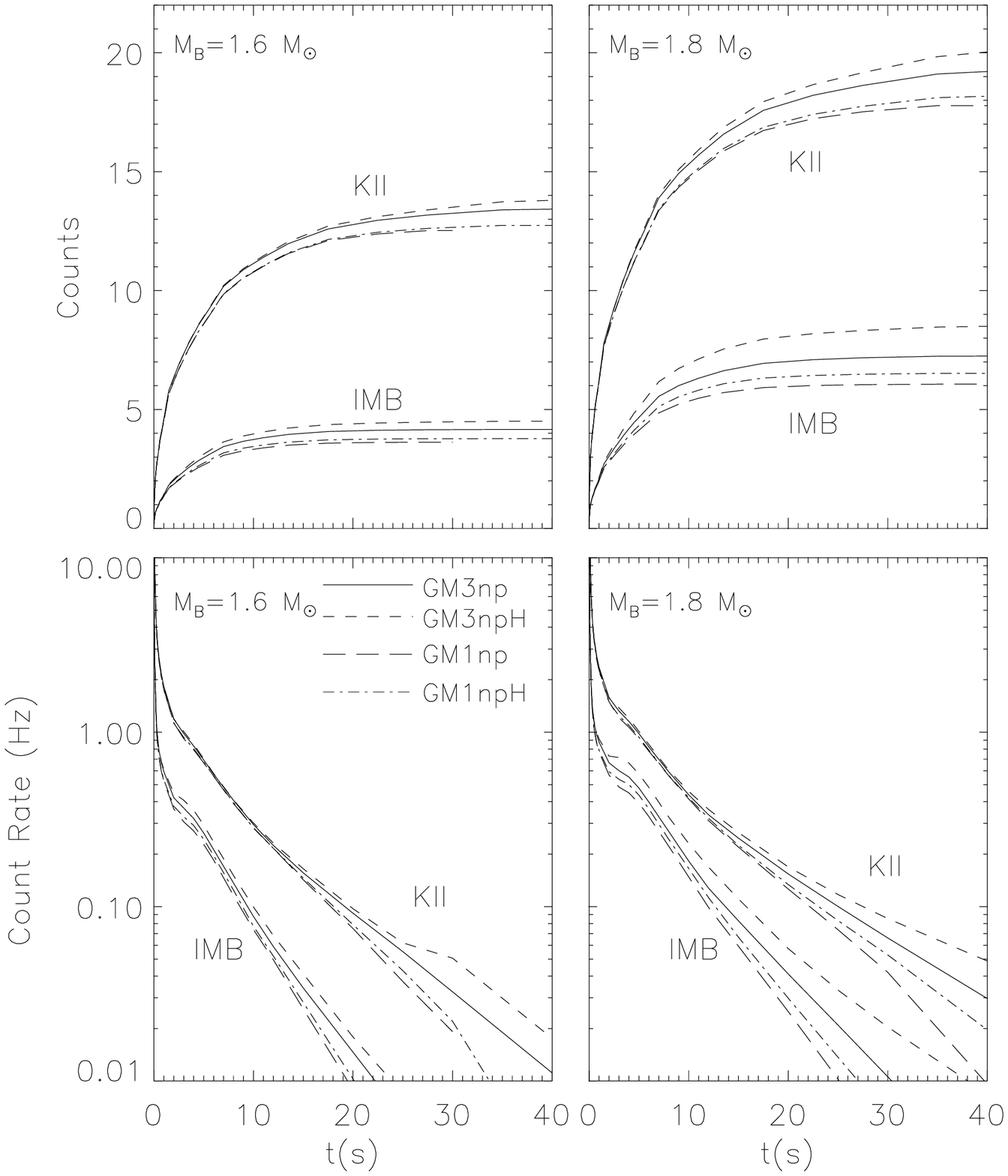}
\end{center}
\caption{}
{\label{count}}
\end{figure}

\noindent FIG. 11.-- Effects of changing initial entropy profiles.  Results
of the baseline GM3 model (solid curves)  with nucleons only with an inital $s
\sim 1$ in the core are compared  to those (dashed curves)  of a model with
reduced initial entropy in  the core ($s \sim 0.5$). Symbols as in
Figure \ref{baseline}.
\vspace*{1cm}

\noindent FIG. 12.-- Effects of changing initial entropy profiles.  Results of
the baseline GM3 model (solid curves)  with nucleons only with an inital $s
\sim 1$ in the core are compared to those (dashed curves) of a model with an
enhanced initial entropy in  the core ($s \sim 2$).  The dotted curves show
effects of enhancing the entropy in the surface regions. Symbols as in
Figure \ref{baseline}.
\vspace*{1cm}

\noindent FIG. 13.-- Effects of changing initial lepton profiles.  Results of
the baseline GM3 model (solid curves)  with nucleons only with an inital
$Y_L\cong 0.38$  in the core are compared to those (dashed curves) of a model
with an enhanced initial lepton content in  the core ($Y_L\cong 0.4$).  Symbols
as in Figure \ref{baseline}.
\vspace*{1cm}

\noindent FIG. 14.-- Comparison of total neutrino energy luminosity versus
time between the baseline model and different initial models.
\vspace*{1cm}


\noindent FIG. 15.-- Temporal variation of the entropy, temperature and baryon
density in matter containing nucleons only (left panels) and in matter that
also contains hyperons (right panels).  Differences arising from the underlying
EOS are  apparent by comparing the solid and dashed curves.  The abscissa shows
the enclosed baryonic mass in a $1.6{\rm M}_\odot$ star.
\vspace*{1cm}

\noindent FIG. 16.-- Temporal variation of the net electron neutrino fraction,
electron neutrino chemical potential, and the net electron concentration
in matter containing nucleons only (left panels) and in matter that
also contains hyperons (right panels).  Differences arising from the underlying
EOS are  apparent by comparing the solid and dashed curves.  The abscissa shows
the enclosed baryonic mass in a $1.6{\rm M}_\odot$ star.
\vspace*{1cm}

\noindent FIG. 17.-- Effects of changing the underlying EOS (GM1 versus GM3) on
the core values of the thermodynamic quantities versus time in matter
containing nucleons only  (thin solid and dashed curves) and in matter that
also contains hyperons (thick solid and dashed curves).
Results are for a $M_B=1.6{\rm M}_\odot$ star.
\vspace*{1cm}

\noindent FIG. 18.-- Comparison of the mean neutrino energy (top panels) and
the total neutrino energy luminosity (bottom panels) versus time for the
different EOS models. The left panels show results for a
baryon mass $M_B=1.6 M_\odot$ and the right panels for a baryon
mass $M_B=1.8 M_\odot$.
\vspace*{1cm}

\noindent FIG. 19.-- Temporal evolution of strangeness per baryon for the
indicated baryon masses.
\vspace*{1cm}

\noindent FIG. 20.-- Mass dependence of the core values of the thermodynamic
quantities versus time in matter containing nucleons only  (solid curves) and
in matter that also contains hyperons (dashed curves).
\vspace*{1cm}

\noindent FIG. 21.-- Mass dependence of the average energy of the
emitted  neutrinos (top panels) and the total neutrino energy
luminosity (bottom panels) in  models with and without hyperons.
\vspace*{1cm}

\noindent FIG. 22.-- Top panel: Evolution of the central baryon number density
for different baryonic mass stars containing hyperons (model GM3npH) which
are metastable. Bottom panel: Time required by stars shown in the top panel
to reach the unstable configuration.
\vspace*{1cm}

\noindent FIG. 23.-- Evolution of  ($GM_G/Rc^2$) in a
$M_B=1.88{\rm M}_\odot$ star containing hyperons (model GM3npH). Inset labels
show time in seconds.
\vspace*{1cm}

\noindent FIG. 24.-- Evolution of convectively unstable regions in a
$M_B=1.6{\rm M}_\odot$ star.  Top panel shows results for a star containing
nucleons only and the bottom panel is for a star that also contains hyperons.
\vspace*{1cm}

\noindent FIG. 25.-- Neutrino signals in terrestrial detectors KII and IMB,
and for a supernova at 50 kpc. Top panels show total number of counts,
and the bottom panels show the count rate. Left panels are for
M$_B$=1.6 M$_\odot$, and right panels show results for M$_B$=1.8 M$_\odot$.

\newpage

\newpage

\newpage

\newpage

\newpage

\newpage

\newpage

\end{document}